\documentclass[11pt]{article}
 \usepackage{benstyle}
 
\newcommand{\Romnum}[1]{\uppercase\expandafter{\romannumeral#1}}
\newcommand{\romnum}[1]{\romannumeral#1\relax}
 
\begin{document}

\title{Uniform Recovery from Subgaussian Multi-Sensor Measurements}
\author{Il Yong Chun \\ Department of Electrical Engineering and Computer Science \\ University of Michigan \\ USA \and Ben Adcock \\ Department of Mathematics \\ Simon Fraser University \\ Canada}
\maketitle

\begin{abstract}
Parallel acquisition systems are employed successfully in a variety of different sensing applications when a single sensor cannot provide enough measurements for a high-quality reconstruction.  In this paper, we consider compressed sensing (CS) for parallel acquisition systems when the individual sensors use subgaussian random sampling.  Our main results are a series of uniform recovery guarantees which relate the number of measurements required to the basis in which the solution is sparse and certain characteristics of the multi-sensor system, known as sensor profile matrices.  In particular, we derive sufficient conditions for optimal recovery, in the sense that the number of measurements required per sensor decreases linearly with the total number of sensors, and demonstrate explicit examples of multi-sensor systems for which this holds.  We establish these results by proving the so-called Asymmetric Restricted Isometry Property (ARIP) for the sensing system and use this to derive both nonuniversal and universal recovery guarantees.  Compared to existing work, our results not only lead to better stability and robustness estimates but also provide simpler and sharper constants in the measurement conditions.  Finally, we show how the problem of CS with block-diagonal sensing matrices can be viewed as a particular case of our multi-sensor framework.  Specializing our results to this setting leads to a recovery guarantee that is at least as good as existing results.
\end{abstract}

\section{Introduction}
In compressed sensing (CS), it has been conventional to consider a single sensor acquiring measurements of signal.  Assuming a finite-dimensional and linear model, this can be viewed as the problem of recovering an unknown vector $f \in \bbC^{N}$ from $m \ll N$ noisy measurements
\bes{
y = \tilde{A} f + n,
}
where $\tilde{A} \in \bbC^{m \times N}$ is a matrix representing the measurements taken by the sensor and $n \in \bbC^{m}$ is noise.  Typically, $f$ has a sparse representation in some known orthonormal sparsifying basis (referred to as a \textit{sparsity basis}), represented as a unitary matrix $U \in \bbC^{N \times N}$.  That is, $f = U x$, where the vector $x \in \bbC^N$ is sparse, or compressible.  In this case, one may replace \R{standard_CS} by
\be{
\label{standard_CS}
y = A x + n,
}
where $A = \tilde{A} U$, and consider the equivalent problem of recovering $x$ from \R{standard_CS}.  Provided the matrix $A$ satisfies an appropriate condition -- for example, the \textit{Restricted Isometry Property (RIP)} -- then $x$ can be recovered stably and robustly from the measurements $y$.  For example, if a bound for the noise is available, i.e.\ $\| n \|_{2} \leq \eta$ for some known $\eta$, then one may solve the $\ell^1$ minimization problem
\be{
\label{recovery_alg}
\min_{z \in \bbC^N} \| z \|_{1} \ \mbox{subject to $\| A z - y \|_2 \leq \eta$}.
}
As is now well known, it is possible to find matrices $A$ which satisfy the RIP with a number of measurements $m$ that scales linearly in the sparsity $s$ of $x$ and logarithmically (or polylogarithmically) in the ambient dimension $N$.  Examples include subgaussian random matrices, randomly subsampled isometries, partial random circulant matrices and so on.

\subsection{Compressed sensing and parallel acquisition}\label{ss:CSPA}
In this paper, we consider the generalization of \R{standard_CS} to a so-called \textit{parallel acquisition} system, introduced recently in \cite{Chun&Adcock:17TIT}, where, rather than a single sensor, $C \geq 1$ sensors simultaneously measure $x$.  Mathematically, one can model the measurement process in such problems as
\be{
\label{parallel_CS}
y_{c} = A_c x + n_c,\qquad c=1,\ldots,C,
}
where $A_{c} \in \bbC^{m_c \times N}$ is the measurement matrix in the $c^{\rth}$ sensor and $n_c \in \bbC^{m_c}$ is noise.  Typically, the matrices $A_1,\ldots,A_C$ are assumed to take the following form:
\bes{
A_c = \tilde{A}_{c} H_c U,
}
where $\tilde{A}_{c} \in \bbC^{m_c \times N}$ are standard CS measurement matrices, $U \in \bbC^{N \times N}$ is the sparsity basis, and the $H_c \in \bbC^{N \times N}$ are certain deterministic matrices, referred to as \textit{sensor profile} matrices \cite{Chun&Adcock:17TIT,Chun&Adcock:16ITW,Chun&Li&Adcock:16MMSPARSE}.  These matrices model environmental conditions specific to the particular sensing problem; for example, a communication channel between the signal $f = U x$ and the sensors, the geometric position of the sensors relative to $f$, or the effectiveness of the sensors to $f$.  In particular, they are usually fixed by the given application.  Letting
\be{
\label{Adef}
A = \left [ \begin{array}{c} A_1 \\ \vdots \\ A_C \end{array} \right ],\qquad y =  \left [ \begin{array}{c} y_1 \\ \vdots \\ y_C \end{array} \right ],\qquad n =  \left [ \begin{array}{c} n_1 \\ \vdots \\ n_C \end{array} \right ],
}
allows one to recast \R{parallel_CS} in the form \R{standard_CS}.  Hence, if $A$ satisfies an RIP, then one can recover $x$ from the multi-sensor measurements using a standard CS procedure (e.g.\ $\ell^1$ minimization \R{recovery_alg}).

In this paper we study the case where the matrices $\tilde{A}_{c}$ in the individual sensors  are subgaussian random matrices.  Our intention is to derive conditions on the total number of measurements $m = m_1+\ldots+m_C$ for the matrix $A$ to satisfy an RIP-like property.  As we shall document, these conditions depend on the both the sensor profile matrices $H_c$ and the sparsity basis $U$.  Yet we will identify broad classes of these matrices for which $A$ satisfies a suitable RIP-like property for a near-optimal number of measurements $m$: that is, growing linearly in $s$ and (poly)logarithmically in $N$, but crucially \textit{independent} of of the number of sensors $C$.

Such independence is critical for applications.  In particular, it implies that that the average number of measurements per sensor $m_{\mathrm{avg}} = C^{-1} \sum^{C}_{c=1} m_{c}$ decreases linearly with increasing $C$.  This demonstrates the benefits of multi-sensor over single-sensor systems: doubling the number of sensors can effectively halve the acquisition time, or power or cost, depending on what the constraining factor is in the given application.  See \S \ref{ss:applications} for further details.  

Within this parallel acquisition setup, we consider two particular classes of problem, both of which arise in applications.  These were introduced in \cite{Chun&Adcock:17TIT} and termed \textit{distinct} and \textit{identical} sampling respectively.  In the former, the matrices $\tilde{A}_1,\ldots,\tilde{A}_C$ are independent subgaussian random matrices with the same subgaussian parameter.  In the latter, we assume that $m_1 = \ldots = m_C = m/C$ and $\tilde{A}_{1} = \ldots = \tilde{A}_C = \tilde{A} \in \bbC^{m/C \times N}$, where $\tilde{A}$ is a subgaussian random matrix.  In other words, the only differences between individual sensors in the identical case are the sensor profile matrices $H_c$.  As one may expect, and as we shall document in this paper, optimal recovery in the identical case requires more stringent conditions on the sensor profile matrices than in the distinct case.

\subsection{Applications}\label{ss:applications}
Parallel acquisitions systems have been used to provide significant benefits in various practical applications.  These benefits include scan time reduction in parallel Magnetic Resonance Imaging (MRI), power consumption reduction in Wireless Sensor Networks (WSN), recovery of a high number of non-zeros in the signal with low-sampling-rate devices in Synthetic Aperture Radar (SAR) imaging, or recovery of higher-resolution or higher-dimensional signals in multi-view imaging or light-field imaging.  For example, the most general system model in parallel MRI can be viewed as an example of identical sampling with diagonal sensor profiles \cite{Chun&Adcock&Talavage:15TMI,Chun&Adcock&Talavage:14EMBS_pMRI,Pruessmann&Weiger&Scheidegger&Scheidegger:99MRM}.  Results presented in \cite[\S \Romnum{4}-B]{Chun&Adcock:17TIT} demonstrate the benefits of parallel over single-coil MRI: the scan time (roughly proportional to the number of measurements per sensor) can be reduced linearly by increasing the number of coils $C$, without affecting the reconstruction fidelity.

Other applications to which the above framework applies are discussed in more depth in \cite{Chun&Adcock:17TIT}.  In passing we mention multiview imaging \cite{Park&Wakin:12EJASP, Traonmilin&etal:15JMIV}, super-resolution imaging \cite{Baboulaz&Dragotti:09TIP, Duarte&Eldar:11SP, Jiang&Huang&Wilford:14APSIP-SIP}, the recovery of sparse signals from low sampling rate sensors (with application to SAR imaging, for example) \cite{Aceska&etal:16Arxiv}, system identification in dynamical systems \cite[Chpt.\ $6$]{Sanandaji:12PhD}, and light-field imaging \cite{Nien:14PhD}.

\subsection{Contributions} \label{sec:contrib}
For a random matrix to satisfy an RIP, it is conventional to require that the expected value $\bbE(m^{-1} A^* A) = I$.  In the case where $A$ is given by \R{Adef}, this is equivalent (after rescaling) to the so-called \textit{joint isometry condition} \cite{Chun&Adcock:17TIT}, given by
\be{
\label{joint_iso}
C^{-1} \sum^{C}_{c=1} H^*_c H_c = I.
}
This condition can be quite restrictive for applications: it may be impossible, or at best difficult, to design the matrices $H_c$ so that it holds.  Hence in this paper we allow a substantial relaxation of \R{joint_iso} to the \textit{joint near-isometry condition}
\be{
\label{joint_near_iso}
\alpha I \preceq C^{-1} \sum^{C}_{c=1} H^*_c H_c\preceq \beta I.
}
Here $\beta \geq \alpha >0$ are constants and the notation $B\preceq A$ means that $A-B$ is positive semi-definite.  Rather than the classical RIP, we shall consider the so-called \textit{Asymmetric Restricted Isometry Property (ARIP)} (see Definition \ref{d:ARIP}).  Like the RIP, this is also sufficient for stable and robust recovery using $\ell^1$ minimization (see Theorem \ref{t:ARIP_stable_robust}).

\subsubsection{Measurement conditions}
Our main contributions are the following conditions, which are sufficient for the matrix $A$ of \R{Adef} to satisfy the ARIP:
\be{
\label{intro_bounds}
\begin{aligned}
m &\gtrsim \delta^{-2}  \cdot \frac{\beta}{\alpha} \cdot  \Gamma_{\mathrm{distinct}}^2 \cdot s \cdot L_1,\qquad \mbox{(distinct sampling)},
\\
m &\gtrsim \delta^{-2} \cdot \frac{\beta}{\alpha} \cdot  \Gamma_{\mathrm{identical}}^2 \cdot s \cdot L_1, \qquad \mbox{(identical sampling)}. 
\end{aligned}
}
See Theorems \ref{t:dist:subgaussRIP} and \ref{t:idt:subgaussRIP} respectively.  Here $L_1$ is a log factor, equal to $L_1 =  \ln^2(2s) \ln(2N) \ln(2m) + \ln(2/\varepsilon)$ where $0 < \varepsilon < 1$ is the failure probability, and $\delta$ is a factor in the ARIP condition.  The terms $\Gamma_{\mathrm{distinct}}$ and $\Gamma_{\mathrm{identical}}$ are functions of the sensor profile matrices $H_{1},\ldots,H_C$ and sparsity basis $U$, given by
\eas{
\Gamma_{\mathrm{distinct}} &= \frac{1}{\sqrt{\alpha}} \max_{c=1,\ldots,C} \max_{j =1,\ldots,N} \| H_c U e_j \|_2,
\\
\Gamma_{\mathrm{identical}} &= \frac{1}{\sqrt{\alpha}} \max_{j=1,\ldots,N} \left\| \left[ \begin{array}{ccc} H_1 U e_j & \cdots & H_C U e_j \end{array} \right] \right\|_{2 \rightarrow 2},
}
respectively.  The bounds \R{intro_bounds} are optimal, provided the factors $\beta/\alpha$ and $\Gamma = \Gamma_{\mathrm{distinct}}$ or  $\Gamma = \Gamma_{\mathrm{identical}}$ are independent of the number of sensors $C$.  We present several classes of examples where this is the case.  Note that these factors are all easily computable, so optimality of a given configuration of sensor profiles $H_c$ and sparsity basis $U$ can be checked numerically. 
We also determine bounds for both $\Gamma_{\mathrm{distinct}}$ and $\Gamma_{\mathrm{identical}}$.  These are
\eas{
1 \leq \Gamma_{\mathrm{distinct}} \leq \sqrt{\beta/\alpha} \sqrt{C},\qquad 1 \leq \Gamma_{\mathrm{distinct}} \leq \Gamma_{\mathrm{identical}} \leq \sqrt{\beta/\alpha} C,
}
(see Propositions \ref{prop:GammaXi:bounds:dist} and \ref{prop:GammaXi:bounds:idt} respectively).  In particular, and as one would expect, the bound for identical sampling in \R{intro_bounds} is always larger than that of distinct sampling.

\subsubsection{Universal measurement conditions}
In the single-sensor model \R{standard_CS}, a key property of subgaussian random matrices is their \textit{universality}: the measurement condition is independent of the choice of sparsity basis $U$.  This is not the case in the bounds \R{intro_bounds}, since $\Gamma_{\mathrm{distinct}}$ and $\Gamma_{\mathrm{identical}}$ both depend on $U$.  However, we also prove the following universal bounds:
\be{
\label{intro_bounds_universal}
\begin{aligned}
m &\gtrsim \delta^{-2} \cdot \frac{\beta}{\alpha} \cdot  \Xi_{\mathrm{distinct}}^2 \cdot s \cdot L_2,\qquad \mbox{(distinct sampling)}
\\
m &\gtrsim \delta^{-2} \cdot \frac{\beta}{\alpha} \cdot \Xi_{\mathrm{identical}}^2 \cdot s \cdot L_2,\qquad \mbox{(identical sampling)}
\end{aligned}
}
where $L_2 =  \ln(2N/s) + s^{-1} \ln(2/\varepsilon)$,
\eas{
\Xi_{\mathrm{distinct}} &= \frac{1}{\sqrt{\alpha}} \max_{c=1,\ldots,C} \| H_c \|_{2 \rightarrow 2}, 
\\
\Xi_{\mathrm{identical}} &= \frac{1}{\sqrt{\alpha}} \left\| \left[ \begin{array}{ccc} H_1 & \cdots & H_C \end{array} \right] \right\|_{2 \rightarrow 2}.
}
See Theorems \ref{t:dist:subgaussRIP_univ} and \ref{t:idt:subgaussRIP_univ}.  Unsurprisingly, the smaller log factor aside, these bounds are more stringent than their nonuniversal counterparts \R{intro_bounds}.  This is demonstrated by the inequalities
\bes{
\Gamma_{\mathrm{distinct}} \leq \Xi_{\mathrm{distinct}} \leq \sqrt{\beta/\alpha} \sqrt{C},\quad \Gamma_{\mathrm{identical}} \leq \Xi_{\mathrm{distinct}} \leq \sqrt{\beta/\alpha} C.
}
See Propositions \ref{prop:GammaXi:bounds:dist} and \ref{prop:GammaXi:bounds:idt} respectively.

\subsubsection{Examples}
In \S \ref{sec:Eg} we illustrate the results \R{intro_bounds} and \R{intro_bounds_universal} with a series of examples.  These include diagonal and circulant sensor profile matrices (both of which arise in applications) and a number of different sparsifying transforms (including the canonical basis, i.e.\ $U = I$, wavelet, Fourier, and cosine sparsity bases).  In all cases we identify examples of sensor profiles matrices which lead to optimal, i.e.\ $C$-independent, measurement conditions in \R{intro_bounds} and \R{intro_bounds_universal}.

\subsection{Application to block-diagonal sensing matrices}\label{ss:blockdiagintro}
In conventional CS, a series of works have sought to design effective sensing matrices that are block diagonal, i.e.
\bes{
\tilde{A} = \left [ \begin{array}{cccc} \bar{A}_{1} & 0 & \cdots & 0 \\  0 & \bar{A}_2 & \ddots & \vdots \\ \vdots & \ddots & \ddots & 0 \\ 0 & \cdots & 0 & \bar{A}_C \end{array} \right ] \in \bbC^{m \times N},
}
where $\bar{A}_{c} \in \bbC^{m/C \times N/C}$ are standard CS matrices (e.g.\ random Gaussians).  These sensing matrices have a variety of uses.  For example, they require less memory and computational expense than unstructured or dense CS matrices (e.g.\ full random Gaussian matrices).      They also arise naturally in a number of applications, including distributed sensing, streaming applications and system identification.  See \cite{Eftekhari&etal:15ACHA} and references therein for further details.

Although not the original motivation for this work, block-diagonal sensing matrices with independent blocks can be viewed as a particular example of our multi-sensor framework.  Specifically, they correspond to so-called perfectly-partitioned diagonal sensor profile matrices $H_c$ (note that the identical and distinct setups lead to exactly the same measurement matrices in this case).  The $c^{\rth}$ such matrix is zero on its diagonal, except for $c^{\rth}$ block where it is a multiple of the identity matrix.  Specializing our main results \R{intro_bounds} to this case gives the measurement condition
\bes{
m \gtrsim \delta^{-2} \cdot \bar{\Gamma}^2(U) \cdot s \cdot L,
}
where $\bar{\Gamma}(U)$ is a particular quantity (see \R{eq:Gamma_dist_perfP}) that satisfies
\be{
\label{barmuboundintro}
\bar{\Gamma}(U) \leq \sqrt{C} \min \left \{ \sqrt{N \mu(U)} , \sqrt{C} \right \},
}
and $\mu(U) = \max_{i,j=1,\ldots,N} | u_{i,j} |^2$ is the coherence of $U$.  
This result also establishes optimal recovery from block-diagonal measurement matrices whenever the sparsity basis is incoherent, i.e.\ $\mu(U) \lesssim 1/N$; for example, a Fourier or cosine basis.

The RIP for block-diagonal sensing matrices with subgaussian blocks was first studied systematically in \cite{WakinEtAlConcMeasBlock,Eftekhari&etal:15ACHA}.
For so-called \textit{distinct block diagonal (DBD)} matrices, which are precisely the matrices discussed above, the measurement condition of \cite{Eftekhari&etal:15ACHA} is
\bes{
m \gtrsim \delta^{-2} \cdot \tilde{\mu}^2(U) \cdot s \cdot L,\qquad \tilde{\mu}(U) = \sqrt{C} \min \left \{ \sqrt{N \mu(U)} , \sqrt{C} \right \}.
}
Our result improves this bound by replacing $\tilde{\mu}(U)$ with the constant $\bar{\Gamma}(U)$ in \R{barmuboundintro}.  See \S \ref{sec:relation} for further details.

\subsection{Relation to previous work} \label{sec:intro:relation}
The parallel acquisition problem was first studied from a CS perspective in \cite{Chun&Adcock:17TIT}.  The results proved therein are nonuniform recovery guarantees, but apply to the more general measurement matrices arising from sampling with random jointly isotropic families of vectors.  In this paper we prove uniform recovery guarantees, which lead to better stability and robustness estimates, but our results only apply to subgaussian random matrices.  However, by specializing to these types of measurement matrices we also obtain simpler and sharper constants $\Gamma_{\mathrm{distinct}}$ and $\Gamma_{\mathrm{identical}}$ than those of \cite{Chun&Adcock:17TIT}, as well as the somewhat better bounds for diagonal sensor profile matrices proved in \cite{Chun&Adcock:16ITW}.  Universal recovery guarantees with subgaussian random matrices were first considered in the parallel acquisition problem in \cite{Chun&Li&Adcock:16MMSPARSE}.  The corresponding guarantees in this paper also improve on those results (see Remark \ref{r:relationICME}).  This aside, another important improvement in this work over \cite{Chun&Adcock:17TIT} is the relaxation of the joint isometry condition \R{joint_iso} to the joint near-isometry condition \R{joint_near_iso}.

As remarked in the previous section, a special case of our setup is DBD subgaussian sensing matrices introduced in \cite{WakinEtAlConcMeasBlock,Eftekhari&etal:15ACHA}.  For earlier work on concentration inequalities for such matrices, see \cite{ParkEtAlBlockDiagConcentration,RozellEtAlConcRepeatedBlock}.  Similar to \cite{Eftekhari&etal:15ACHA}, the proofs of our main results make use of the techniques of Krahmer, Rauhut \& Mendelson on suprema of chaos processes \cite{Krahmer:14CPAM}.

\subsection{Notation}
Throughout, we write $\nm{\cdot}_{p}$ for the vector $p$-norm and $\| \cdot \|_{p \rightarrow p}$ for the matrix $p$-norm (i.e.\ $\| A \|_{p \rightarrow p} = \sup_{\|x\|_p = 1} \| A x \|_p$).  The Frobenius norm of a matrix is denoted by $\| \cdot \|_F$. We write $\ip{\cdot}{\cdot}$ for the standard inner product on $\bbC^N$.  
As is conventional, we write $\nm{\cdot}_{0}$ for the $\ell^0$-norm, i.e.\ the number of nonzeros of a vector.  The canonical basis on $\bbC^N$ will be denoted by $e_1,\ldots,e_N$.  If $\Delta \subseteq \{1,\ldots,N\}$ then we use the notation $P_{\Delta}$ for both the orthogonal projection $P_{\Delta} \in \bbC^{N \times N}$ with
\bes{
(P_{\Delta}x)_{j} = \left \{ \begin{array}{cc} x_j & j \in \Delta \\ 0 & \mbox{otherwise} \end{array} \right .,\qquad x \in \bbC^N,
}
and the matrix $P_{\Delta} \in \bbC^{|\Delta| \times N}$ with
\bes{
(P_{\Delta}x)_{j} = x_j,\quad j \in \Delta,\qquad x \in \bbC^N.
}
The precise meaning will be clear from the context.  
Distinct from the index $i$, we denote the imaginary unit by $\I$.  In addition, we use the notation $a \lesssim b$ or $a \gtrsim b$ to mean there exists a constant $c>0$ independent of all relevant parameters (in particular, the number of sensors $C$) such that $a \leq c b$ or $a \geq c b$ respectively.  For self-adjoint matrices $A,B \in \bbC^{N \times N}$, the notation $B \preceq A$ denotes that $A-B$ is a positive semi-definite matrix. The condition number of a matrix $A$ is denoted by $\kappa(A)$.

\section{Main results}

\subsection{Preliminaries}

First we recall the definition of sparsity:

\defn{[Sparsity]
\label{d:sparsity}
A vector $z \in \bbC^N$ is $s$-sparse for some $1 \leq s \leq N$ if $\| z \|_0 \leq s$.
}
We shall write 
\bes{
\Sigma_{s} = \left\{ z \in \bbC^N : \| z \|_0 \leq s \right\}.
}
for the set of $s$-sparse vectors, and
\bes{
B_{s} = \left\{ z \in \bbC^N : \| z \|_0 \leq s,  \| z \|_2 \leq 1 \right\}.
}
for the intersection of $\Sigma_{s}$ with the unit Euclidean ball.

We also recall the definition of restricted isometry property:

\defn{[Restricted isometry property, RIP]
\label{d:RIP}
A matrix $A \in \bbC^{m \times N}$ satisfies the Restricted Isometry Property (RIP) of order $s$ if there exists $0<\delta<1$ such that
\be{
\label{eq:RIP}
(1-\delta) \| x \|_1^2 \leq \| A x \|^2_2 \leq (1+\delta) \| x \|^2_2,\qquad \forall x \in \Sigma_{s}.
}
If $\delta = \delta_s$ is the smallest constant such that \R{eq:RIP} holds, then we refer to $\delta_s$ as the $s^{\textmd{th}}$ Restricted Isometry Constant (RIC) of $A$.
}

In this paper, we will also consider a more general notion (see, for example, \cite{FoucartLaiEllq}):
\defn{[Asymmetric RIP]
\label{d:ARIP}
A matrix $A \in \bbC^{m \times N}$ satisfies the Asymmetric Restricted Isometry Property (ARIP) of order $s$ if there exists $\beta \geq \alpha > 0$ such that
\be{
\label{eq:ARIP}
\alpha \| x \|^2_2 \leq \| A x \|^2_2 \leq \beta \| x \|^2_2,\qquad \forall x \in \Sigma_{s}.
}
If $\alpha = \alpha_{s}$ and $\beta = \beta_{s}$ are the largest and smallest constants respectively such that \R{eq:ARIP} holds then we refer to $(\alpha_{s}, \beta_{s})$ as the $s^{\textmd{th}}$ Asymmetric Restricted Isometry Constants (ARICs) of $A$.
}

We incorporate this asymmetry because it places less stringent conditions on the sensor profile matrices $H_c$ -- see \S \ref{sec:setupDist} and \S \ref{sec:setupIdt}.  Observe that if $(\alpha,\beta) = (1-\delta,1+\delta)$ then this is just the standard RIP for the sparse signal model (Definition \ref{d:RIP}).  
We additionally remark that the ARIP of order $2s$ implies stable and robust recovery, uniform in $x \in \bbC^N$, when solving \R{recovery_alg}.  The following result is standard.  We include a short proof for completeness.

\thm{
\label{t:ARIP_stable_robust}
Suppose that a matrix $A \in \bbC^{m \times N}$ has the ARIP of order $2s$ with ARICs $(\alpha_{2s},\beta_{2s})$ satisfying
\be{
\label{ARIPsuffic}
\frac{\beta_{2s}}{\alpha_{2s}} < \frac{\sqrt{2}+1}{\sqrt{2}-1}.
}
Let $x \in \bbC^N$ and $y = A x + n$ with $\| n \|_2 \leq \eta$.  Then for any minimizer $\hat{x} \in \bbC^N$ of
\be{
\label{recovery_alg2}
\min_{z \in \bbC^N} \| z \|_{1}\ \mbox{subject to $\| A z - y \|_2 \leq \eta$},
}
we have
\be{
\label{l1bound}
\| \hat{x} - x \|_2 \lesssim \frac{\sigma_s(x)}{\sqrt{s}} + \frac{1}{\sqrt{\alpha_{2s}}}\eta,
}
where $\sigma_{s}(x) = \min \left \{ \| x - z \|_{1} : z \in \Sigma_{s} \right \}$.
}
\prf{
Let $t = \sqrt{\frac{2}{\alpha_{2s} + \beta_{2s}}} \leq \frac{1}{\sqrt{\alpha_{2s}}}$.  It is easy to check that the matrix $\tilde{A} = t A$ satisfies the standard RIP with constant 
\be{
\label{delta2sbound}
\delta_{2s} \leq \frac{\beta_{2s} - \alpha_{2s}}{\beta_{2s} + \alpha_{2s}}.
}
Also, the minimization problem \R{recovery_alg2} is equivalent to
\bes{
\min_{z \in \bbC^N} \| z \|_{1}\ \mbox{subject to $\| \tilde{A} z - \tilde{y} \|_2 \leq t\eta$},
}
where $\tilde{y} = t y = \tilde{A} x + \tilde{n}$ where $\tilde{n} = t n$ satisfies $\| \tilde{n} \|_2 \leq t \eta$.  Due to a result of \cite{Cai&Zhang:14TIT}, the bound \R{l1bound} holds, provided the RIC $\delta_{2s} < 1/\sqrt{2}$.  Rearranging \R{delta2sbound} now gives \R{ARIPsuffic}. 
}

This aside, let us recall that a random variable is subgaussian with parameter $\phi$ if 
\bes{
\bbP(|X| \geq \phi t ) \leq 2 \exp \! \left( - t^2/2 \right),
}
for every $t \geq 1$.  A random vector in $\bbC^N$ is $\phi$-subgaussian if its elements are independent, zero-mean, unit-variance, and $\phi$-subgaussian random variables (see \cite{Vershynin:bookCh, Krahmer:14CPAM}), and a matrix $A$ is $\phi$-subgaussian if its entries independent, zero-mean, unit-variance, and $\phi$-subgaussian random variables.

\subsection{Distinct sampling} \label{sec:setupDist}

Let $\tilde{A}_{1},\ldots,\tilde{A}_{C} \in \bbC^{m/C \times N}$ be independent subgaussian random matrices with the same subgaussian parameter $\phi$.
We assume that the matrices $H_c$ satisfy the \textit{joint near-isometry condition} 

\be{
\label{eq:joint_iso_dist}
\alpha I \preceq C^{-1} \sum^{C}_{c=1} H^*_c H_c \preceq \beta I ,
}
for constants $0 < \alpha \leq \beta$.  Note that such constants always exist if the matrix $C^{-1} \sum^{C}_{c=1} H^*_c H_c$ is nonsingular, and are equal to its minimal and maximal eigenvalues respectively.  We write it in the form \R{eq:joint_iso_dist} since it will be useful later.  Note that we will primarily be interested in the case where the ratio $\beta / \alpha$ is independent of $C$.  Note also that the condition number
\be{
\label{eq:condNumb}
\kappa \left( C^{-1} \sum^{C}_{c=1} H^*_c H_c \right) = \frac{\beta}{\alpha},
}
so this is equivalent to stipulating that the matrix $C^{-1}  \sum^{C}_{c=1} H^*_c H_c$ has a condition number independent of $C$.  As we see below, the constants $\alpha$ and $\beta$ will relate to the ARIP of the corresponding measurement matrix $A$.  Had we sought the classical RIP, we would have required the much more stringent condition $C^{-1} \sum^{C}_{c=1} H^*_c H_c = I$.  The relaxed condition \R{eq:joint_iso_dist} allows for substantially more flexibility in the design of the sensor profile matrices $H_c$.

The measurement matrix $A$ is now formed by
\be{
\label{def:A_dist}
A = \frac{1}{\sqrt{m}} \left [ \begin{array}{c} A_1 \\ \vdots \\ A_C \end{array} \right ], \qquad A_c = \tilde{A}_c H_c U \in \bbC^{m/C \times N}, \qquad c=1,\ldots,C.
}
Due to \R{eq:joint_iso_dist} and the fact that $U$ is unitary, we have
\be{
\label{eq:rIso_dist}
\alpha I \preceq \bbE A^* A \preceq \beta I,
}
hence we shall seek to establish the ARIP for $A$ (as opposed to the RIP, which would be conventional had $\bbE(A^*A) = I$).  The following two theorems are our main results in this case:

\thm{[ARIP for distinct sampling] \label{t:dist:subgaussRIP}
For $0 < \delta, \varepsilon < 1$ and $1 \leq s \leq N$, let $A$ be as in \R{def:A_dist}, where
\be{
\label{dist:subgaussRIP:measurement}
m \gtrsim \delta^{-2} \cdot \frac{\beta}{\alpha} \cdot  \Gamma_{\mathrm{distinct}}^2 \cdot s \cdot L_1,
}
$\alpha,\beta$ are as in \R{eq:joint_iso_dist},
\be{
\label{Gammadistinct}
\Gamma_{\mathrm{distinct}} = \frac{1}{\sqrt{\alpha}} \max_{c=1,\ldots,C} \max_{j=1,\ldots,N} \| H_c U e_j \|_2,
}
and $L_1 =  \ln^2(2s) \ln(2N) \ln(2m) + \ln(2/\varepsilon)$.
Then with probability at least $1-\varepsilon$, the ARICs $(\alpha_{s}, \beta_{s})$ of $A$ satisfy 
\bes{
\alpha_s \geq (1-\delta) \alpha,\qquad \beta_s \leq (1+\delta ) \beta.
}
}

\thm{[ARIP for distinct sampling -- Universal bound] \label{t:dist:subgaussRIP_univ}
For $0 < \delta, \varepsilon < 1$ and $1 \leq s \leq N$, let $A$ be as in \R{def:A_dist}, where
\be{
\label{dist:subgaussRIP_univ:measurement}
m \gtrsim \delta^{-2} \cdot \frac{\beta}{\alpha} \cdot  \Xi_{\mathrm{distinct}}^2 \cdot L_2,
}
$\alpha,\beta$ are as in \R{eq:joint_iso_dist},
\be{
\label{Xidistinct}
\Xi_{\mathrm{distinct}} = \frac{1}{\sqrt{\alpha}} \max_{c=1,\ldots,C} \| H_c \|_{2 \rightarrow 2},
}
and $L_2 =  \ln(2N/s) + s^{-1} \ln(2/\varepsilon)$.
Then with probability at least $1-\varepsilon$, the ARICs $(\alpha_{s}, \beta_{s})$ of $A$ satisfy 
\bes{
\alpha_s \geq (1-\delta) \alpha,\qquad \beta_s \leq (1+\delta ) \beta.
}
}
Since $\Xi_{\mathrm{distinct}}$ is independent of the sparsity basis $U$, this latter bound is universal.  We note also that it is possible to deduce a universal bound directly from Theorem \ref{t:dist:subgaussRIP}.  Since $\| H_c U e_j \|_2 \leq \| H_c \|_{2 \rightarrow 2}$, a direct application of Theorem \ref{t:dist:subgaussRIP} gives
\be{
\label{eq:dist:subgaussRIP_univ}
m \gtrsim \delta^{-2} \cdot \frac{\beta}{\alpha} \cdot  \Xi_{\mathrm{distinct}}^2 \cdot s \cdot L_1.
}
However, the log factors in Theorem \ref{t:dist:subgaussRIP_univ} are smaller than those in \R{eq:dist:subgaussRIP_univ}.

\rem{
\label{r:relationICME}
The recovery guarantee in Theorem \ref{t:dist:subgaussRIP_univ} improves a previous result based on concentration inequalities.
The universal recovery guarantee in \cite[Thm.\ 3.2]{Chun&Li&Adcock:16MMSPARSE} depends on the larger constant $\Xi_{\mathrm{distinct}}^4$ and applies only to the case $\alpha,\beta=1$.
}

Theorems \ref{t:dist:subgaussRIP} and \ref{t:dist:subgaussRIP_univ} assume the same number of measurements per sensor, equal to $m/C$.  More generally, let $m_1,\ldots,m_C \in \bbN$ so that $m = m_1 + \ldots + m_C$ and consider independent subgaussian random matrices $\tilde{A}_{c} \in \bbC^{m_c \times N}$, $c=1,\ldots,C$, with the same subgaussian parameter $\phi$.  Define the overall measurement matrix
\be{
\label{def:A_dist_diffmc}
A = \left [ \begin{array}{c} \frac{1}{\sqrt{C m_1}} A_1 \\ \vdots \\ \frac{1}{\sqrt{C m_C}} A_C \end{array} \right ], \qquad A_c = \tilde{A}_c H_c U \in \bbC^{m_c \times N}, \qquad c=1,\ldots,C.
}
We now have the following generalizations of Theorem \ref{t:dist:subgaussRIP} and \ref{t:dist:subgaussRIP_univ}:

\thm{[ARIP for distinct sampling with different $m_c$'s] \label{t:dist:subgaussRIP_diff_mc}
For $0 < \delta, \varepsilon < 1$ and $1 \leq s \leq N$, let $A$ be as in \R{def:A_dist_diffmc}, where
\be{
\label{dist:subgaussRIP:measurement_diff_mc}
\min_{c=1,\ldots,C} m_c \gtrsim \delta^{-2} \cdot C^{-1} \cdot \frac{\beta}{\alpha} \cdot  \Gamma_{\mathrm{distinct}}^2 \cdot s \cdot L_1,
}
$\alpha,\beta$ are as in \R{eq:joint_iso_dist}, $\Gamma_{\mathrm{distinct}}$ is as in \R{Gammadistinct}, and $L_1 =  \ln^2(2s) \ln(2N) \ln(2m) + \ln(2/\varepsilon)$.
Then with probability at least $1-\varepsilon$, the ARICs $(\alpha_{s}, \beta_{s})$ of $A$ satisfy 
\bes{
\alpha_s \geq (1-\delta) \alpha,\qquad \beta_s \leq (1+\delta ) \beta.
}
}

\thm{[ARIP for distinct sampling with different $m_c$'s -- Universal bound] \label{t:dist:subgaussRIP_univ_diff_mc}
For $0 < \delta, \varepsilon < 1$ and $1 \leq s \leq N$, let $A$ be as in \R{def:A_dist_diffmc}, where
\be{
\label{dist:subgaussRIP_univ:measurement_diff_mc}
\min_{c=1,\ldots,C} m_c \gtrsim \delta^{-2} \cdot C^{-1} \cdot \frac{\beta}{\alpha} \cdot  \Xi_{\mathrm{distinct}}^2 \cdot L_2,
}
$\alpha,\beta$ are as in \R{eq:joint_iso_dist}, $\Xi_{\mathrm{distinct}}$ is as in \R{Xidistinct}, and $L_2 =  \ln(2N/s) + s^{-1} \ln(2/\varepsilon)$.
Then with probability at least $1-\varepsilon$, the ARICs $(\alpha_{s}, \beta_{s})$ of $A$ satisfy 
\bes{
\alpha_s \geq (1-\delta) \alpha,\qquad \beta_s \leq (1+\delta ) \beta.
}
}

As one would expect, these two theorems imply that the ARIP is satisfied for the measurement matrix $A$, provided every sensor takes sufficiently many measurements.  Note that the overall measurement condition in the case of Theorem \ref{t:dist:subgaussRIP_diff_mc} is
\be{
\label{overall_meas}
m = m_1+\ldots+m_C \gtrsim \delta^{-2} \cdot \frac{\beta}{\alpha} \cdot \Gamma^2_{\mathrm{distinct}} \cdot s \cdot L_1,
}
which is equivalent to that of Theorem \ref{t:dist:subgaussRIP} (an identical statement applies to Theorems \ref{t:dist:subgaussRIP_univ_diff_mc} and \ref{t:dist:subgaussRIP_univ}).  However, we caution the reader that \R{overall_meas}, while necessary for the ARIP, is not sufficient since it does not guarantee \R{dist:subgaussRIP_univ:measurement_diff_mc} will hold.  We also remark in passing that Theorems \ref{t:dist:subgaussRIP} and \ref{t:dist:subgaussRIP_univ} are simple corollaries of Theorems \ref{t:dist:subgaussRIP_diff_mc} and \ref{t:dist:subgaussRIP_univ_diff_mc} corresponding to the case $m_1 = \ldots = m_C = m/C$.

Let us now consider the question of optimal recovery in the sense defined in \S \ref{ss:CSPA}.  These results imply that optimal recovery is possible if $U$ and the $H_c$ are such that $\Gamma_{\mathrm{distinct}}$ or $\Xi_{\mathrm{distinct}}$ are independent of $C$.  Hence, we now examine these quantities in more detail:

\prop{
\label{prop:GammaXi:bounds:dist}
The quantities $\Gamma_{\mathrm{distinct}}$ and $\Xi_{\mathrm{distinct}}$ defined in \R{Gammadistinct} and \R{Xidistinct} satisfy
\bes{
1 \leq \Gamma_{\mathrm{distinct}} \leq  \Xi_{\mathrm{distinct}} \leq \sqrt{\beta / \alpha} \sqrt{C}.
}
Moreover, the inequalities are sharp.
}

This result has several implications.  First, there are choices of the $H_c$ which yield optimal universal and nonuniversal recovery guarantees.  Second, as is to be expected, the bounds \R{dist:subgaussRIP:measurement} and \R{dist:subgaussRIP_univ:measurement} depend on the ratio $\beta/\alpha$, not on the individual factors themselves.  We remark also that the $\sqrt{C}$ dependence in the upper inequality is also reasonable.  The resulting worst-case bound $m_c \gtrsim \beta/\alpha \cdot s  \cdot L$ implies that at worst each sensor should take enough measurements to recover the signal from those measurements only.

\subsection{Identical sampling} \label{sec:setupIdt}

The setup for identical sampling is rather different to that of \S \ref{sec:setupDist}.  Let $\tilde{A} \in \bbC^{m/C \times N}$ be a subgaussian random matrix.	
As before, we assume that $H_c \in \bbC^{N \times N}$ satisfy the \textit{joint near-isometry condition} condition
\be{
\label{eq:joint_iso_idt}
\alpha I \preceq C^{-1} \sum^{C}_{c=1} H^*_c H_c \preceq \beta I,
}
and form the matrix
\be{
\label{def:A_idt}
A = \frac{1}{\sqrt{m}} \left [ \begin{array}{c} A_1 \\ \vdots \\ A_C \end{array} \right ],\qquad A_c = \tilde{A} H_c U \in \bbC^{m/C \times N}, \qquad c=1,\ldots,C.
}
For similar reasons to \R{eq:rIso_dist}, we have
\be{
\label{eq:rIso_idt}
\alpha I \preceq \bbE A^* A \preceq \beta I.
}
The follow two theorems are our main results in this case:

\thm{[ARIP for identical sampling] \label{t:idt:subgaussRIP}
For $0 < \delta, \varepsilon < 1$ and $1 \leq s \leq N$ let $A$ be as in \R{def:A_idt}, where
\be{
\label{idt:subgaussRIP:measurement}
m \gtrsim \delta^{-2} \cdot \frac{\beta}{\alpha} \cdot  \Gamma_{\mathrm{identical}}^2 \cdot s \cdot L_1,
}
$\alpha,\beta$ are as in \R{eq:joint_iso_dist}, 
\be{
\label{Gammaidentical}
\Gamma_{\mathrm{identical}} = \frac{1}{\sqrt{\alpha}} \max_{j=1,\ldots,N} \left\| \left[ \begin{array}{ccc} H_1 U e_j & \cdots & H_C U e_j \end{array} \right] \right\|_{2 \rightarrow 2},
}
and $L_1 =  \ln^2(2s) \ln(2N) \ln(2m) + \ln(2/\varepsilon)$.
Then with probability at least $1-\varepsilon$, the ARICs $(\alpha_{s}, \beta_{s})$ of $A$ satisfy 
\bes{
\alpha_s \geq (1-\delta) \alpha,\qquad \beta_s \leq (1+\delta ) \beta.
}
}

\thm{[ARIP for identical sampling -- Universal bound] \label{t:idt:subgaussRIP_univ}
For $0 < \delta, \varepsilon < 1$ and $1 \leq s \leq N$, let $A$ be as in \R{def:A_idt}, where
\be{
\label{idt:subgaussRIP_univ:measurement}
m \gtrsim \delta^{-2} \cdot \frac{\beta}{\alpha} \cdot \Xi_{\mathrm{identical}}^2 \cdot L_2,
}
$\alpha,\beta$ are as in \R{eq:joint_iso_idt},
\be{
\label{Xiidentical}
\Xi_{\mathrm{identical}} = \frac{1}{\sqrt{\alpha}} \left\| \left[ \begin{array}{ccc} H_1 & \cdots & H_C \end{array} \right] \right\|_{2 \rightarrow 2},
}
and $L_2 =  \ln(2N/s) + s^{-1} \ln(2/\varepsilon)$.
Then with probability at least $1-\varepsilon$, the ARICs $(\alpha_{s}, \beta_{s})$ of $A$ satisfy 
\bes{
\alpha_s \geq (1-\delta) \alpha,\qquad \beta_s \leq (1+\delta ) \beta.
}
}

Similar to the distinct case, if $\Gamma_{\mathrm{identical}}$ or $\Xi_{\mathrm{identical}}$ are independent of $C$ then one obtains an optimal recovery guarantee.  The following result provides bounds for these quantities:

\prop{ 
\label{prop:GammaXi:bounds:idt}
The quantities $\Gamma_{\mathrm{identical}}$ and $\Xi_{\mathrm{identical}}$ defined in \R{Gammaidentical} and \R{Xiidentical} respectively satisfy the sharp inequalities
\be{
\label{Gamma:idt:bound}
\Gamma_{\mathrm{distinct}} \leq \Gamma_{\mathrm{identical}}  \leq \sqrt{\beta/\alpha} C,
}
where $\Gamma_{\mathrm{distinct}}$ is as in \R{Gammadistinct},
\be{
\label{Xi:idt:bound1}
 \Gamma_{\mathrm{identical}} \leq \Xi_{\mathrm{identical}} \leq \sqrt{\beta/\alpha} C,
}
and
\be{
\label{Xi:idt:bound2}
\Xi_{\mathrm{distinct}} \leq \Xi_{\mathrm{identical}},
}
where $\Xi_{\mathrm{distinct}}$ is as in \R{Xidistinct}.
}

This proposition implies firstly that the bounds \R{idt:subgaussRIP:measurement} and \R{idt:subgaussRIP_univ:measurement} are determined by the ratio $\beta/\alpha$, not on the factors themselves, and secondly, as one would expect, the bounds for identical sampling (whether universal or nonuniversal) are always at least as large as those for distinct sampling.  Note that, for general $H_c$'s, the upper bounds in \R{Gamma:idt:bound} and \R{Xi:idt:bound1} depend on $C$, as opposed to $\sqrt{C}$ as is the case for distinct sampling (see Proposition \ref{prop:GammaXi:bounds:dist}).  This implies a worst-case bound of the form $m \gtrsim C^2 \cdot s \cdot L$.  Fortunately,  if the matrices $H_c$ are normal -- that is, if $H_c H^*_c = H^*_c H_c$ $\forall c$ -- then this bound reduces to $\sqrt{C}$:

\prop{ 
\label{prop:GammaXi:bounds:idt:2}
If the $H_{c}$ are normal, then one has the sharp bounds
\bes{
\Gamma_{\mathrm{identical}} \leq \Xi_{\mathrm{identical}} \leq \sqrt{\beta/\alpha} \sqrt{C}.
}
}

\section{Examples} \label{sec:Eg}

\begin{figure}[!t]
\centering
\begin{tabular}{ccc}
\includegraphics[scale=0.455, trim=0.5em 0.45em 2.2em 1em, clip]{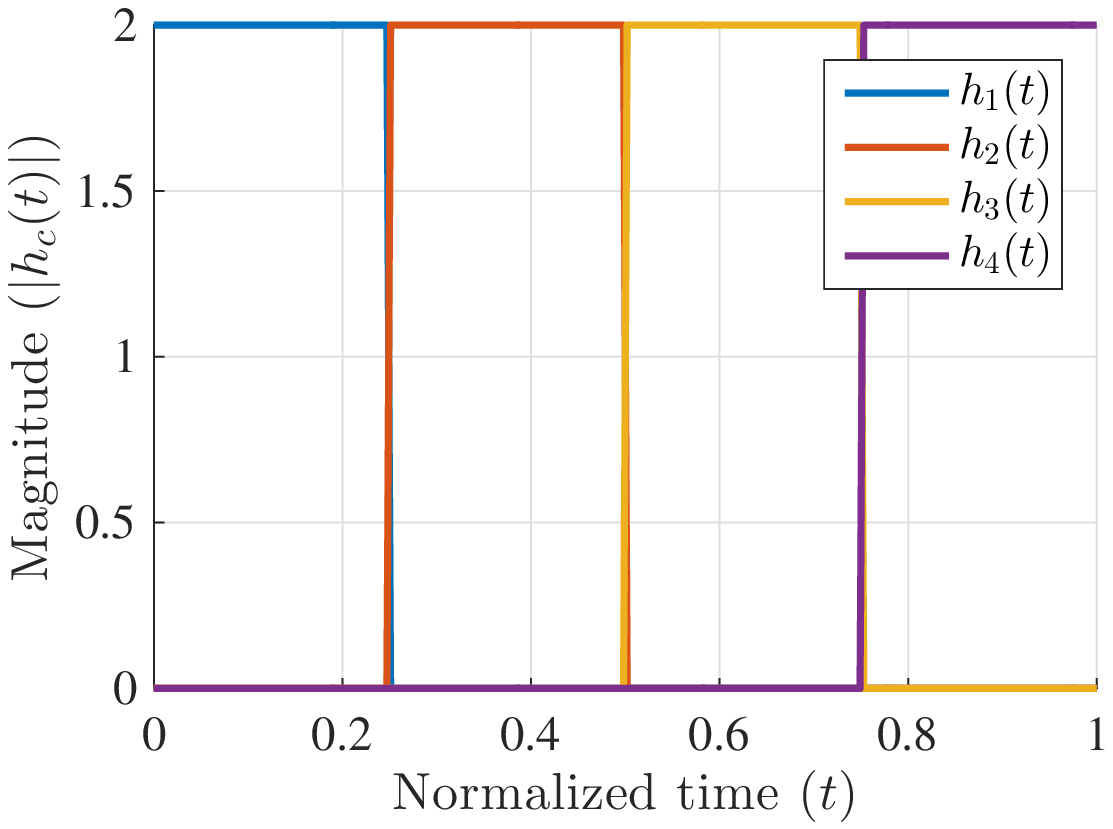} &
\includegraphics[scale=0.455, trim=0.5em 0.45em 2.2em 1em, clip]{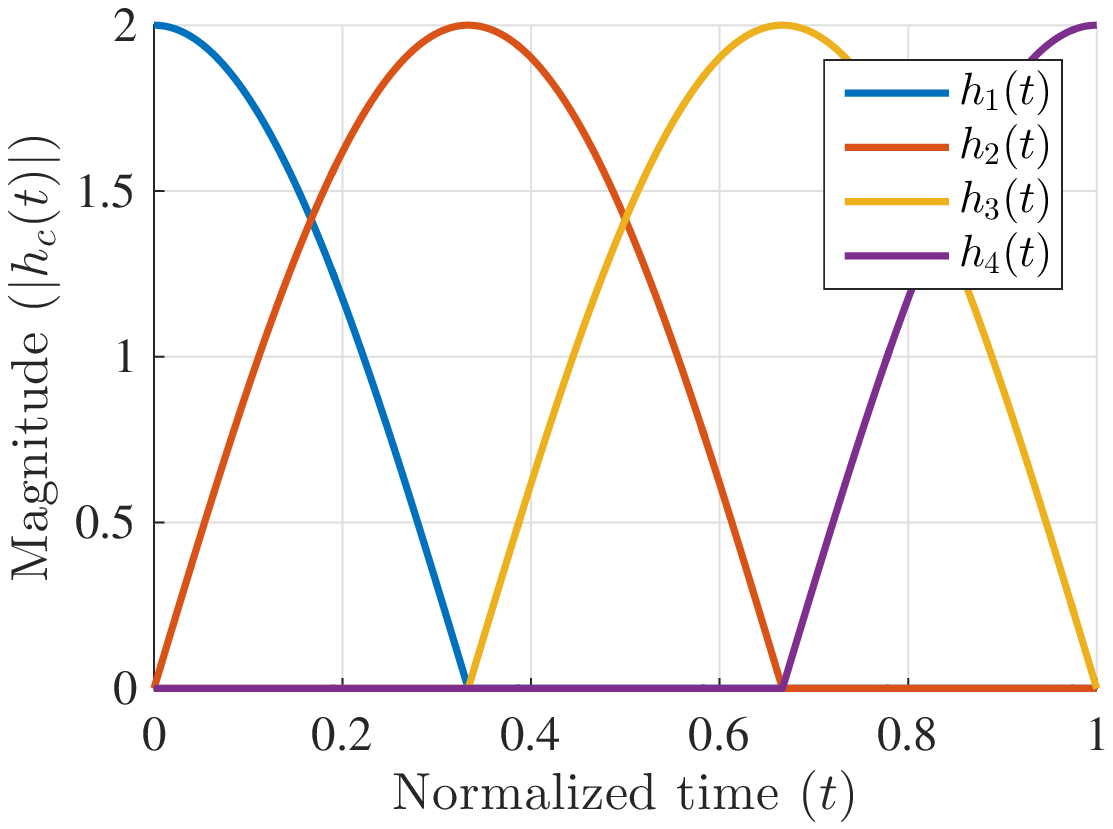} &
\includegraphics[scale=0.455, trim=0.5em 0.45em 2.2em 1em, clip]{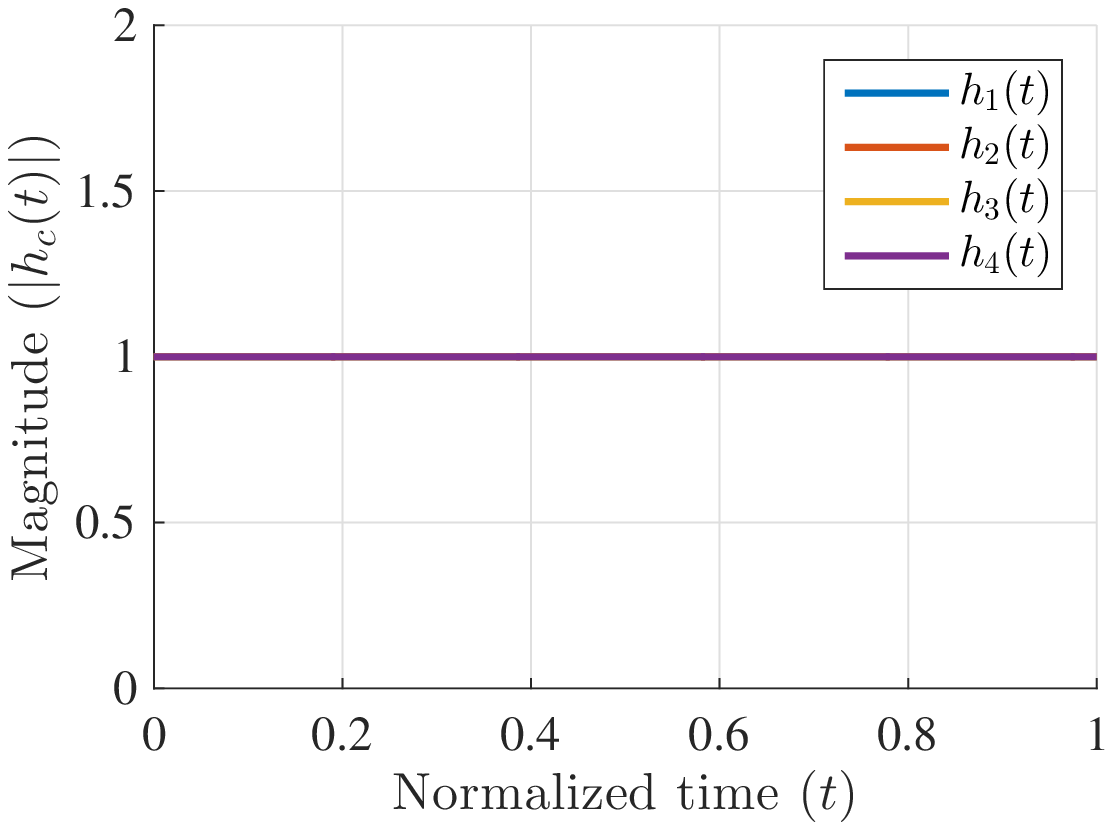} 
\\
{\small (a) Perfectly-partitioned $H_c$} & {\small (b) Banded $H_c$} & {\small (c) Globally-spread $H_c$} \\
\end{tabular}

\caption{Examples of diagonal sensor profiles $H_c = \diag(h_c)$ ($\alpha,\beta=1$ and $C=4$): Eigenvalues of circulant sensor profiles are constructed in a similar way to construct $h_c$'s.}
\label{fig:sensorProf}
\end{figure}

We now illustrate our main results by looking at several different classes of sensor profile matrices $H_c$.  We are interested in understanding which types of sensor profiles lead to optimal recovery guarantees.  In other words, we wish to identify classes of sensor profiles for which the constants $\Gamma_{\mathrm{distinct}}$ and $\Gamma_{\mathrm{identical}}$ -- or even, wherever possible, their universal counterparts $\Xi_{\mathrm{distinct}}$ and $\Xi_{\mathrm{identical}}$ -- are independent of the number of sensors $C$.  In this case, Theorems \ref{t:dist:subgaussRIP} and \ref{t:idt:subgaussRIP} (respectively Theorems \ref{t:dist:subgaussRIP_univ} and \ref{t:idt:subgaussRIP_univ}) yield measurement conditions that are optimal with respect to the number of sensors $C$.

We consider two main classes of sensor profiles -- diagonal and circulant -- which are closely related to the practical applications.
For example, diagonal sensor profiles can be used to model the spatial profiles of coils in a parallel MRI \cite{Chun&Adcock&Talavage:15TMI} or model the wireless channel between a source and sensors in WSN applications \cite{Choi&Park&Lee:bookCh, Oliver&Lee:11SPARS}.  Circulant sensor profiles can be used to model geometric features of the scene captured by cameras in a multi-view imaging \cite{Park&Wakin:12EJASP} or to model antenna beam patterns in SAR imaging \cite{Aceska&etal:16Arxiv}.

Within the class of diagonal sensor profile matrices $H_c = \diag(h_c) \in \bbC^{N \times N}$, we consider the following three examples:
\begin{enumerate}
\item[(\romnum{1})] \textit{Perfectly-partitioned profiles}.\  These are nonoverlapping sensor profiles, defined by $H_c = \sqrt{C} P_{I_c}$, where 
\be{
\label{Icdef}
I_c = \{ (c-1) N/C + 1,\ldots, c N/C \},\qquad c=1,\ldots,C,
}
is a partition of $\{1,\ldots,N\}$ into equally-sized subintervals (for simplicity we assume $N/C$ is an integer).  See Fig.\ \ref{fig:sensorProf}(a).
\item[(\romnum{2})] \textit{Banded profiles}.\ Define $1 \leq q \leq C$ by
\be{
\label{qdef}
q = \max_{c=1,\ldots,C} \left | \left \{ d : \supp(h_c) \cap \supp(h_d),\ d=1,\ldots,C \right \} \right |,
}
where $\supp(h_c) = \{ i : (h_c)_i \neq 0 \}$ denotes the support of $h_c$.\footnote{In other words, the quantity $q$ is the number of times that different sensor profiles overlap.}  We say the sensor profile matrices are \textit{banded} if $q$ is independent of $C$.  Outside of example (i), which is banded with $q = 1$, the specific example of this setup that we shall consider are smooth sensor profiles with compact support. This example is constructed by a truncated cosine function multiplied with a phase vector $\{ \exp( \I ( \frac{(c-1) 2\pi}{C}  + \frac{2\pi}{NC}  ) ),\ldots, \exp(\I \frac{c 2 \pi}{C} ) \}$ and $\sqrt{C}$ \cite{Chun&Adcock:16ITW}; see Fig.\ \ref{fig:sensorProf}(b).  Note that $q = \ord{1}$ holds for this example even if $C$ increases.

\item[(\romnum{3})] \textit{Globally-spread sensor profiles}. Unlike the previous two examples, globally-spread sensor profiles are nonzero in most (if not all) their entries.  The particular example we consider is the case where the entries of $h_c$ are unit-magnitude complex numbers drawn uniformly at random from the unit circle.  See Fig.\ \ref{fig:sensorProf}(c).
\end{enumerate}
Note that in all the above examples the sensor profile matrices are, for simplicity, chosen so that $C^{-1} \sum_{c=1}^C H_c^* H_c = I$.  In particular, \R{eq:condNumb} holds with $\alpha = \beta = 1$.

To construct examples of circulant sensor profile matrices $H_c \in \bbC^{N \times N}$, we first note that since such matrices are unitarily diagonalizable with the discrete Fourier transform, the joint isometry condition \R{eq:joint_iso_idt} becomes
\be{
\label{eq:joint_iso_idt:circ}
\alpha I \preceq C^{-1} \sum^{C}_{c=1} \Lambda^*_c \Lambda_c \preceq \beta I.
}
where $\Lambda_c = \diag(\lambda_c)$ is the diagonal matrix of eigenvalues of $H_c$; that is, $H_c = F^* \Lambda_c F$, where $F \in \bbC^{N \times N}$ is unitary discrete Fourier transform (DFT) matrix.  Hence, examples of circulant sensor profiles can be constructed by defining the diagonal matrices $\Lambda_c$ in the same way as in the schemes (i)--(iii) introduced above to generate the diagonal sensor profile matrices.

Having constructed sensor profiles, we also consider a number of sparsity bases $U$.  Fourier, cosine, wavelet, and canonical sparsity bases are constructed by inverse DFT, the inverse discrete cosine transform (i.e.\ DCT-\Romnum{3}), inverse discrete wavelet transform, and $U=I$, respectively.
In particular, the corresponding wavelet transform is constructed using the Haar transform with a $4$-level-decomposition (i.e.\ the wavelet expansion from the finest resolution level of $\log_2(N)-1$ to the coarsest resolution level of $4$).

\subsection{Distinct sampling}

Throughout this section we use the notation $\mu(V) = \max_{i,j=1,\ldots,N} | v_{i,j} |^2$ to denote the coherence of a matrix $V \in \bbC^{N \times N}$.

\begin{figure}[!t]
\centering
\begin{tabular}{ccc}
{\small (a-1) Perfectly-partitioned $H_c$} & {\small (a-2) Banded $H_c$} & {\small (a-3) Globally-spread $H_c$} \\
\includegraphics[scale=0.455, trim=0.5em 0.45em 2.2em 1em, clip]{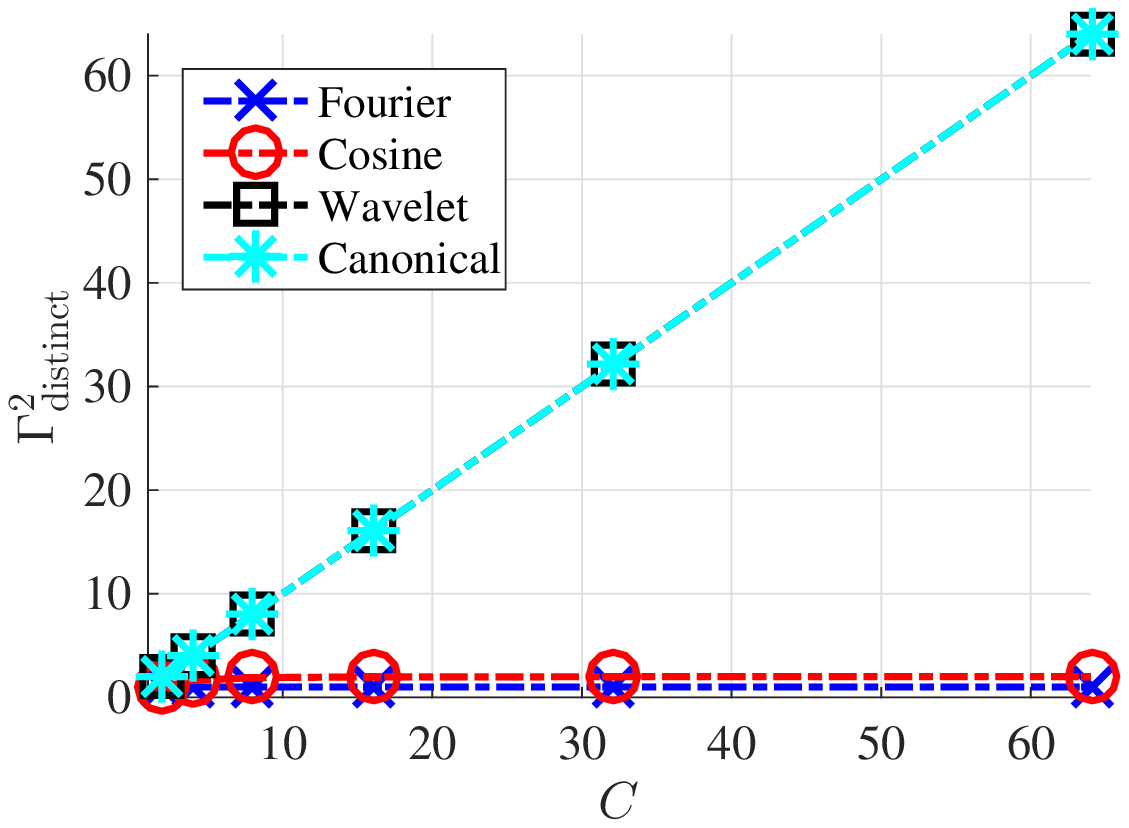} &
\includegraphics[scale=0.455, trim=0.5em 0.45em 2.2em 1em, clip]{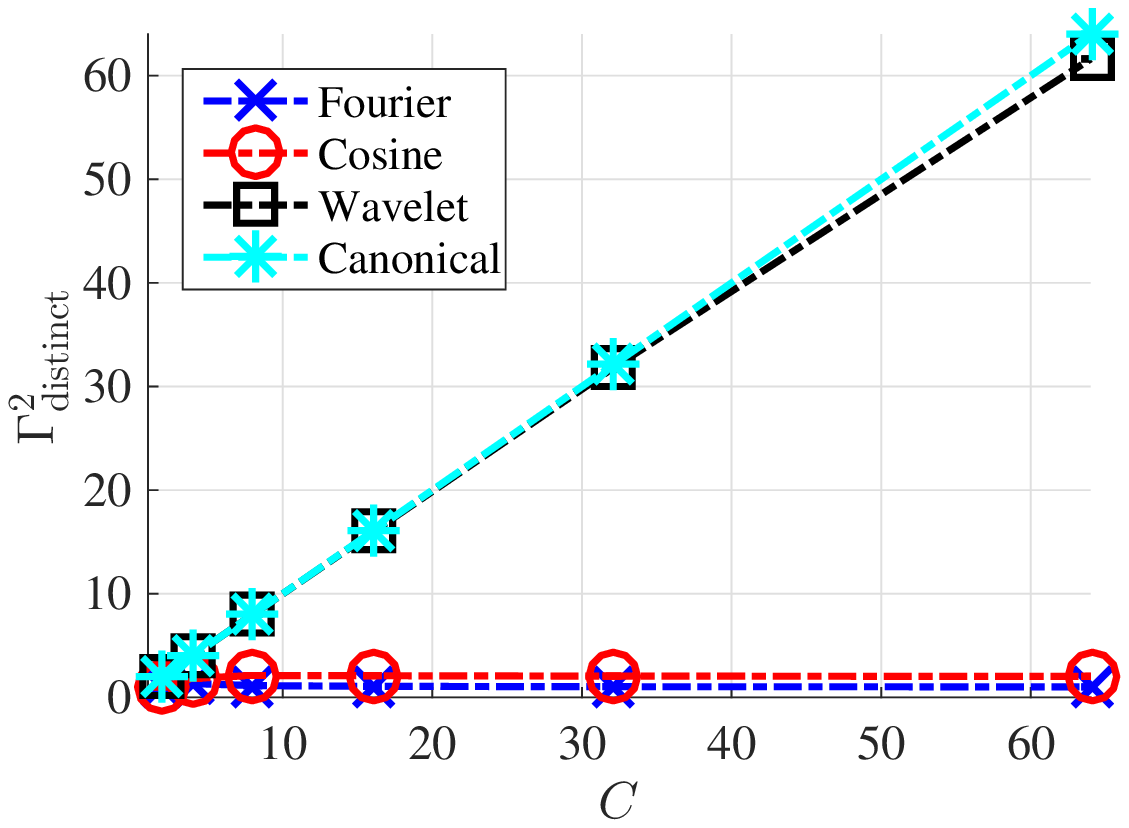} &
\includegraphics[scale=0.455, trim=0.5em 0.45em 2.2em 1em, clip]{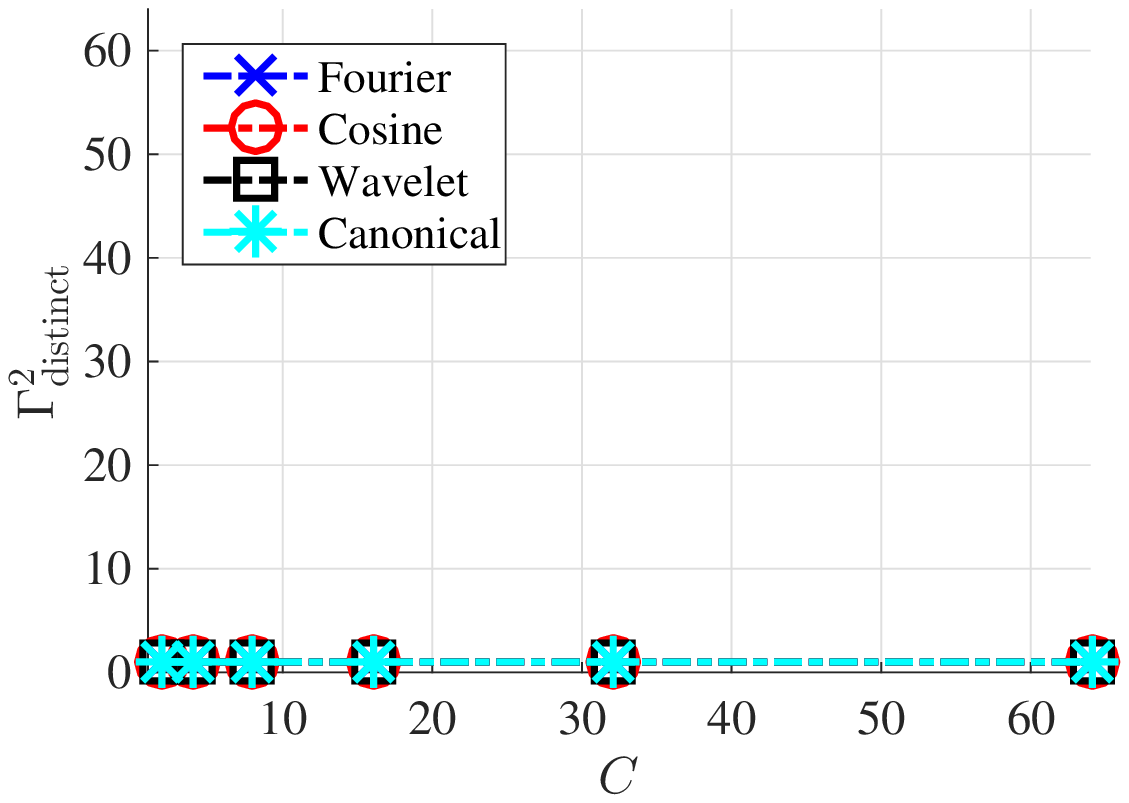} 
\\
\multicolumn{3}{ c }{\small{(a) Computed $\Gamma_{\mathrm{distinct}}^2$ for diagonal sensor profile matrices $H_c$}} \\
\\
{\small (b-1) Perfectly-partitioned $\Lambda_c$} & {\small (b-2) Banded $\Lambda_c$} & {\small (b-3) Globally-spread $\Lambda_c$} \\
\includegraphics[scale=0.455, trim=0.5em 0.45em 2.2em 1em, clip]{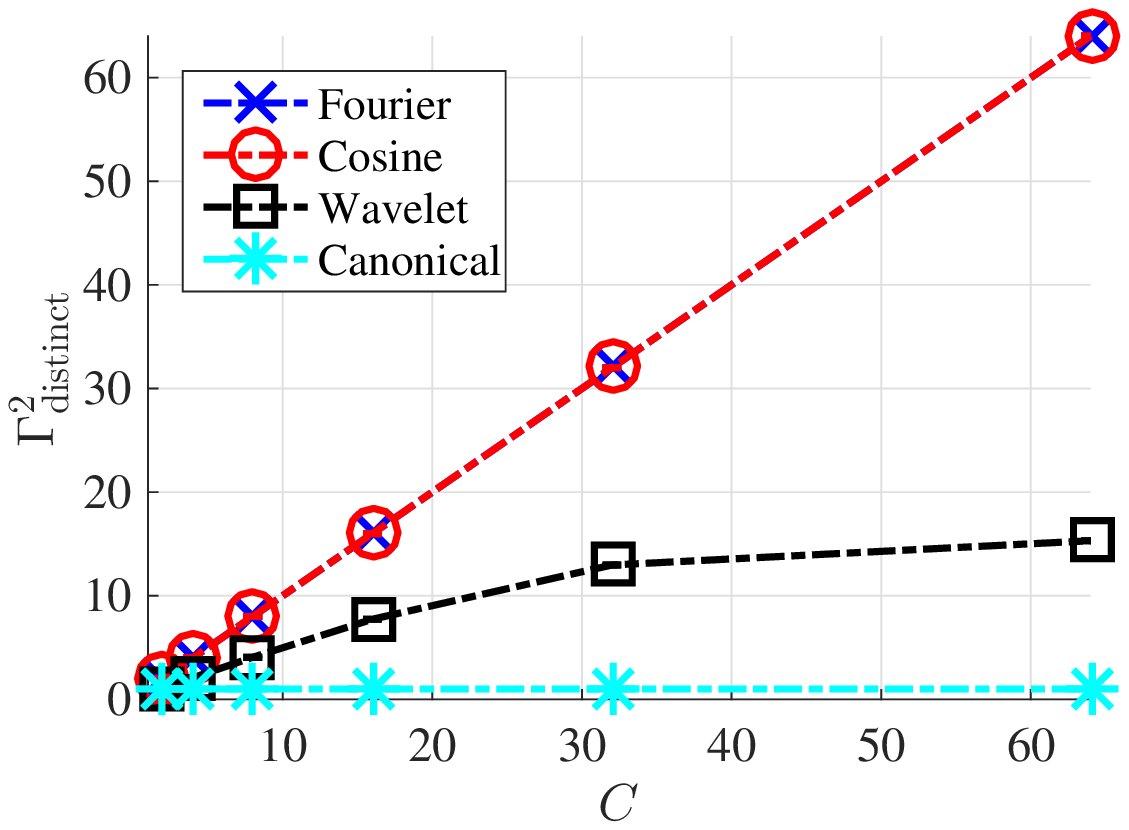} &
\includegraphics[scale=0.455, trim=0.5em 0.45em 2.2em 1em, clip]{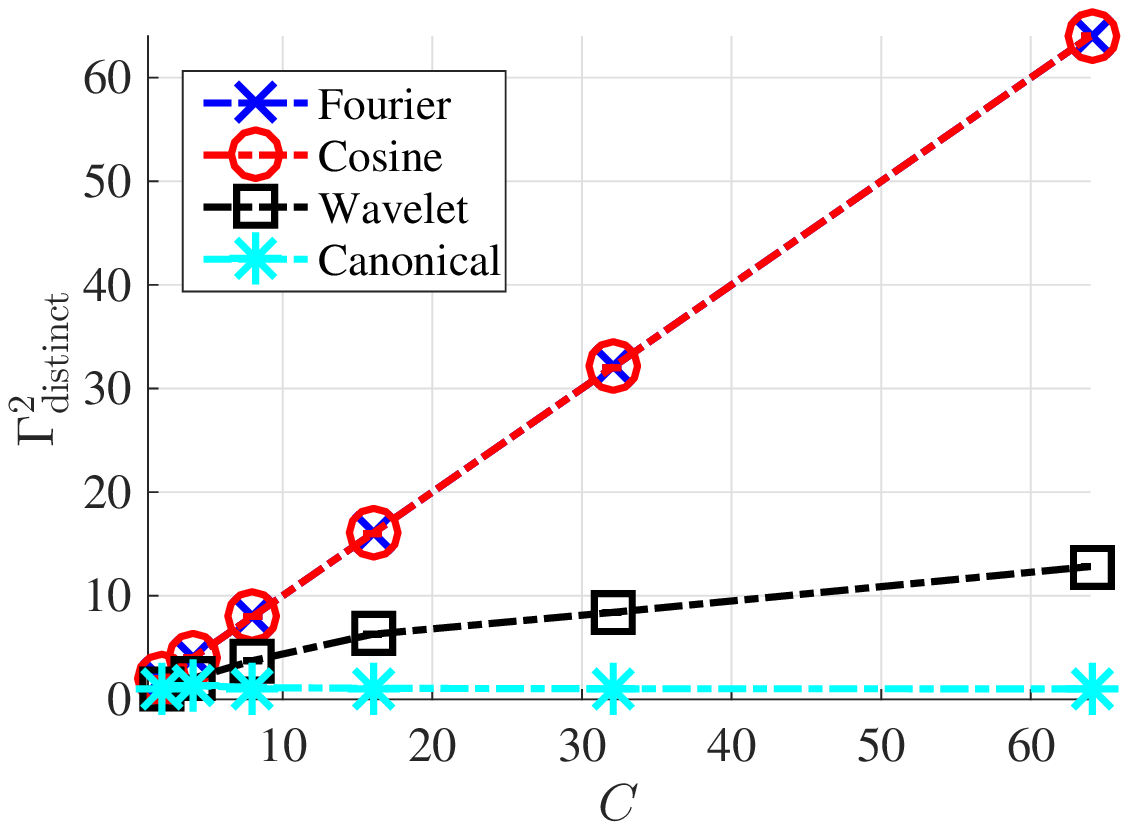} &
\includegraphics[scale=0.455, trim=0.5em 0.45em 2.2em 1em, clip]{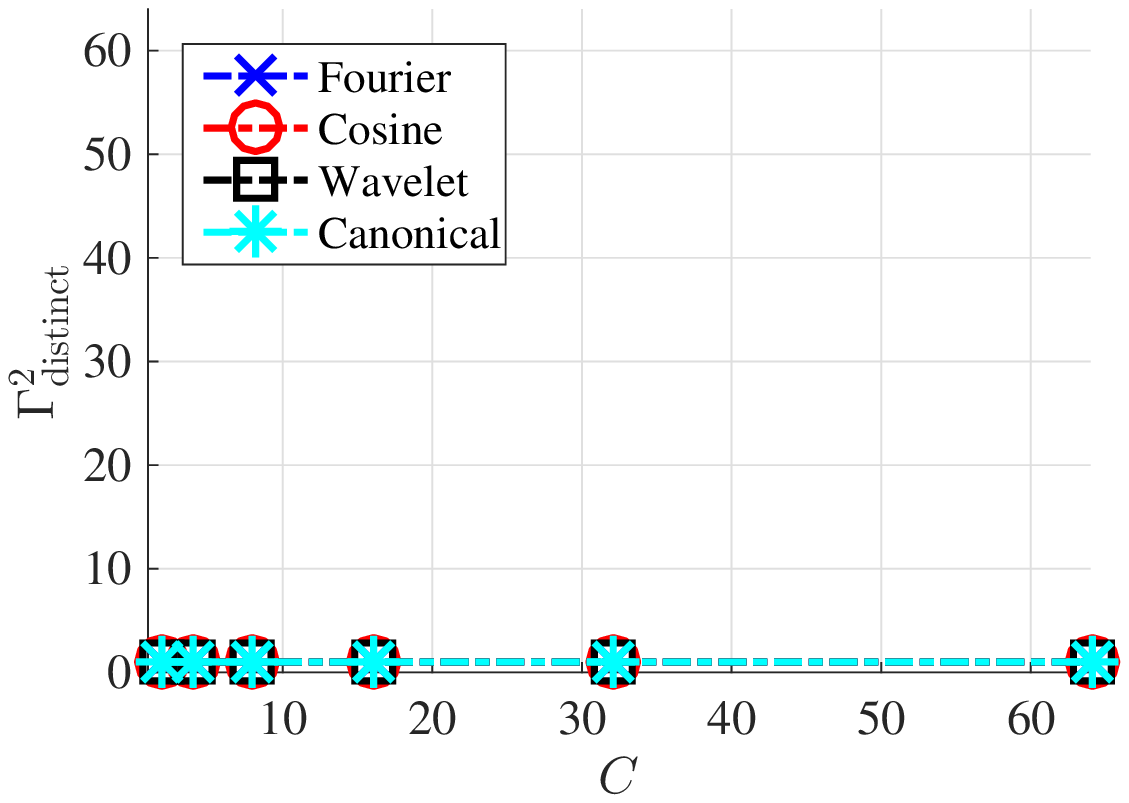} \\
\multicolumn{3}{ c }{\small{(b) Computed $\Gamma_{\mathrm{distinct}}^2$ for circulant sensor profile matrice $H_c = F^* \Lambda_c F$}} 
\end{tabular}

\caption{Computed values of $\Gamma_{\mathrm{distinct}}^2$ against $C$ for different sparsity bases and sensor profile matrices $H_c$.  The value $N = 256$ was used throughout, and for (a-3) and (b-3) the results were averaged over $50$ trials.}
\label{fig:GammaDist}
\end{figure}

\subsubsection{Diagonal sensor profile matrices} \label{sec:Eg:distDiag}

We commence with the following general result:
\lem{
\label{l:diagprofilebounds}
Suppose that the sensor profile matrices $H_{c} = \diag(h_{c})$, where $h_{c} \in \bbC^N$.  If $\Gamma_{\mathrm{distinct}}$ and $\Xi_{\mathrm{distinct}}$ are as in \R{Gammadistinct} and \R{Xidistinct} respectively, then
\bes{
\Gamma_{\mathrm{distinct}} \leq \frac{1}{\sqrt{\alpha}} \sqrt{\mu(U)} \max_{c=1,\ldots,C} \| h_c \|_{2},
}
and
\bes{
\Xi_{\mathrm{distinct}} = \frac{1}{\sqrt{\alpha}} \max_{c=1,\ldots,C} \| h_c \|_{\infty}.
}
}
\prf{
For the first result, we note that
\bes{
\Gamma_{\mathrm{distinct}} = \frac{1}{\sqrt{\alpha}} \max_{c=1,\ldots,C} \max_{j=1,\ldots,N} \sqrt{\sum_{i=1}^N | (h_c)_i u_{i,j} |^2} \leq \frac{1}{\sqrt{\alpha}} \sqrt{\mu(U)} \max_{c=1,\ldots,C} \nm{ h_c }_2.
}
The second result is follows immediately from the diagonal structure of $H_c$, which implies that $\| H_c \|_{2 \rightarrow 2} = \| h_c \|_{\infty}$.
}

Consider examples (\romnum{1})--(\romnum{3}) and recall that $\alpha = 1$ in these cases.  For the perfectly-partitioned sensor profiles (example (\romnum{1})), one has $\| h_c \|_{\infty} = \sqrt{C}$ and $\| h_c \|_{2} \leq \sqrt{N}$.  Hence, by the previous result, $\Xi_{\mathrm{distinct}} = \sqrt{C}$ whereas $\Gamma_{\mathrm{distinct}} \leq \sqrt{\mu(U) N}$.
On the one hand, this means that universal bound of Theorem \ref{t:dist:subgaussRIP_univ} always gives the worst-case measurement condition, scaling linearly with $C$.  This is to be expected.  If $U = I$ for example, then one can easily construct an $s$-sparse vector $x$ for which $H_1 x = \sqrt{C} x$ and $H_{c} x =0$ for $c=2,\ldots,C$.  Thus, only the first sensor provides any nonzero measurements.  On the other hand, if the sparsity basis is incoherent, i.e.\ $\mu(U) \lesssim1/N$, then the nonuniversal bound of Theorem \ref{t:dist:subgaussRIP} is optimal with respect to $C$.  Incoherence means that $Ux$ is spread out when $x$ is sparse, thus all sensors typically provide some useful information about the signal.   Numerical verification of these results are shown in Fig.\ \ref{fig:GammaDist}(a-1).

Similar results apply in the case of banded sensor profiles (example (\romnum{2})).  For coherent sparsity bases (in our case, the canonical and wavelet bases) we expect the worst-case scaling $\Gamma_{\mathrm{distinct}} = \ordu{\sqrt{C}}$ as $C \rightarrow \infty$ (and therefore $\Xi_{\mathrm{distinct}} = \ordu{\sqrt{C}}$ as well), but for incoherent bases we find that $\Gamma_{\mathrm{distinct}} = \ordu{1}$.  See Fig.\ \ref{fig:GammaDist}(a-2).  On the other hand, the global sensor profiles (example (\romnum{3})) provide optimal scalings.  Since $\| h_{c} \|_{\infty} = 1$ in this case, we have $\Gamma_{\mathrm{distinct}}  = \Xi_{\mathrm{distinct}} = 1$ (for the first equality we use the {bounds $\Xi_{\mathrm{distinct}}  \geq \Gamma_{\mathrm{distinct}} $ and  $\Gamma_{\mathrm{distinct}}  \geq 1$ -- see Proposition \ref{prop:GammaXi:bounds:dist}).  Thus, global sensor profiles provide optimal universal and nonuniversal recovery guarantees in the case of distinct sampling.  This is verified in Fig.\ \ref{fig:GammaDist}(a-3).

\subsubsection{Circulant sensor profile matrices} \label{sec:Eg:distCirc}

Recall that for circulant sensor profile matrices we have $H_c = F^* \Lambda_c F$, where $\Lambda_c = \diag(\lambda_c)$ is the diagonal matrix of eigenvalues $\lambda_{c} \in \bbC^N$ of $H_c$ and $F$ is the unitary DFT matrix.

\lem{
Suppose that the sensor profile matrices $H_{c} = F^* \diag( \lambda_c ) F$, where $\lambda_c \in \bbC^N$.  If $\Gamma_{\mathrm{distinct}}$ and $\Xi_{\mathrm{distinct}}$ are as in \R{Gammadistinct} and \R{Xidistinct} respectively, then
\bes{
\Gamma_{\mathrm{distinct}} \leq \frac{1}{\sqrt{\alpha}} \sqrt{\mu(F U)} \max_{c=1,\ldots,C} \| \lambda_c \|_{2},
}
and
\bes{
\Xi_{\mathrm{distinct}} = \frac{1}{\sqrt{\alpha}} \max_{c=1,\ldots,C} \| \lambda_c \|_{\infty}.
}
}
\prf{
We argue as in the proof of Lemma \ref{l:diagprofilebounds}, using the fact that $F$ is unitary.
}

This result is very similar to that for diagonal sensor profiles (Lemma \ref{l:diagprofilebounds}), the only differences being the change from $h_c$ to $\lambda_c$ and $U$ to $F U$.  Thus, similar conclusions apply in the case of examples (i)--(iii).  For examples (i) and (ii), one obtains an optimal scaling for $\Gamma_{\mathrm{distinct}}$ if $F U$ is incoherent, as opposed to just $U$ in the diagonal case.  This is the case for example with the canonical and wavelet bases (of $4$-level decomposition).  Example (\romnum{3}) is explained by a similar reasoning to that of the globally-spread diagonal sensor profiles, since $ \| \lambda_c \|_{\infty} = 1$ in this case as well.  A numerical illustration is shown in Fig.\ \ref{fig:GammaDist}(b).

\subsection{Identical sampling} \label{sec:Eg:idt}

We now consider examples (i)--(iii) in the context of identical sampling.

\begin{figure}[!t]
\centering
\begin{tabular}{ccc}
{\small (a-1) Perfectly-partitioned $H_c$} & {\small (a-2) Banded $H_c$} & {\small (a-3) Globally-spread $H_c$} \\
\includegraphics[scale=0.455, trim=0.5em 0.45em 2.2em 1em, clip]{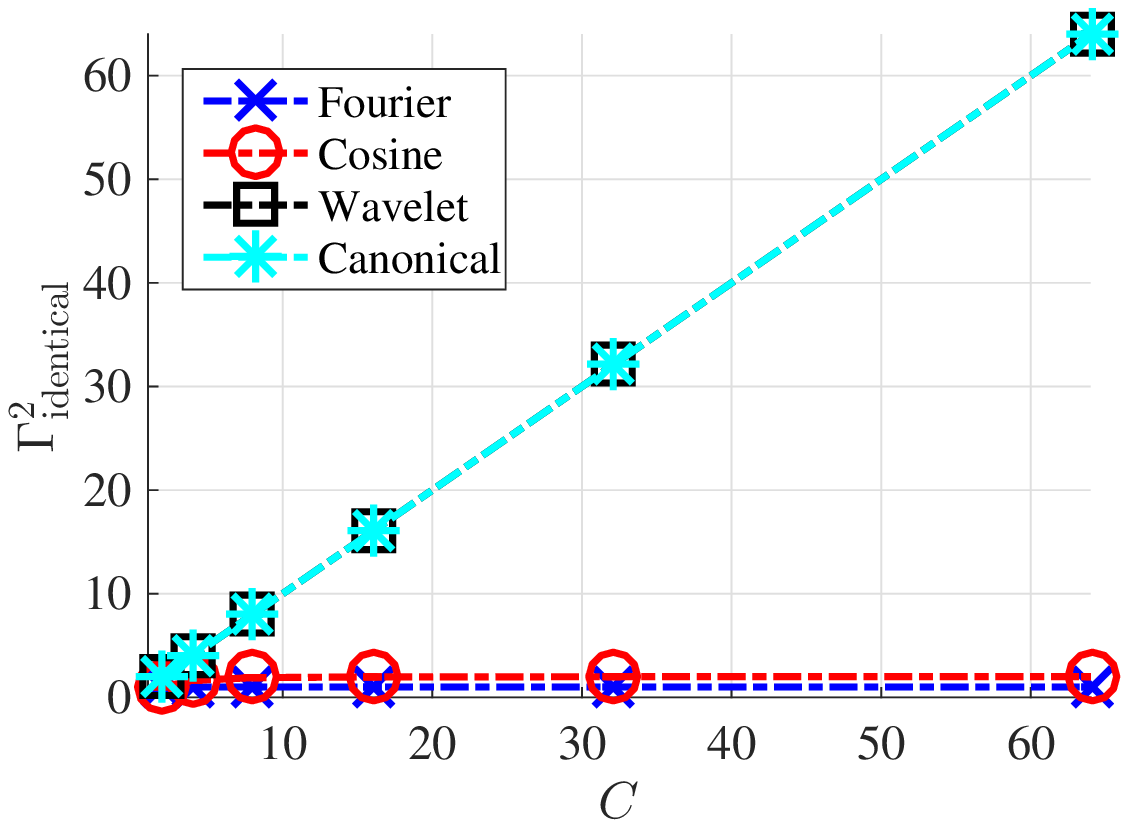} &
\includegraphics[scale=0.455, trim=0.5em 0.45em 2.2em 1em, clip]{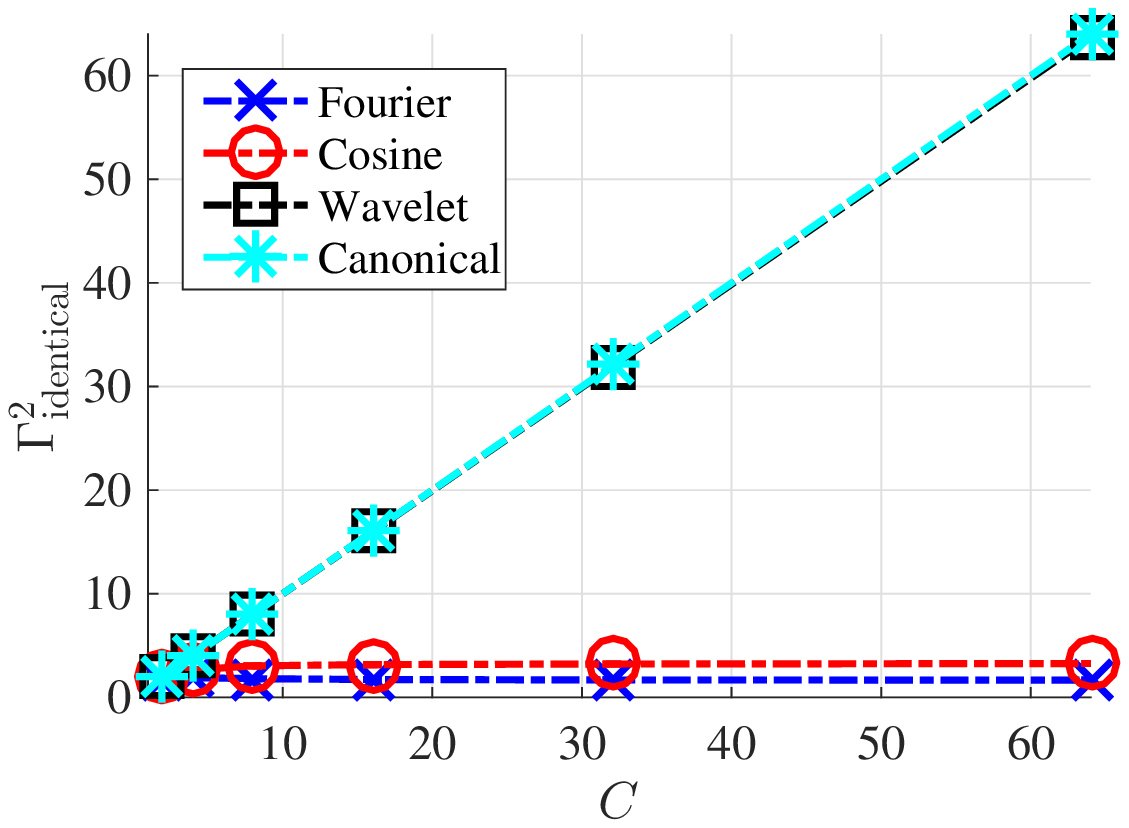} &
\includegraphics[scale=0.455, trim=0.5em 0.45em 2.2em 1em, clip]{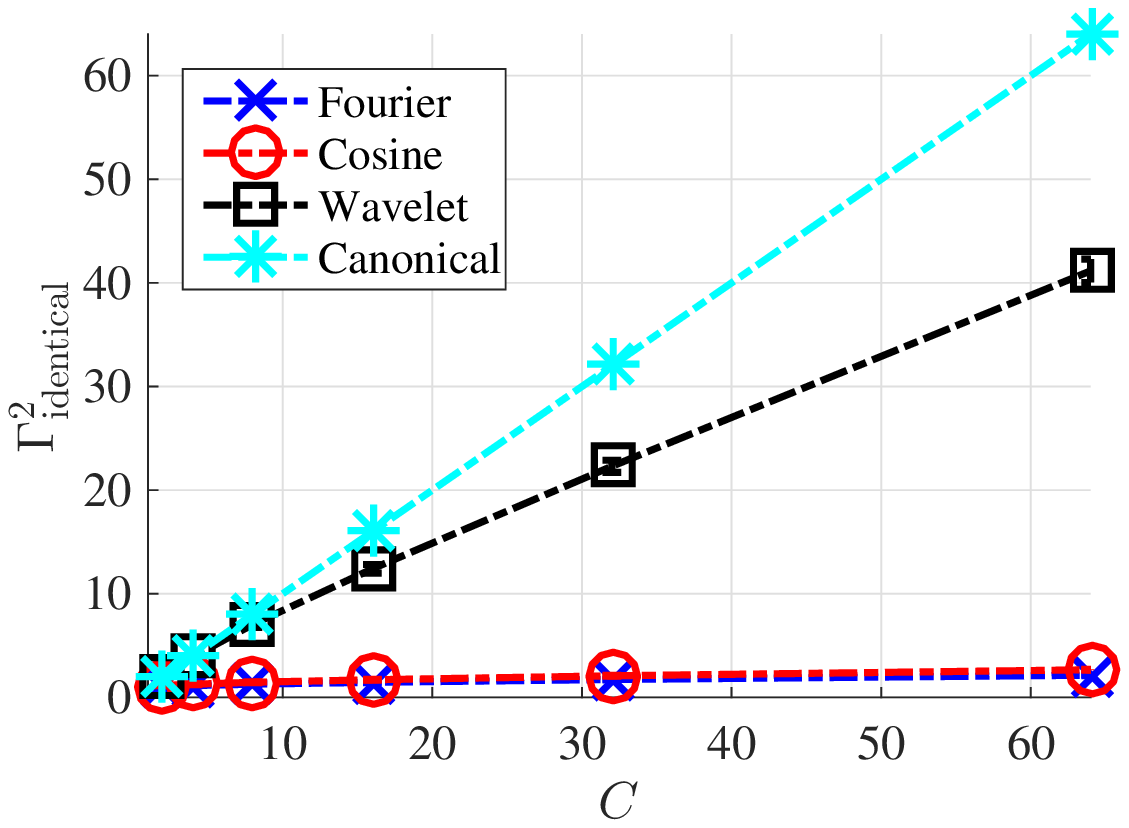} 
\\
\multicolumn{3}{ c }{\small{(a) Computed $\Gamma_{\mathrm{identical}}^2$ for diagonal sensor profile matrices $H_c$}} \\
\\
{\small (b-1) Perfectly-partitioned $\Lambda_c$} & {\small (b-2) Banded $\Lambda_c$} & {\small (b-3) Globally-spread $\Lambda_c$} \\
\includegraphics[scale=0.455, trim=0.5em 0.45em 2.2em 1em, clip]{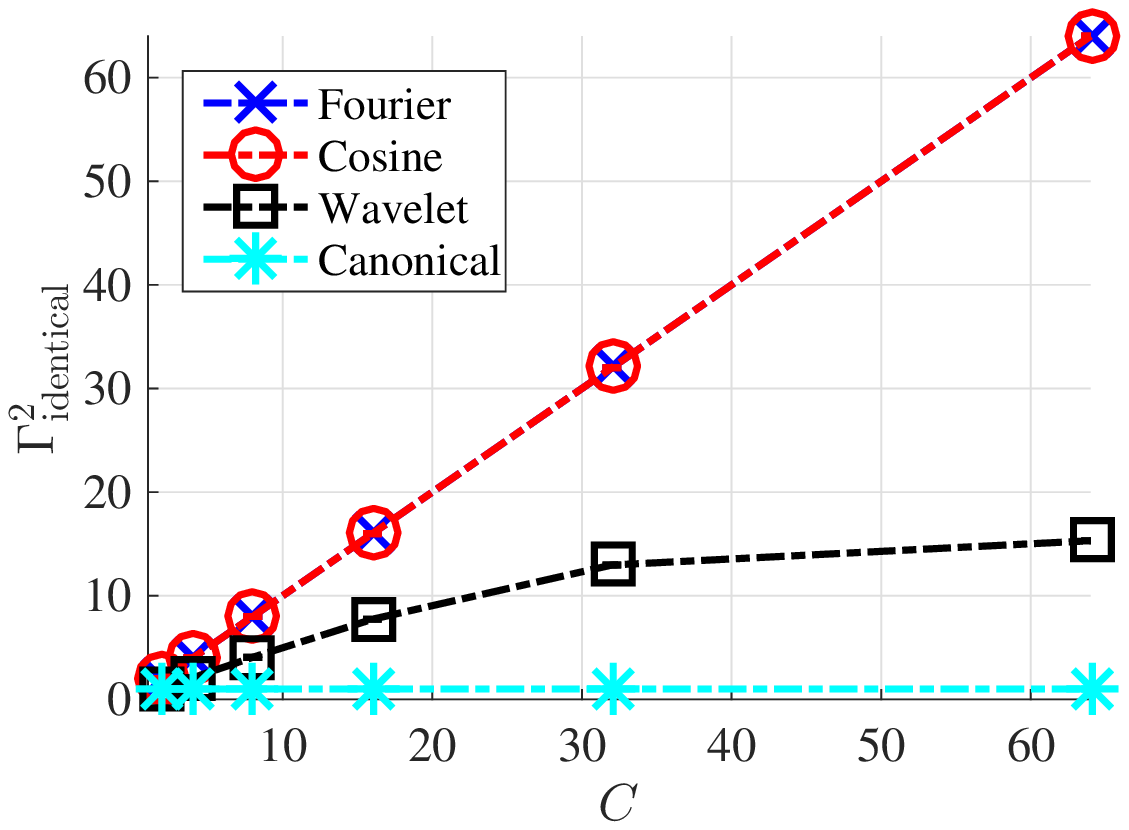} &
\includegraphics[scale=0.455, trim=0.5em 0.45em 2.2em 1em, clip]{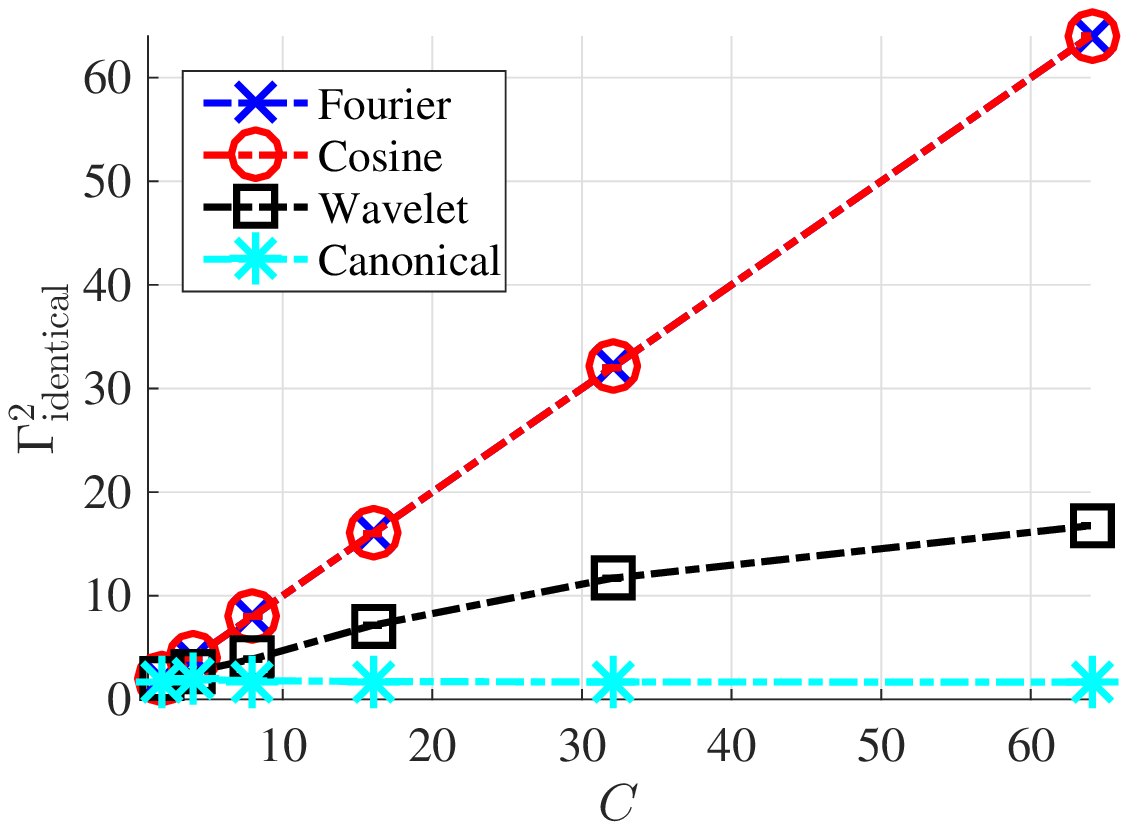} &
\includegraphics[scale=0.455, trim=0.5em 0.45em 2.2em 1em, clip]{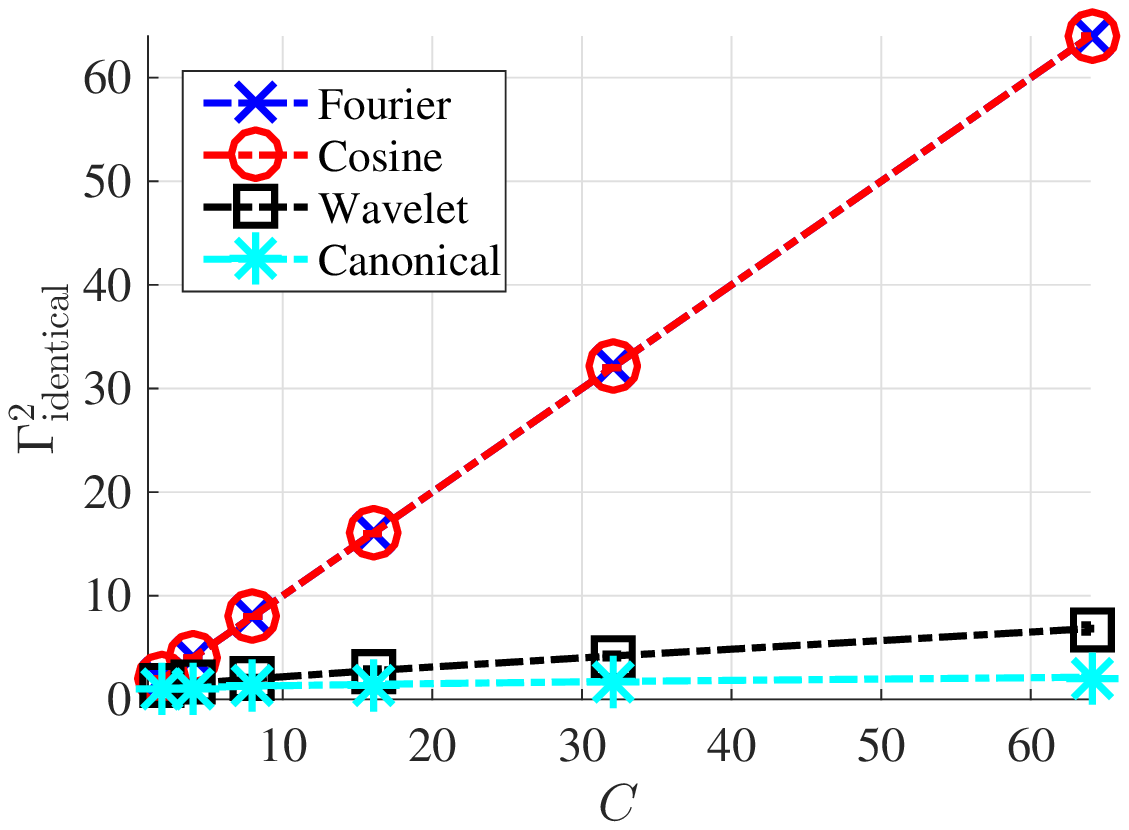} \\
\multicolumn{3}{ c }{\small{(b) Computed $\Gamma_{\mathrm{identical}}^2$ for circulant sensor profile matrice $H_c = F^* \Lambda_c F$}} 
\end{tabular}

\caption{Computed values of $\Gamma_{\mathrm{identical}}^2$ against $C$ for different sparsity bases and sensor profile matrices $H_c$.  The value $N = 256$ was used throughout, and for (a-3) and (b-3) the results were averaged over $50$ trials.}
\label{fig:GammaIdt}
\end{figure}

\subsubsection{Diagonal sensor profile matrices}

We commence with the following lemma:
\lem{
\label{l:Gammaidtworstcase}
Suppose the sensor profile matrices $H_{c} = \diag(h_{c})$, where $h_{c} \in \bbC^N$.  If $\Gamma_{\mathrm{identical}}$ and $\Xi_{\mathrm{identical}}$ are as in \R{Gammaidentical} and \R{Xiidentical} respectively, then
\bes{
\sqrt{\mu(U) C} \leq \Gamma_{\mathrm{identical}} \leq \sqrt{\beta/\alpha} \sqrt{C}
\qquad \mathrm{and} \qquad 
\sqrt{C} \leq \Xi_{\mathrm{identical}} \leq \sqrt{\beta/\alpha} \sqrt{C}.
}
}
\prf{
Note that both upper bounds were proved in Proposition \ref{prop:GammaXi:bounds:idt:2}.  For the lower bound for $\Gamma_{\mathrm{identical}}$, we note that
\eas{
 \left\| \left[ \begin{array}{ccc} H_1 U e_j  \cdots  H_C U e_j \end{array} \right] \right\|^2_{2 \rightarrow 2} = \max_{\substack{ \| z \|_2 = 1 \\ z \in \bbC^C}} \nm{\sum^{C}_{c=1} z_c H_c U e_j }_2^2 = \max_{\substack{ \| z \|_2 = 1 \\ z \in \bbC^C}} \sum^{N}_{i=1} | u_{i,j} |^2 \left | \sum^{C}_{c=1} (h_c)_i z_c \right |^2
 }
 and
 \bes{
 \max_{\substack{ \| z \|_2 = 1 \\ z \in \bbC^C}} \sum^{N}_{i=1} | u_{i,j} |^2 \left | \sum^{C}_{c=1} (h_c)_i z_c \right |^2
 \geq |u_{i,j} |^2 \max_{\substack{ \| z \|_2 = 1 \\ z \in \bbC^C}}  \left | \sum^{C}_{c=1} (h_c)_i z_c \right |^2
  = |u_{i,j} |^2 \sum^{C}_{c=1} | (h_{c})_i |^2 
 \geq | u_{i,j} |^2 C \alpha,
}
where the first inequality holds for all $i$, and in the last step we use the fact that $\alpha \leq C^{-1} \sum^{C}_{c=1} | (h_c)_i |^2 \leq \beta$, due to the joint near-isometry condition \R{eq:joint_iso_idt}.  Since $i$ and $j$ are arbitrary, we deduce the lower bound.   Finally, for the lower bound for $\Xi_{\mathrm{identical}} $, we recall first from Proposition \ref{prop:GammaXi:bounds:idt:2} that $\Xi_{\mathrm{identical}}  \geq \Gamma_{\mathrm{identical}} $ for any choice of sparsity basis $U$, then apply the lower bound for $\Gamma_{\mathrm{identical}}$ with the choice $U = I$.
}

This primary usefulness of this lemma is in establishing worst-case recovery guarantees.  In particular, if $U$ is coherent, i.e.\ $\mu(U) \approx 1$, then the measurement condition necessarily scales linearly with $C$, regardless of the choice of diagonal sensor profiles.  This is illustrated in Fig.\ \ref{fig:GammaIdt}(a), where the coherent canonical and wavelet bases both exhibit the worst-case scaling for $\Gamma_{\mathrm{identical}}$.  Note that this lemma also implies that an optimal universal bound cannot be achieved with diagonal sensor profile matrices.

On the other hand, Fig.\ \ref{fig:GammaIdt}(a) indicates that optimal nonuniversal bounds are possible, at least for the incoherent Fourier and cosine bases.  We shall now establish this result theoretically.  The following lemma applies to examples (\romnum{1}) and (\romnum{2}):

\lem{
\label{l:Gammaidt:bound:eg}
Let $H_c = \diag(h_c)$ and $q$ be as in \R{qdef}.  Then
\bes{
\sqrt{\mu(U) C} \leq \Gamma_{\mathrm{identical}} \leq \frac{1}{\sqrt{\alpha}} \sqrt{\mu(U) q}\max_{c=1,\ldots,C} \| h_c \|_2.
}
In particular, for the perfectly partitioned sensor profiles (example (i)) one has
\bes{
\sqrt{\mu(U) C} \leq \Gamma_{\mathrm{identical}} \leq \sqrt{\mu(U) N},
}
and for the banded sensor profiles (example (ii)) one has
\bes{
\sqrt{\mu(U) C} \leq \Gamma_{\mathrm{identical}} \lesssim \sqrt{\mu(U) N}.
}
}
\prf{
The lower bounds are due to Lemma \ref{l:Gammaidtworstcase}.
For the upper bound, we note that
\bes{
\alpha \Gamma^2_{\mathrm{identical}}  = \max_{j=1,\ldots,N} \max_{\substack{ \| z \|_2 = 1 \\ z \in \bbC^C}} \sum^{N}_{i=1} | u_{ij} |^2 \left | \sum^{C}_{c=1} (h_c)_i z_c \right |^2  \leq \mu(U) \max_{\substack{ \| z \|_2 = 1 \\ z \in \bbC^C}}  \sum^{N}_{i=1} \left | \sum^{C}_{c=1} (h_c)_i z_c \right |^2
}
and therefore
\eas{
\alpha \Gamma^2_{\mathrm{identical}}  & \leq \mu(U)\max_{\substack{ \| z \|_2 = 1 \\ z \in \bbC^C}} \sum^{C}_{c,d=1} z_c \overline{z_d} h^*_d h_c
= \mu(U)\max_{\substack{ \| z \|_2 = 1 \\ z \in \bbC^C}} \sum^{C}_{c=1} \sum_{d : \supp(h_d) \cap \supp(h_c) \neq \emptyset} z_c \overline{z_d} h^*_d h_c
\\
& \leq \frac12 \mu(U) \max_{c=1,\ldots,C} \| h_c \|^2_2 \max_{\substack{ \| z \|_2 = 1 \\ z \in \bbC^C}} \sum^{C}_{c=1} \sum_{d : \supp(h_d) \cap \supp(h_c) \neq \emptyset} \left ( |z_c|^2 + |z_d|^2 \right )
\\
& \leq \mu(U) q \max_{c=1,\ldots,C} \| h_c \|^2_2,
}
as required.  Note that in the last step we use the definition of $q$.  This gives the first result.  For the other two results we merely observe that $\alpha = \beta = 1$ in both cases, and $q = 1$ for example (i) and $q = \ord{1}$ as $C \rightarrow \infty$ for example (ii).
}
This lemma confirms that for incoherent sparsity bases, much like for distinct sampling, one has an optimal recovery guarantee with examples (\romnum{1})--(\romnum{2}).  See Fig.\ \ref{fig:GammaIdt}(a-1).

Finally, we consider example (\romnum{3}):
\lem{
\label{l:idt_global_diag}
Let $0 <  \varepsilon < 1$ and suppose that $H_{c} = \diag(h_{c})$, where the $h_{c} \in \bbC^N$ are independent Rademacher sequences.  Then
\bes{
\Gamma_{\mathrm{identical}} \leq \frac{1}{\sqrt{\alpha}} \left( \sqrt{N} + \sqrt{C} + \sqrt{2 c^{-1} \log (2 \varepsilon^{-1}) } \right) \sqrt{\mu(U)},
}
with probability at least $1-\varepsilon$, where $c$ is a universal constant.
}
\prf{
Observe that
\bes{
\alpha \Gamma^2_{\mathrm{identical}} = \max_{j=1,\ldots,N}  \max_{\substack{ \| z \|_2 = 1 \\ z \in \bbC^C}} \sum^{N}_{i=1} | u_{i,j} |^2 \left | \sum^{C}_{c=1} (h_c)_i z_c \right |^2 \leq \mu(U) \sigma_{\max}^2(G),
}
where $G \in \bbC^{N \times C}$ is the matrix with entries $G_{i,c} = (h_c)_i$.  Thus $G$ is a Bernoulli random matrix.  
According to \cite[Thm.\ 5.39]{Vershynin:bookCh}, we have the following union bound for $\sigma_{\max}(G)$:
\bes{
\bbP \left\{ \sigma_{\max}(G) > \sqrt{N} + c \sqrt{C} + t \right\} \leq 2 \exp(-c t^2/2),\qquad t \geq 0,
}
for some universal constant $c>0$.  Taking $t=\sqrt{2 c^{-1} \log(2 / \varepsilon)}$ gives
\bes{
\bbP \left\{ \sigma_{\max}(G) > \sqrt{N} + \sqrt{C} + \sqrt{2 \log(2 / \varepsilon)} \right\} \leq \varepsilon,
}
which completes the proof.
}
}

We note that Lemma \ref{l:idt_global_diag} is the first result to consider randomness in the sensor profile matrix. The concentration inequality technique used in \cite[Lem.\ 3]{Chun&Li&Adcock:16MMSPARSE} cannot be exploited because randomness in the $H_c$'s typically breaks the independence assumption of measurements.  We remark also that  we are currently unaware of a deterministic construction of diagonal sensor profiles in the identical sampling case which achieves a bound similar to the one presented in this lemma.

\subsubsection{Circulant sensor profile matrices}
As before, suppose that $H_c = F^* \Lambda_c F$ where $F$ is the unitary DFT matrix and $\Lambda = \diag(\lambda_c)$ is the diagonal matrix of eigenvalues.  Observe that
\eas{
\Gamma_{\mathrm{identical}} &= \frac{1}{\sqrt{\alpha}} \max_{j=1,\ldots,N} \left\| \left[ \begin{array}{ccc} H_1 U e_j & \cdots & H_C U e_j \end{array} \right] \right\|_{2 \rightarrow 2}
\\
& = \frac{1}{\sqrt{\alpha}} \max_{j=1,\ldots,N} \left\| \left[ \begin{array}{ccc} \Lambda_1 F U e_j & \cdots & \Lambda_C F U e_j \end{array} \right] \right\|_{2 \rightarrow 2}.
}
This is precisely the $\Gamma_{\mathrm{identical}} $ for a diagonal sensor profile setup with matrices $\hat{H}_{c} = \Lambda_c$ and with sparsity basis $\hat{U} = F U$.  Thus, we may apply all the bounds proved in the previous section to this case.  In particular, for examples (i)--(iii) we have an optimal recovery guarantee whenever $F U$ is incoherent, i.e.\ for the wavelet (of $4$-level-decomposition) and canonical bases respectively.  
Conversely, for the Fourier and cosine bases we get the worst recovery guarantee.  See Fig.\ \ref{fig:GammaIdt}.

\subsection{Block diagonal measurement matrices and relation to previous work} \label{sec:relation}

As discussed in \S \ref{ss:blockdiagintro}, block-diagonal sensing matrices are a special case of our framework that are of independent interest in CS.  This case corresponds to the perfectly-partitioned sensor profiles of example (i), since for these profiles the overall measurement matrix $A$ is block diagonal, i.e.\ 
\be{
\label{def:blkDiagSys}
A = \sqrt{\frac{C}{m}} \left [ \begin{array}{c} A_1 \\ \vdots \\ A_C \end{array} \right ] = \sqrt{\frac{C}{m}} \left [ \begin{array}{ccc} \Phi_1 \\ & \ddots & \\ & & \Phi_C \end{array} \right ] U,
}
where $\Phi_{c} =  \tilde{A}_c P_{I_c} \in \bbC^{m/C \times N/C}$, $c = 1,\ldots,C$, are independent subgaussian random matrices.  Note that the distinct and identical setups give exactly the same measurement matrix in this case.
The following result gives our main recovery guarantee for such sensing matrices.  For convenience, we now introduce the matrices $U_{1},\ldots,U_{C} \in \bbC^{N/C \times N}$ so that
\be{
\label{eq:def:Uc}
U = \left [ \begin{array}{c} U_1 \\ \vdots \\ U_C \end{array} \right ].
}
\cor{[The RIP for subgaussian block-diagonal measurement matrices] \label{c:RIP_blkDiag}
For $0 < \delta, \varepsilon < 1$ and $1 \leq s \leq N$ let $A$ be measurement matrix \R{def:blkDiagSys}.  If
\be{
\label{blockdiagmeascond}
m \gtrsim \delta^{-2} \cdot  \bar{\Gamma}^2(U) \cdot s \cdot \left(  \ln^2(2s) \ln(2N) \ln(2m) + \ln(2/\varepsilon) \right),
}
where
\be{
\label{eq:Gamma_dist_perfP}
\bar{\Gamma}(U) =  \sqrt{C} \max_{c=1,\ldots,C} \max_{j=1,\ldots,N} \| U_c e_j \|_{2},
}
then with probability at least $1-\varepsilon$, the RIC of $A$ satisfies $\delta_s \leq \delta$.  The constant $\bar{\Gamma}(U)$ satisfies
\be{
\label{barmubound}
\bar{\Gamma}(U) \leq \min \left \{ \sqrt{N \mu(U)} , \sqrt{C} \right \},
}
where $\mu(U)$ is the coherence of $U$.
}
\prf{
As noted above, it suffices to consider the distinct setup.  By definition one has
\bes{
\Gamma_{\mathrm{distinct}} = \sqrt{C} \max_{c=1,\ldots,C} \max_{j=1,\ldots,N} \| U_c e_j \|_{2} = \bar{\Gamma}(U).
}
Hence Theorem \ref{t:dist:subgaussRIP} gives the first result.
Also, \R{barmubound} immediately follows from the definition of $U_c$ in \R{eq:def:Uc} and $\mu(U)$.
}

This result shows that the recovery guarantee for \R{def:blkDiagSys} is determined by the constant $\bar{\Gamma}(U)$.  As shown in \R{barmubound}, is $\ord{1}$ as $C \rightarrow \infty$ whenever the sparsity basis is incoherent.

As discussed in \S \ref{ss:blockdiagintro}, the RIP for block-diagonal sensing matrices with subgaussian blocks was studied systematically in \cite{Eftekhari&etal:15ACHA}.  
Therein the authors refer to DBD matrices to describe the system model in \R{def:blkDiagSys}.
Their main results for the DBD case is the measurement condition\footnote{These bounds imply the RIC of the measurement matrix $\delta_s \leq \delta$, except with probability $\lesssim N^{-\ln(N) \log^2(s)}$.  The slight difference in the log factors between these and our result \R{blockdiagmeascond} is due to our choice in leaving the failure probability as a parameter $\varepsilon$ and a minor improvement in one of the log terms from $\ln(N)$ to $\ln(m)$.  Replacing $\ln(m)$ by $\ln(N)$ and setting $\varepsilon$ so that the two log terms in \R{blockdiagmeascond} matched would give the same log factors as these bounds.}
\bes{
m \gtrsim \delta^{-2} \cdot \tilde{\mu}^2(U) \cdot s \cdot \ln^2(s) \cdot \ln^2(N),
}
where
\be{
\label{eq:DBDconst}
\tilde{\mu}(U) = \min \{ \sqrt{\mu(U) N}, \sqrt{C} \}.
}
Note that
\bes{
\bar{\Gamma}(U) \leq \tilde{\mu}(U),
}
by inequality in \R{barmubound}.
Hence, in general, Corollary \ref{c:RIP_blkDiag} gives a better recovery guarantee than the DBD subgaussian sensing results in \cite{Eftekhari&etal:15ACHA}.  Note that \cite{Eftekhari&etal:15ACHA} also consider \textit{repeated block diagonal (RBD)} subgaussian matrices, where the matrices $\Phi_1 = \ldots = \Phi_C = \Phi$ in \R{def:blkDiagSys} are repeats of a single subgaussian matrix $\Phi$.  This may also be viewed as an instance of our identical sampling setup with matrix $\tilde{A} = [\hat{A} \cdots \hat{A}]$, where $\hat{A}^{m/C \times N/C}$ is a subgaussian random matrix.  This generalization is omitted here for simplicity.

\section{Empirical phase transition results and discussion}

In this section, we present empirical validation of theoretical results using phase transitions  (see, for example, \cite{Monajemi&etal:13PNAS} and references therein). 

\subsection{Simulation setup}
The overall simulation setup is as follows. For an $s$-sparse signal $x \in \bbC^{128}$, the positions of $s$ non-zero elements were chosen uniformly at random without replacement, and the non-zero elements were chosen randomly and uniformly distributed on the unit circle.  
Subgaussian random matrices were constructed using i.i.d.\ Gaussian random variables having zero mean and unit variance.
For the phase transition graph of resolution $50 \times 50$, the horizontal and vertical axes are defined by $m/CN \in (0,1]$ and $s/N \in (0,1]$ respectively.  The empirical success fraction was calculated as $\#\{\textmd{successes}\}/\#\{\textmd{trials}\}$ using $20$ trials, where success corresponds to a relative recovery error $\| x - \hat{x} \|_{2} / \| x\|_2 < \mathrm{tol}$ for $\mathrm{tol} = 0.001$.  Throughout, we use CVX with the MOSEK solver \cite{cvx, Grand&Boyd:08cvx}.  The empirical phase transition point is obtained by finding the closest (and at least) $50\%$ empirical success point.

Based on this setup, phase transitions curves were computed for the following four cases:
\bulls{
\item[(a)] Diagonal sensor profile matrices with the canonical sparsity basis and perfectly-partitioned sensor profiles (see example (\romnum{1}) in \S \ref{sec:Eg}).
\item[(b)] Diagonal sensor profile matrices with the Fourier sparsity basis and perfectly-partitioned sensor profiles (see example (\romnum{1}) in \S \ref{sec:Eg}).
\item[(c)] Circulant sensor profile matrices with the canonical sparsity basis and globally-spread sensor profiles (see example (\romnum{3}) in \S \ref{sec:Eg}).
\item[(d)] Circulant sensor profile matrices with the Fourier sparsity basis and globally-spread sensor profiles (see example (\romnum{3}) in \S \ref{sec:Eg}).
}

\subsection{Results and discussion}
\begin{figure}[!t]
\centering
\begin{tabular}{cccc}
{\small \shortstack{(a) Perfectly-partitioned diagonal \\ sensor and canonical sparsity basis}} 
& {\small \shortstack{(b) Perfectly-partitioned diagonal \\ sensor and Fourier sparsity basis}} \\
\includegraphics[scale=0.475, trim=0.5em 0.45em 2.5em 2em, clip]{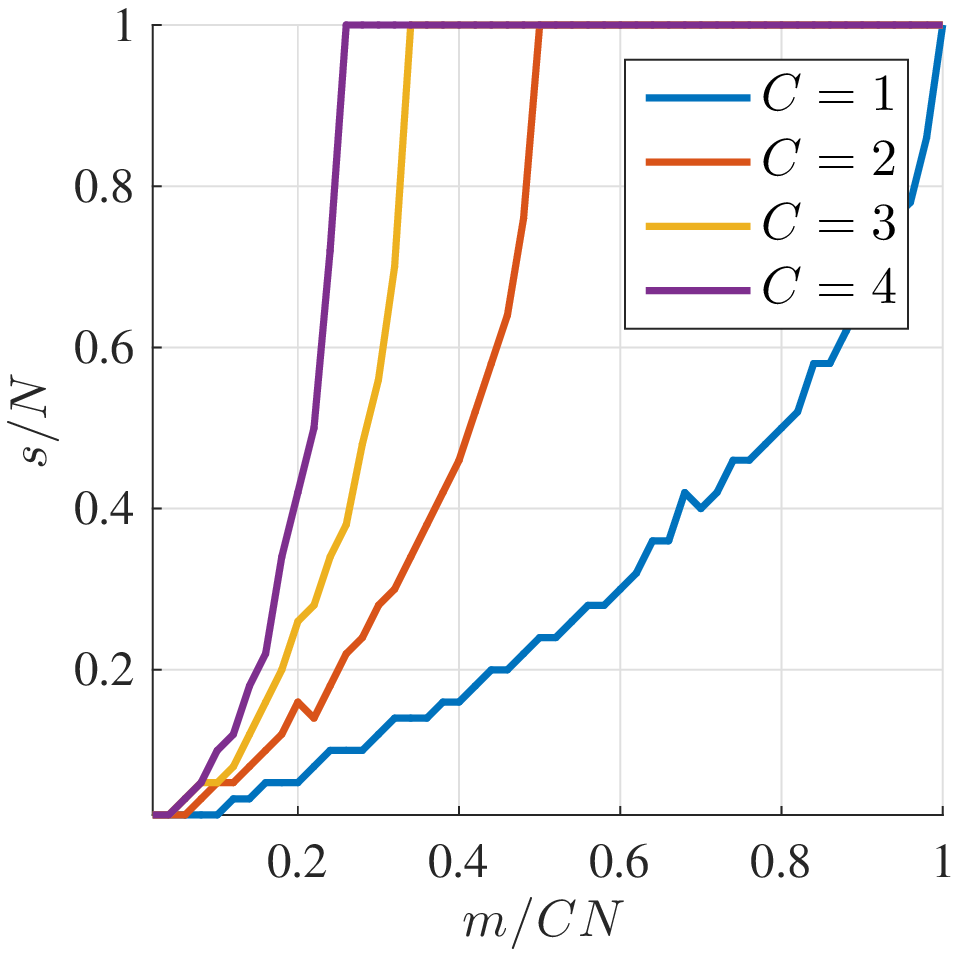} &
\includegraphics[scale=0.475, trim=0.5em 0.45em 2.5em 2em, clip]{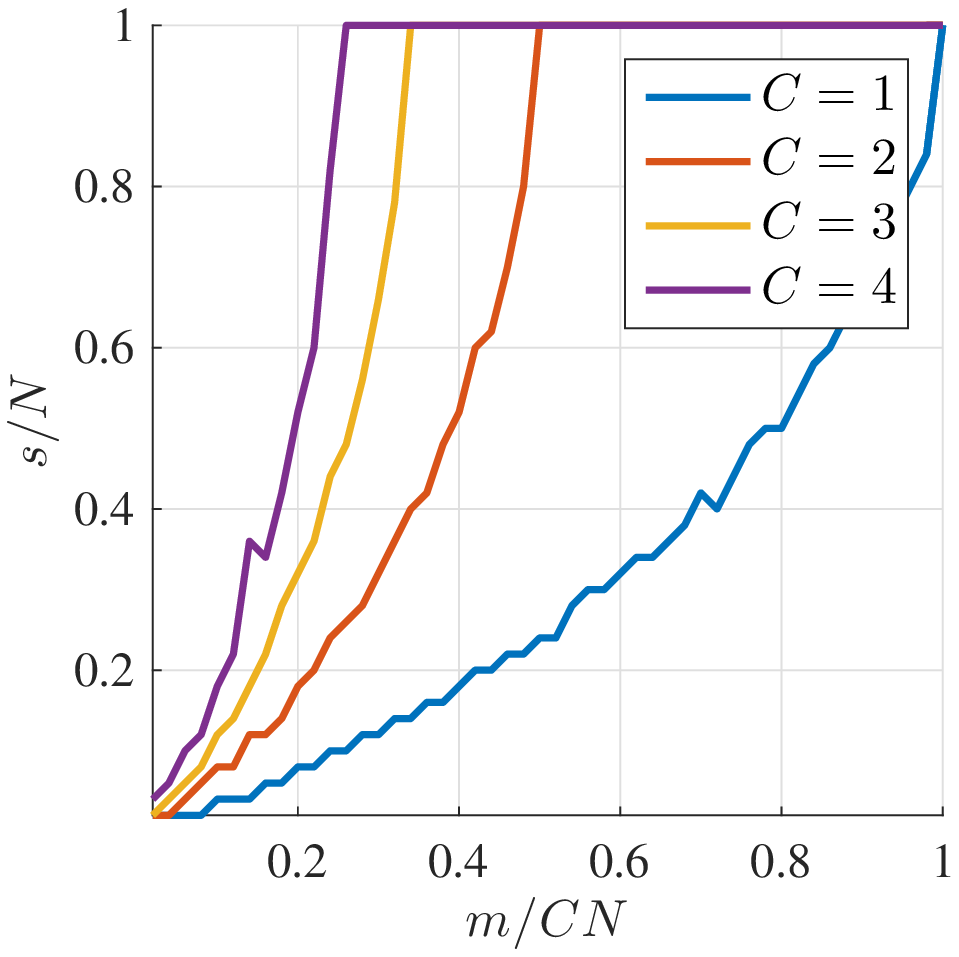} \\

{\small \shortstack{(c) Globally-spread spectrum circulant \\ sensor and canonical sparsity basis}} 
&{\small \shortstack{(d) Globally-spread spectrum circulant \\ sensor and Fourier sparsity basis}} \\
\includegraphics[scale=0.475, trim=0.5em 0.45em 2.5em 2em, clip]{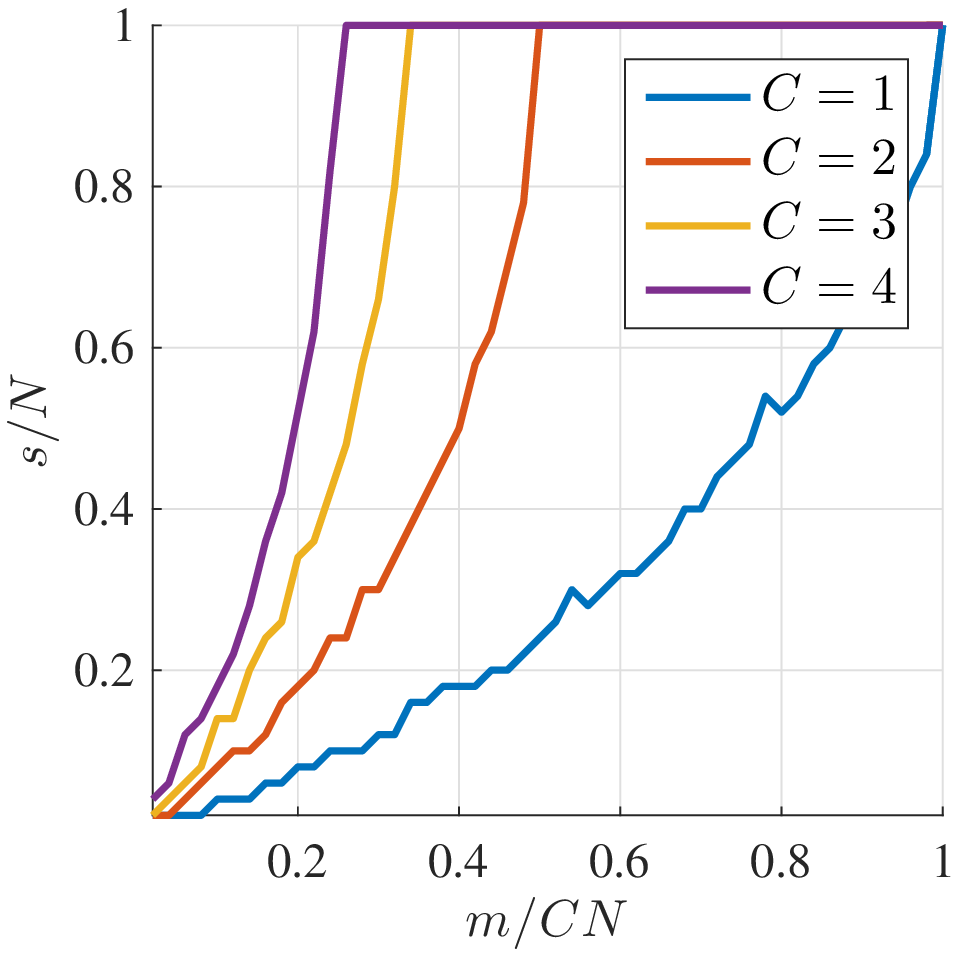} &
\includegraphics[scale=0.475, trim=0.5em 0.45em 2.5em 2em, clip]{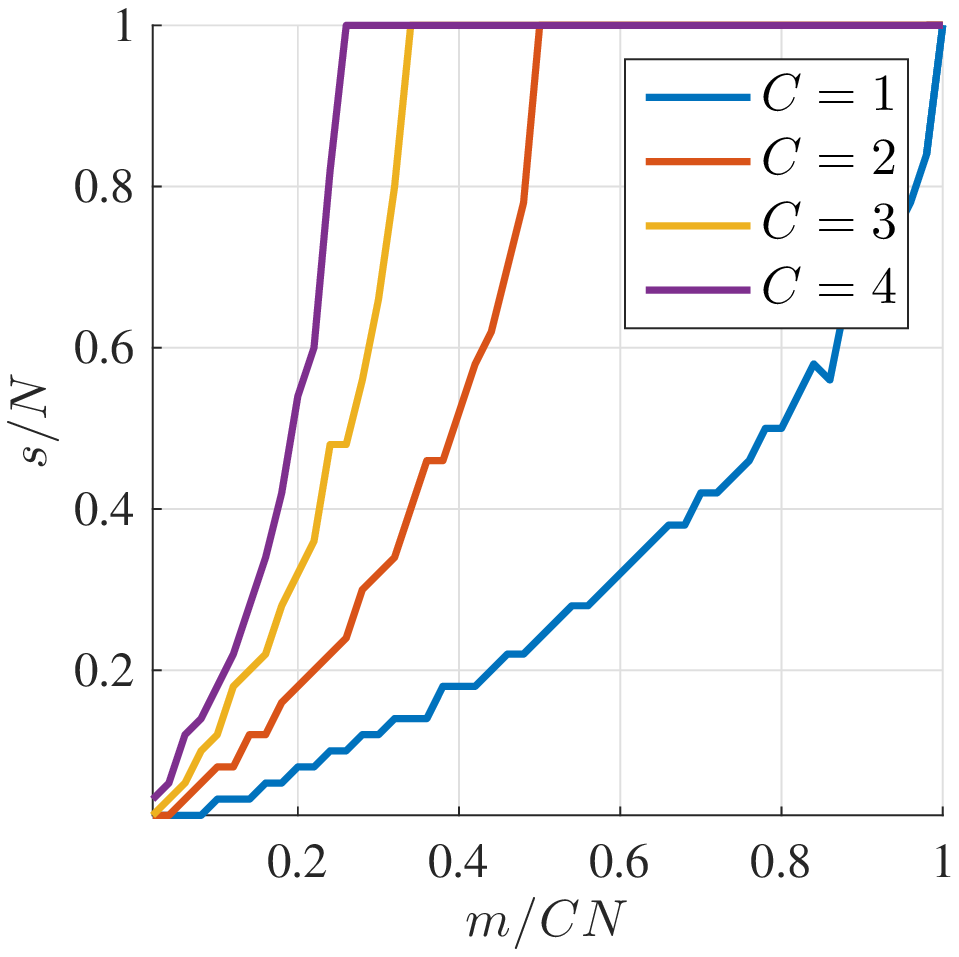} 
\end{tabular}

\caption{Empirical phase transition results for the distinct sampling scenario.
For both sampling scenarios, the empirical probability of successful recovery increases as $C$ increases. The results in (b) and (c) are in agreement with our theoretical results.
}
\label{fig:PT_dist}
\end{figure}

\begin{figure}[!t]
\centering
\begin{tabular}{cccc}
{\small \shortstack{(a) Perfectly-partitioned diagonal \\ sensor and canonical sparsity basis}} 
& {\small \shortstack{(b) Perfectly-partitioned diagonal \\ sensor and Fourier sparsity basis}} \\
\includegraphics[scale=0.475, trim=0.5em 0.45em 2.5em 2em, clip]{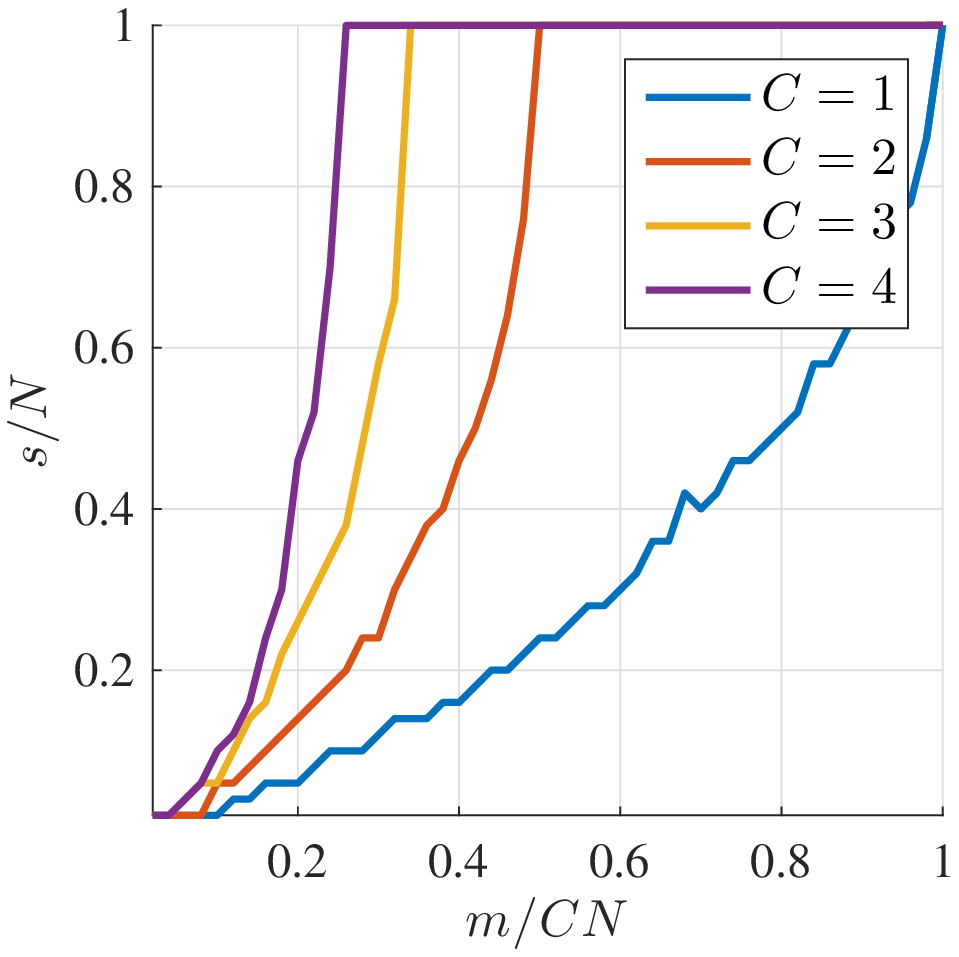} &
\includegraphics[scale=0.475, trim=0.5em 0.45em 2.5em 2em, clip]{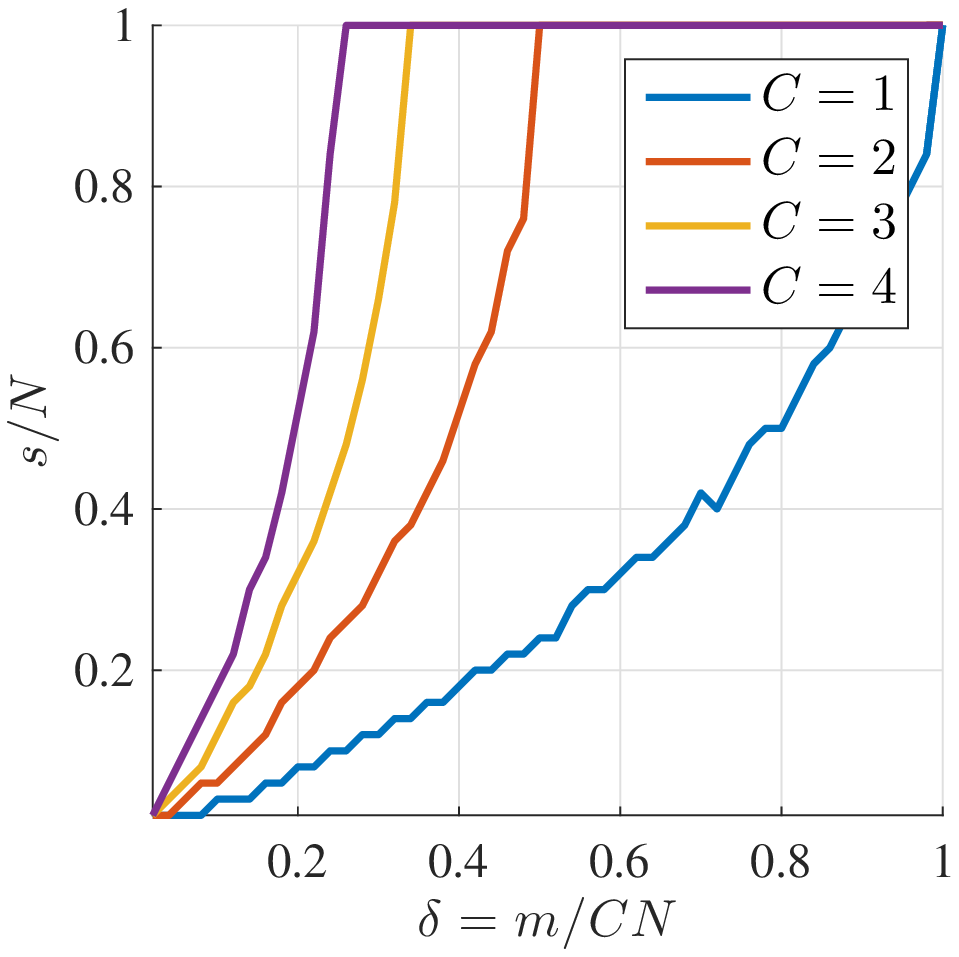} \\

{\small \shortstack{(c) Globally-spread spectrum circulant \\ sensor and canonical sparsity basis}} 
&{\small \shortstack{(d) Globally-spread spectrum circulant \\ sensor and Fourier sparsity basis}} \\
\includegraphics[scale=0.475, trim=0.5em 0.45em 2.5em 2em, clip]{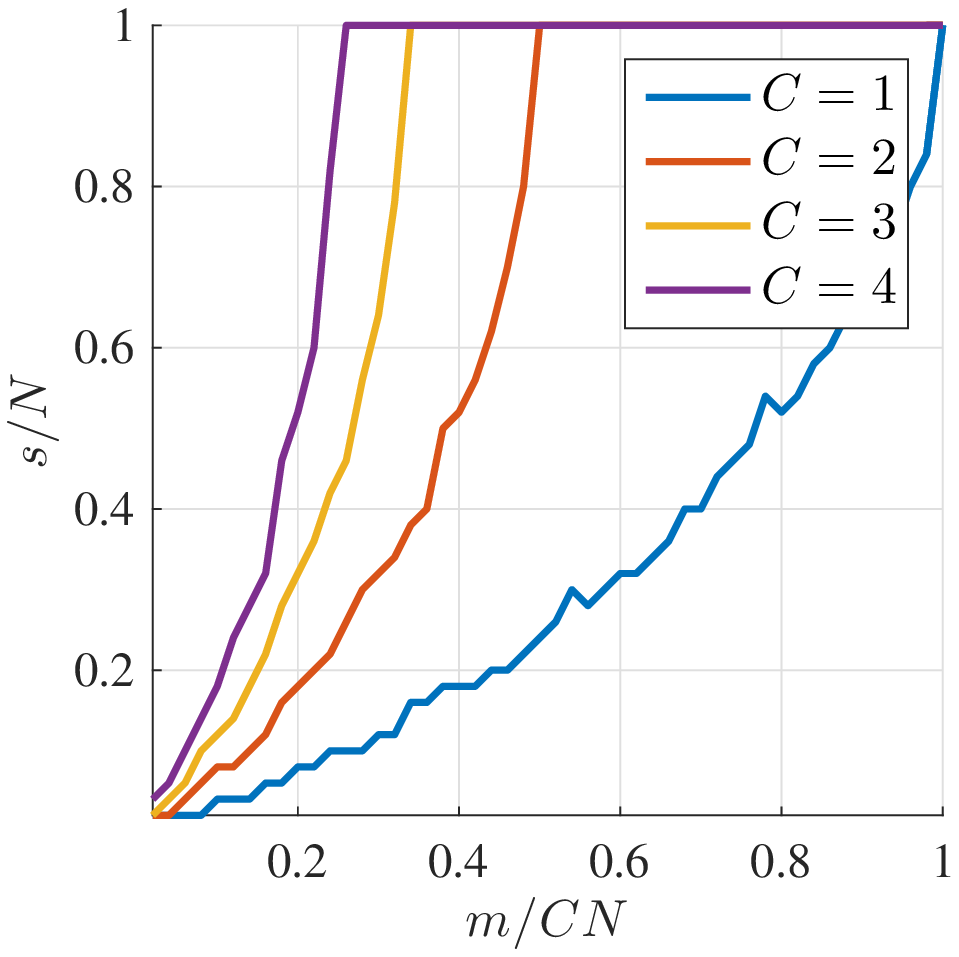} &
\includegraphics[scale=0.475, trim=0.5em 0.45em 2.5em 2em, clip]{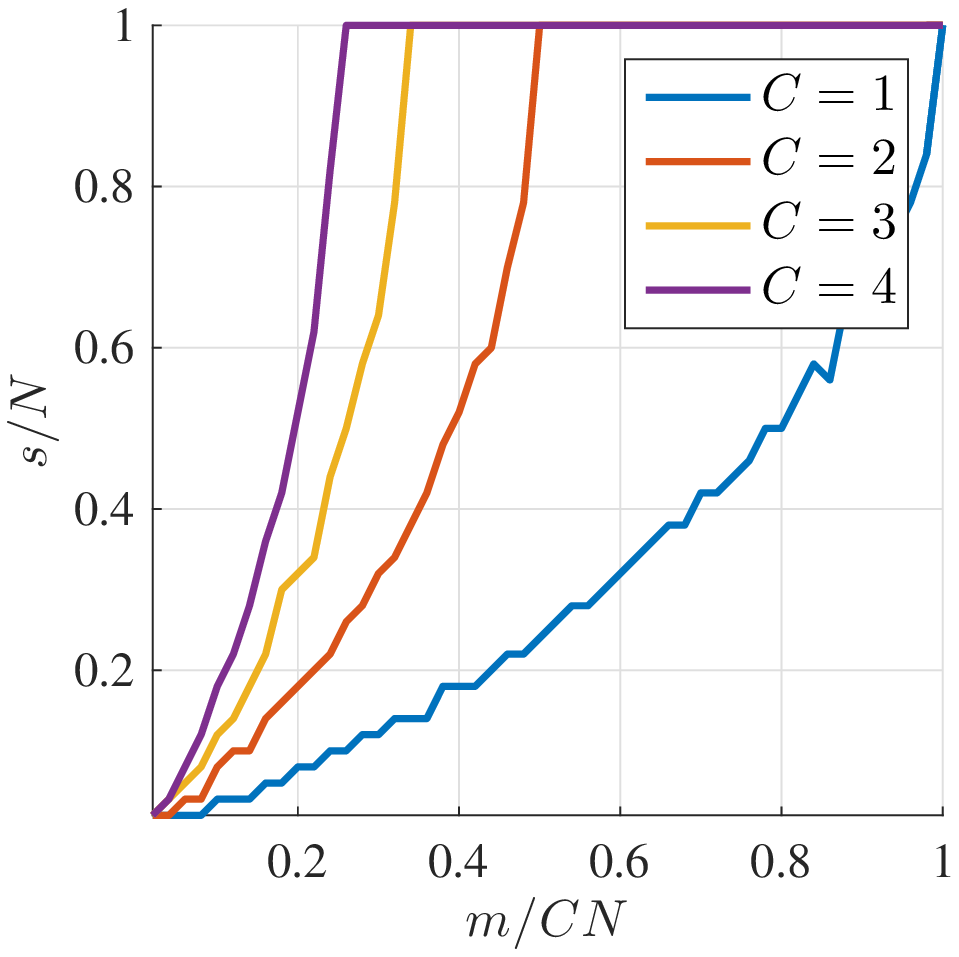} 
\end{tabular}

\caption{Empirical phase transition results for the identical sampling scenario.
For both sampling scenarios, the empirical probability of successful recovery increases as $C$ increases. The results in (b) and (c) are in agreement with our theoretical results.
}
\label{fig:PT_idt}
\end{figure}

Figs.\ \ref{fig:PT_dist}--\ref{fig:PT_idt} show phase transition curves for two diagonal and two circulant sensor profile matrices.  
The phase transition results in (b) and (c) are in good agreement with our theoretical results on optimal recovery guarantees.
Note that (a) and (d) are cases for which our theoretical results predict a worst-case recovery guarantee, i.e.\ $m \gtrsim C \cdot s \cdot L$ -- see the discussion in \S \ref{sec:Eg}.  However, these phase transition curves appear to show optimal recovery for these cases, i.e.\ $m \gtrsim s \cdot L$ independent of $C$.  We remark that this phenomenon has also been observed in previous works \cite{Chun&Adcock:17TIT, Chun&Adcock:16ITW, Chun&Li&Adcock:16MMSPARSE, Eftekhari&etal:15ACHA}.

The reason for this dissonance is that phase transition experiments tend to generate random sparse vectors where the nonzero entries are reasonably spread out.  Take, for example, case (a).  As remarked in \S \ref{sec:Eg:distDiag}, to recover an arbitrary $s$-sparse vector one requires at least $C \cdot s$ measurements, since for any $c$, there exists an $s$-sparse vector for which $\supp(x) \subseteq I_c$.  However, such worst-case vectors are generated in a phase transition experiment with low probability.  The `average' vectors generated in such experiments are more adequately described as \textit{sparse and distributed}, i.e.\ $s$-sparse but having roughly $s/C$ of their nonzero entries in each interval $I_c$.

Optimal (i.e.\ independent of $C$) recovery guarantees for the so-called sparse and distributed signal model with the canonical sparsity basis and diagonal sensor profiles have been presented in \cite{Chun&Adcock:17TIT, Chun&Adcock:16ITW, Chun&Li&Adcock:16MMSPARSE}.  This is based on the so-called \textit{sparsity in levels} signal model introduced in \cite{Adcock&etal:16FMS} (see also \cite{Roman&Hansen&Adcock:14arXiv}).  These results are nonuniform, however.  We expect though that one can prove uniform recovery guarantees for subgaussian random matrices for the sparse and distributed signal model -- such as was done in this paper for the sparse signal model -- thus verifying the results seen in Fig.\ \ref{fig:PT_dist}. For some work in this direction for sampling with subsampled isometries, see \cite{Chen&Adcock:18ACHA}.

\section{Outline of the proofs}
Given a matrix $A$ satisfying
\bes{
\alpha I \preceq \bbE A^* A \preceq \beta I,
}
our aim is to estimate the quantity $\sup_{z \in B_s} \left | \| A z \|^2_2 - \bbE \| A z \|^2_2 \right |$.  Indeed, if
\bes{
\sup_{z \in B_s} \left | \| A z \|^2_2 - \bbE \| A z \|^2_2 \right | \leq \alpha \delta,
}
for some $0 < \delta < 1$, then it follows that $A$ satisfies the ARIP with ARICs
\bes{
\alpha_{s} \geq (1-\delta) \alpha,\qquad \beta_s \leq (1+\delta) \beta.
}
To estimate $\sup_{z \in B_s} \left | \| A z \|^2_2 - \bbE \| A z \|^2_2 \right |$, we shall follow the ideas of \cite{Krahmer:14CPAM} (see also \cite{Eftekhari&etal:15ACHA}) and relate this quantity to the supremum of a certain chaos process.  In particular, we will use the following theorem:

\thm{[\mbox{\cite[Thm. 3.1]{Krahmer:14CPAM}}]
\label{t:Krahmer}
Let $\cA \subset \bbC^{m \times N}$ be a set of matrices and let $\xi$ be a random vector whose entries are i.i.d., zero-mean, unit-variance, and $\phi$-subgaussian random variables.
Set
\eas{
E_1 &= \gamma_2(\cA, \| \cdot \|_{2 \rightarrow 2} ) \left( \gamma_2(\cA, \| \cdot \|_{2 \rightarrow 2}) + d_F(\cA) \right) + d_F(\cA) d_{2 \rightarrow 2}(\cA) 
\\
E_2 &= d_{2 \rightarrow 2}(\cA) \left( \gamma_2(\cA, \| \cdot \|_{2 \rightarrow 2} ) + d_F(\cA) \right)
\\
E_3 &= d_{2 \rightarrow 2}^2(\cA)
}
where $d_F(\cA) = \sup_{A \in \cA} \| A \|_F$ is the radius of $\cA$ in Frobenius norm, $d_{2 \rightarrow 2}(\cA) = \sup_{A \in \cA} \| A \|_{2 \rightarrow 2}$ is the radius of $\cA$ in the spectral norm, and $\gamma_2(\cA, \| \cdot \|_{2 \rightarrow 2})$ is as in Definition \ref{d:gamma2}.  Then, for $t > 0$,
\bes{
\bbP \! \left\{ \sup_{A \in \cA} \left| \| A \xi \|^2_2 - \bbE  \| A \xi \|^2_2 \right| \geq c_1 E_1 + t \right\} \leq 2 \exp \! \left( - c_2 \min \! \left\{ \frac{t^2}{E_2^2}, \frac{t}{E_3} \right\}  \right),
}
where $c_1, c_2$ depend only on $\phi$.
}

The $\gamma_2$ functional in this theorem is defined as follows:

\defn{[$\gamma_2$ functional \cite{Talagrand:05book}] 
\label{d:gamma2}
For a metric space $(T,d)$, a sequence of subsets of $T$, $\{ T_i: i \geq 0 \}$, is called \textit{admissible} if $| T_0 | = 1$ and $ | T_i | \leq 2^{2i}$ for $r \geq 1$.  
The $\gamma_2$ functional is defined by
\bes{
\gamma_{2} (T,d) = \inf \sup_{t \in T} \sum_{i=0}^{\infty} 2^{i/2} d(t,T_i)
}
where the infimum is taken over all admissible sequences $\{ T_i \}$.
}

In practice, we will estimate $\gamma_2$ using covering numbers.  Recall that for a subset $T$ of a metric space $(\hat{T},d)$ and $r>0$, the \textit{covering number} $\cN (T,d,r)$ is the smallest integer $\cN$ such that $T$ can be covered with balls $B(x_l,r) = \{ x \in \hat{T}, d(x,x_l) \leq r \}$, $x_l \in T$, $l \in \{ 1, \ldots, \cN \}$, i.e.\
\bes{
T \subseteq \bigcup_{l=1}^{\cN} B(x_l,r).
}
In words, $T$ can be covered with $\cN$ balls of radius $r$ in the metric $d$.
The $\gamma_2$ functional can be bounded in terms of covering numbers through the Dudley's entropy integral (e.g.\ \cite{Talagrand:05book}).
More specifically, the $\gamma_2$ functional of set of matrices $\cA$ endowed with the operator norm $\| \cdot \|_{2 \rightarrow 2}$ satisfies
\be{
\label{eq:gamma2_Dudley}
\gamma_2(\cA, \| \cdot \|_{2 \rightarrow 2}) \lesssim \int_0^{d_{2 \rightarrow 2} (\cA)} \! \sqrt{\ln \cN (\cA, \| \cdot \|_{2 \rightarrow 2}, \nu)} \D \nu,
}
where $d_{2 \rightarrow 2} (\cA)$ is the radius of $\cA$ in the spectral norm.

In the next two sections, we estimate three quantities involved in Theorem \ref{t:Krahmer} for both the distinct and identical sampling scenarios.  Specifically, we will prove that
\bes{
\bbP \left \{  \sup_{x \in B_s} \left | \| A x \|^2_2 - \bbE \| A x \|^2_2 \right | \geq \alpha \delta \right \} \leq \varepsilon,
}
by showing the following inequalities for $E_{1}$, $E_2$ and $E_3$:
\ea{
\label{E1}
E_1 &\leq  \alpha\delta / (2 c_1)
\\
\label{E2}
E^2_2 &\leq \frac{ \alpha^2 \delta^2 c_2}{4 \ln(2/\varepsilon)}
\\
\label{E3}
E_3 &\leq \frac{ \alpha \delta c_2}{2 \ln(2/\varepsilon)}.
}
Our proofs use similar arguments to those of \cite{Eftekhari&etal:15ACHA}, but with the key generalization to arbitrary sensor profile matrices.

\section{Proofs I -- Distinct sampling}

\subsection{Reformulation as a chaos process} \label{sec:Reform_dist}

Let $A$ be as in \S \ref{sec:setupDist}. We first show that the quantity $\sup_{z \in B_s} \left | \| A z \|^2_2 - \bbE \| A z \|^2_2 \right |$ can be viewed as the supremum of a chaos process.  
We have
\ea{
\| A z \|^2_2 
&= \frac{1}{m} \sum_{c=1}^C \| \tilde{A}_c H_c U z \|^2_2 
= \frac{1}{m} \sum_{c=1}^C \sum_{i=1}^{m/C} \left| \ip{H_c U z}{\tilde{a}_{c,i}} \right|^2
\nn \\
&= \sum_{c=1}^C \Bigg\| \frac{1}{\sqrt{m}} \underbrace{ \left[ \begin{array}{cccc} z^* U^* H_c^* & & \\ & \ddots & \\ & & z^* U^* H_c^* \end{array} \right] }_{\mathrm{\mbox{$=: Z_c \in \bbC^{m/C \times mN/C}$}}} \!\! \underbrace{ \left[ \begin{array}{c} \tilde{a}_{c,1} \\ \vdots \\ \tilde{a}_{c,m/C} \end{array} \right] }_{\mathrm{\hbox{$=: \xi_c \in \bbC^{mN/C}$}}} \!\!\!\!\! \Bigg\|^2_2
= \| A_{z} \xi \|^2_2 
\label{def:Zc_dist}
}
where $a^*_{c,i}$ is the $i\rth$ row of $\tilde{A}_c$,
\be{
\label{def:Z_dist}
A_{z} :=  \frac{1}{\sqrt{m}} \left[ \begin{array}{cccc} Z_1 & & \\ & \ddots & \\ & & Z_C \end{array} \right] \in \bbC^{m \times m N},\qquad \mbox{and} \qquad \xi := \left[ \begin{array}{c} \xi_1 \\ \vdots \\ \xi_C \end{array} \right] \in \bbC^{m N}.
}
Observe that $\xi \in \bbC^{m N}$ is a random vector whose entries are i.i.d., zero-mean, unit-variance, and $\phi$-subgaussian random variables.  
If we define the set of $m \times m N$ matrices
\be{
\label{process_dist}
\cA = \left\{ A_{z} : z \in B_{s} \right\},
}
then it follows that
\bes{
\sup_{z \in B_{s}} \left| \| A z \|^2_2 - \bbE \|  A  z \|^2_2 \right| = \sup_{A_z \in \cA} \left| \| A_z \xi \|^2_2 - \bbE \| A_z \xi \|^2_2 \right|.
}

\subsection{Proof of Theorem \ref{t:dist:subgaussRIP}} \label{sec:prf:dist:subgaussRIP}
Seeking to apply Theorem \ref{t:Krahmer}, we first estimate $d_F(\cA)$, $d_{2\rightarrow2}(\cA)$, and $\gamma_2(\cA)$ for the chaos process \R{process_dist}.

\subsubsection{Estimation of $d_F(\cA)$}  
\label{sec:prf:dist:subgaussRIP:univ_dF}
We have
\bes{
\sup_{A_{z} \in \cA} \| A_{z} \|_F = \sup_{z \in B_{s}} \sqrt{ \frac{1}{m} \frac{m}{C}  z^* U^* \left( \sum_{c=1}^C H_c^* H_c \right) U z } \leq \sqrt{\beta},
}
due to \R{eq:joint_iso_dist} and the fact that $U$ is unitary.   Therefore $d_F(\cA) =  \sup_{A_{z} \in \cA} \| A_{z} \|_F \leq \sqrt{\beta}$.

\subsubsection{Estimation of $d_{2 \rightarrow 2}(\cA)$} \label{sec:proof:d2_dist}
We estimate $d_{2 \rightarrow 2}(\cA) = \sup_{A_{z} \in \cA} \| A_{z} \|_{2 \rightarrow 2}$ as follows.
First, observe that 
\eas{
\| A_{z} \|_{2 \rightarrow 2} &= m^{-1/2} \max_{c=1,\ldots,C} \| Z_c \|_{2\rightarrow2},
\\
\| Z_c \|_{2 \rightarrow 2} &= \max_{\substack{x \in \bbC^N \\ \nm{x}_{2} = 1}} | z^* U^* H^*_c x | = \| H_c U z \|_2,
}
by the block diagonal structure of $A_z$ in \R{def:Z_dist} and the block diagonal structure of $Z_c$ in \R{def:Zc_dist}, respectively.
Using the observations, we obtain
\eas{
\sup_{A_{z} \in \cA} \| A_{z} \|_{2 \rightarrow 2}
&= \frac{1}{\sqrt{m}} \sup_{z \in B_{s}} \max_{c=1,\ldots,C}  \| H_c U z \|_2 
\\
& \leq \frac{1}{\sqrt{m}} \sup_{z \in B_s} \max_{c=1,\ldots,C} \sum^{N}_{j=1} | z_j | \| H_c U e_j \|_2
\\
& \leq \sqrt{\frac{\alpha}{m}} \cdot \Gamma_{\mathrm{distinct}} \cdot  \sup_{z \in B_s} \| z \|_1
\\
& \leq \sqrt{\frac{s \alpha}{m}} \cdot \Gamma_{\mathrm{distinct}} 
}
where $\Gamma_{\mathrm{distinct}} = \alpha^{-1/2} \max_{c=1,\ldots,C} \max_{j=1,\ldots,N} \| H_c U e_j \|_2$.

\subsubsection{Estimation of $\gamma_2(\cA, \| \cdot \|_{2 \rightarrow 2})$} \label{sec:proof:gamma2_dist}

To estimate $\gamma_2(\cA, \| \cdot \|_{2 \rightarrow 2})$ we shall use the bound \R{eq:gamma2_Dudley} involving the covering number $\cN (\cA, \| \cdot \|_{2 \rightarrow 2}, \nu)$.  This covering number is estimated in the following lemma.  This lemma is similar to \cite[Lem. 6]{Eftekhari&etal:15ACHA}, albeit with a slightly tighter estimate (i.e.\ $\ln(N) \ln(m)$ rather than $\ln^2(N)$).  We include a proof in \S \ref{a:RozellCoveringProof} for completeness.

\lem{
\label{l:RozellCovering}
Let $\cF : \bbC^N \rightarrow \bbC^{m \times N'}$ for some $N' \geq m \geq 2$ be a linear map satisfying
\be{
\label{Fnormineq}
\nm{z}_{\cF} := \| \cF(z) \|_{2 \rightarrow 2} \leq \frac{\theta}{\sqrt{m}} \| z \|_{1},\quad \forall z \in \bbC^N,
}
where $\| \cdot \|_{\cF}$ is a semi-norm on $\bbC^N$.
Then, for $0 < \nu < \theta / \sqrt{m}$, we have
\bes{
\sqrt{\ln(\cN(B_{s},\nm{\cdot}_{\cF} , \nu ))} \lesssim \min \left \{  \frac{\theta \sqrt{2s \ln(2m) \ln(2N)} }{\sqrt{m}}  \nu^{-1} ,   \sqrt{2s} \left( \sqrt{\ln \! \left( \frac{eN}{s} \right)} + \sqrt{\ln \! \left(1 + 2 \sqrt{\frac{s}{m}} \frac{\theta}{\nu} \right)} \right) \right \}.
}
where $B_{s}$ is as in Definition~\ref{d:sparsity}.
}

Note that the first term will in general be used when $\nu$ is large and the second term will be used when $\nu$ is small.

Let $\cF : \bbC^{N} \rightarrow \bbC^{m \times m N}$ be the mapping $z \mapsto A_z$, where $A_z$ is as in \R{def:Z_dist}.  As shown in the previous section, note that $\| \cF(z) \|_{2 \rightarrow 2} \leq  \sqrt{\alpha/m} \Gamma_{\mathrm{distinct}} \| z \|_{1}$.  Hence, setting $\theta = \sqrt{\alpha} \Gamma_{\mathrm{distinct}}$ the above lemma gives
\be{
\label{eq:dist:Dudley_int_large}
\sqrt{ \ln ( \cN(\cA, \| \cdot \|_{2 \rightarrow 2}, \nu) ) } \leq \sqrt{\frac{2s \alpha \ln(2m) \ln(2N)}{m}} \Gamma_{\mathrm{distinct}}  \nu^{-1},\qquad \nu > 0,
}
and 
\be{
\label{eq:dist:Dudley_int_small}
\sqrt{ \ln ( \cN(\cA, \| \cdot \|_{2 \rightarrow 2}, \nu) ) } \leq \sqrt{2s} \left( \sqrt{\ln \! \left( \frac{eN}{s} \right)} + \sqrt{\ln \! \left(1 + 2 \sqrt{\frac{s \alpha}{m}} \frac{\Gamma_{\mathrm{distinct}}}{\nu} \right)} \right),\qquad \nu > 0.
}
Fix $0 < \lambda < \sqrt{s \alpha/m} \Gamma_{\mathrm{distinct}}$.  We now get
\eas{
\gamma_2(\cA, \| \cdot \|_{2 \rightarrow 2}) 
 \lesssim &\int_0^{ \sqrt{\frac{s\alpha}{m}} \Gamma_{\mathrm{distinct}} } \! \sqrt{\ln \cN (B_{s}, \| \cF(\cdot) \|_{2 \rightarrow 2}, \nu)} \D \nu
\\
 = &\int_0^{ \lambda } \! \sqrt{\ln \cN (B_{s}, \| \cF(\cdot) \|_{2 \rightarrow 2}, \nu)} \D \nu + \int_{\lambda}^{ \sqrt{\frac{s\alpha}{m}} \Gamma_{\mathrm{distinct}} } \! \sqrt{\ln \cN (B_{s}, \| \cF(\cdot) \|_{2 \rightarrow 2}, \nu)} \D \nu
\\
 \leq &\lambda \sqrt{2s} \left( \sqrt{\ln \! \left( \frac{eN}{s} \right)} + \sqrt{\ln \! \left( e \! \left( 1 + 2 \sqrt{\frac{s\alpha}{m}} \frac{\Gamma_{\mathrm{distinct}}}{\lambda} \right) \right)} \right) 
\\
& + \sqrt{2s \alpha} \sqrt{ \frac{\ln(2m) \ln(2N)}{m} } \Gamma_{\mathrm{distinct}} \ln \! \left( \sqrt{\frac{s\alpha}{m}} \frac{\Gamma_{\mathrm{distinct}}}{\lambda} \right)
}
(for the second inequality, we use the bound $\int_0^a \! \sqrt{\ln \left( 1 + b/\nu \right) } \D \nu \leq a \sqrt{\ln ( e( 1 + b/a ) )}$ for $a,b>0$).
With the choice of $\lambda = \sqrt{\alpha/m}\Gamma_{\mathrm{distinct}} $ we now deduce the overall bound
\eas{
\gamma_2(\cA, \| \cdot \|_{2 \rightarrow 2}) 
& \lesssim \sqrt{ \frac{2s \alpha}{m} } \Gamma_{\mathrm{distinct}} \left( \sqrt{\ln \left ( \frac{eN}{s} \right )}  + \sqrt{\ln\left( e (1 + 2\sqrt{s})\right)} + \sqrt{\ln(2m) \ln(2N)} \ln(\sqrt{s}) \right)
\\
& \lesssim \sqrt{ \frac{2s\alpha}{m} } \cdot \Gamma_{\mathrm{distinct}} \cdot \sqrt{\ln(2m) \ln(2N)} \ln (2s) .
}

\subsubsection{Estimates for $E_1$, $E_2$, and $E_3$} \label{sec:dist:E1-E2-E3}
In the previous three subsections we have shown that
\bes{
d_{F}(\cA) \lesssim \sqrt{\beta},\quad d_{2 \rightarrow 2}(\cA) \lesssim \sqrt{\frac{s \alpha}{m}} \Gamma_{\mathrm{distinct}},\quad \gamma_2(\cA,\nm{\cdot}_{2 \rightarrow 2}) \lesssim \sqrt{\frac{s \alpha}{m}} \Gamma_{\mathrm{distinct}} L,
}
where $L = \sqrt{\ln(2m) \ln(2N)} \ln(2s)$.  From their definitions (see Theorem \ref{t:Krahmer}), it now follows that
\bes{
E_{1} \lesssim \frac{s \alpha}{m} \Gamma^2_{\mathrm{distinct}} L^2 + \sqrt{\frac{s \alpha \beta}{m}} \Gamma_{\mathrm{distinct}} L + \sqrt{\frac{s \alpha \beta}{m}} \Gamma_{\mathrm{distinct}}.
}
Hence \R{E1} holds, provided
\bes{
m \gtrsim \delta^{-1} \cdot s \cdot \Gamma^2_{\mathrm{distinct}} \cdot L^2\quad \mbox{and}\quad m \gtrsim \delta^{-2} \cdot s \cdot \frac{\beta}{\alpha} \cdot \Gamma^2_{\mathrm{distinct}} \cdot L^2.
}
Since $0 < \delta < 1$ and $\beta/\alpha \geq 1$, this reduces to a single inequality
\be{
\label{E1ineq_distinct}
m \gtrsim \delta^{-2} \cdot s \cdot \frac{\beta}{\alpha} \cdot \Gamma^2_{\mathrm{distinct}} \cdot L^2.
}
Now consider $E_2$.  We have
\bes{
E_2 \lesssim \frac{s \alpha}{m} \Gamma^2_{\mathrm{distinct}} L + \sqrt{\frac{s \alpha \beta}{m}} \Gamma_{\mathrm{distinct}}.
}
Hence \R{E2} holds, provided
\bes{
m \gtrsim \delta^{-1} \cdot s \cdot \Gamma^2_{\mathrm{distinct}} \cdot L \cdot \sqrt{\ln(2/\varepsilon)}\quad \mbox{and} \quad m \gtrsim \delta^{-2} \cdot s \cdot \frac{\beta}{\alpha} \cdot \Gamma^2_{\mathrm{distinct}} \cdot \ln(2/\varepsilon).
}
Via Young's inequality, this reduces to the single inequality
\be{
\label{E2ineq_distinct}
m \gtrsim \delta^{-2} \cdot s \cdot \frac{\beta}{\alpha} \cdot \Gamma^2_{\mathrm{distinct}} \cdot \left ( L^2 + \ln(2/\varepsilon) \right ).
}
Finally, for $E_3$ we have
\bes{
E_{3} \lesssim \frac{s \alpha}{m} \Gamma^2_{\mathrm{distinct}},
}
and therefore \R{E3} holds provided
\be{
\label{E3ineq_distinct}
m \gtrsim \delta^{-1} \cdot s \cdot \Gamma^2_{\mathrm{distinct}} \cdot \ln(2/\varepsilon).
}
To complete the proof of Theorem \ref{t:dist:subgaussRIP} we note that \R{E1ineq_distinct}--\R{E3ineq_distinct} are implied by the inequality \R{dist:subgaussRIP:measurement}.

\subsection{Proof of Theorem \ref{t:dist:subgaussRIP_univ}} \label{sec:prf:dist:subgaussRIP_univ}

We follow the same setup as in \S \ref{sec:prf:dist:subgaussRIP}.  Since $d_F(\cA) \leq \sqrt{\beta}$ (see \S \ref{sec:prf:dist:subgaussRIP:univ_dF}), in this section we need only provide different estimates for the quantities $d_{2 \rightarrow 2}(\cA)$ and $\gamma_2(\cA, \| \cdot \|_{2 \rightarrow 2})$.

\subsubsection{Estimation of $d_{2 \rightarrow 2}(\cA)$} 
\label{sec:prf:dist:subgaussRIP:univ_d2}

We estimate $d_{2 \rightarrow 2}(\cA) = \sup_{A_{z} \in \cA} \| A_{z} \|_{2 \rightarrow 2}$ as follows
\eas{
\sup_{A_{z} \in \cA} \| A_{z} \|_{2 \rightarrow 2}
&= \frac{1}{\sqrt{m}} \sup_{z \in B_{s}} \max_{c=1,\ldots,C} \| H_c U z \|_2
\\
&\leq \frac{1}{\sqrt{m}} \max_{c=1,\ldots,C} \| H_c \|_{2 \rightarrow 2} \sup_{z \in B_{s}} \| U z \|_2
\\
&= \frac{1}{\sqrt{m}} \cdot \max_{c=1,\ldots,C} \| H_c \|_{2 \rightarrow 2} \cdot \sup_{z \in B_{s}} \| z \|_2
\\
&\leq \sqrt{\frac{\alpha}{m}} \cdot  \Xi_{\mathrm{distinct}}
}
where $\Xi_{\mathrm{distinct}} =  \alpha^{-1/2}\max_{c=1,\ldots,C} \| H_c \|_{2 \rightarrow 2}$.

\subsubsection{Estimation of $\gamma_2(\cA, \| \cdot \|_{2 \rightarrow 2})$}
\label{sec:prf:dist:subgaussRIP:univ_gamma2}

Note that, as shown above
\bes{
\| A_{z} \|_{2 \rightarrow 2} \leq \sqrt{\frac{\alpha}{m}} \cdot \Xi_{\mathrm{distinct}} \cdot \| z \|_2,
}
and thus $\| A_z - A_{z'} \|_{2 \rightarrow 2} = \| A_{z-z'} \|_{2 \rightarrow 2} \leq \sqrt{\alpha/m}  \Xi_{\mathrm{distinct}} \| z - z' \|_2$.
Therefore, for every $\nu > 0$, $\cN (\cA, \| \cdot \|_{2 \rightarrow 2}, \nu) \leq \cN (B_{s}, \sqrt{\alpha/m}  \Xi_{\mathrm{distinct}} \| \cdot \|_2, \nu)$. 
Thus, the Dudley-type integral \R{eq:gamma2_Dudley} yields
\eas{
\gamma_2(\cA, \| \cdot \|_{2 \rightarrow 2}) 
& \lesssim \int_0^{\sqrt{\frac{\alpha}{m}} \Xi_{\mathrm{distinct}} } \! \sqrt{\ln \cN (B_{s}, \sqrt{\frac{\alpha}{m}} \Xi_{\mathrm{distinct}} \| \cdot \|_2, \nu)} \D \nu
\\
&= \int_0^{\sqrt{\frac{\alpha}{m}} \Xi_{\mathrm{distinct}}} \! \sqrt{\ln \cN ( B_{s}, \| \cdot \|_2, ( \sqrt{\frac{\alpha}{m}}  \Xi_{\mathrm{distinct}} )^{-1} \nu )} \D \nu
\\
&= \sqrt{\frac{\alpha}{m}} \Xi_{\mathrm{distinct}} \int_0^{1} \! \sqrt{\ln \cN (B_{s}, \| \cdot \|_2, \nu)} \D \nu.
}
A simple volumetric argument (see, for example, \cite[Appx. C]{Krahmer:14CPAM}) now gives
\bes{
\cN (B_{s}, \| \cdot \|_2, \nu) \leq {N \choose s} (1 + 2/\nu)^{2s} \leq (eN/s)^s (1 + 2/\nu)^{2s},
}
(noting that the $s$-dimensional complex unit ball can be treated as the real $2s$-dimensional unit ball by isometry).  Therefore, we have
\eas{
\gamma_2(\cA, \| \cdot \|_{2 \rightarrow 2}) 
& \leq \sqrt{\frac{2s \alpha}{m}} \Xi_{\mathrm{distinct}} \left( \sqrt{ \ln (eN/s)  }  +  \int_0^{1} \! \sqrt{ \ln(1+2/\nu ) }  \D \nu \right)
\\
& \lesssim \sqrt{\frac{s \alpha \ln (eN/s)}{m}} \cdot \Xi_{\mathrm{distinct}},
}
where in the second step we use the inequality $\int_0^a \! \sqrt{\ln \left( 1 + b/\nu \right) } \D \nu \leq a \sqrt{\ln ( e( 1 + b/a ) )}$ for $a,b>0$.

\subsubsection{Estimation of $E_1$, $E_2$, and $E_3$} \label{sec:distUniv:E1-E2-E3}

Using these bounds, we now deduce the following inequalities for $E_1$, $E_2$ and $E_3$:
\eas{
E_{1} &\lesssim \frac{s \alpha}{m} \ln(\E N/s) \Xi^2_{\mathrm{distinct}} + \sqrt{\frac{s \alpha \beta \ln(\E N/s)}{m}} \Xi_{\mathrm{distinct}}
\\
E_2 &\lesssim \frac{\sqrt{s \ln(\E N/s)} \alpha}{m} \Xi^2_{\mathrm{distinct}}  + \sqrt{\frac{\alpha \beta}{m}} \Xi_{\mathrm{distinct}}
\\
E_{3} &\lesssim \frac{\alpha}{m} \Xi^2_{\mathrm{distinct}}.
}
Hence \R{E1} holds, provided
\bes{
m \gtrsim \delta^{-2} \cdot s \cdot \frac{\beta}{\alpha} \cdot \Xi^2_{\mathrm{distinct}} \cdot \ln(e N/s),
}
\R{E2} holds, provided
\bes{
m \gtrsim \delta^{-1} \cdot \sqrt{s \ln(\E N/s) \ln(2/\varepsilon)} \cdot \Xi^2_{\mathrm{distinct}}\quad \mbox{and}\quad m \gtrsim \delta^{-2} \cdot \frac{\beta}{\alpha} \cdot \Xi^2_{\mathrm{distinct}}\cdot\ln(2/\varepsilon),
}
and \R{E3} holds, provided
\bes{
m \gtrsim \delta^{-1} \cdot \Xi^2_{\mathrm{distinct}}\cdot \ln(2/\varepsilon).
}
To complete the proof, we note that these inequalities are implied by \R{dist:subgaussRIP_univ:measurement}.

\subsection{Proofs of Theorem \ref{t:dist:subgaussRIP_diff_mc} and \ref{t:dist:subgaussRIP_univ_diff_mc}}

Similar techniques to those used in \S \ref{sec:prf:dist:subgaussRIP} for the proof of Theorem \ref{t:dist:subgaussRIP} can be applied to prove Theorem \ref{t:dist:subgaussRIP_diff_mc}.  The normalization factors in \R{def:Z_dist} are first replaced by $(m_c C)^{-1/2}$, i.e.\
\bes{
A_{z} = \left [ \begin{array}{ccc} (m_1 C)^{-1/2} Z && \\ & \ddots \\ && (m_C C)^{-1/2} Z \end{array} \right ],
}
so that the following holds:
\bes{
\sup_{x \in B_{s}} \left| \| A x \|^2_2 - \| x \|^2_2 \right| = \sup_{A_z \in \cA} \left| \| A_z \xi \|^2_2 - \bbE \| A_z \xi \|^2_2 \right|,
}
where $A$ is as in \R{def:A_dist_diffmc}.  Then, the parameters in Theorem \ref{t:Krahmer} are estimated as
\eas{
d_F(\cA) &\lesssim \sqrt{\beta},
\\
d_{2 \rightarrow 2}(\cA) &\lesssim \sqrt{\frac{s}{C}} \max_{c=1,\ldots,C} \frac{\max_{j=1,\ldots,N} \| H_c U e_j \|_2}{\sqrt{m_c}}
\leq \frac{\sqrt{s \alpha} \cdot \Gamma_{\mathrm{distinct}}}{\sqrt{C} \min_{c=1,\ldots,C} \sqrt{m_c}},
\\
\gamma_2(\cA,\| \cdot \|_{2 \rightarrow 2}) &\lesssim \frac{ \sqrt{s \alpha} \cdot \Gamma_{\mathrm{distinct}} \cdot \sqrt{\ln(2m) \ln(2N)} \ln (2s) }{ \sqrt{C} \min_{c=1,\ldots,C} \sqrt{m_c} },
}
where $\Gamma_{\mathrm{distinct}} = \alpha^{-1/2} \max_{c=1,\ldots,C} \max_{j=1,\ldots,N} \| H_c U e_j \|_2$.
Note that the estimate for $\gamma_2(\cA,\| \cdot \|_{2 \rightarrow 2})$ follows from Lemma \ref{l:RozellCovering}.  To complete the proof, we repeat the procedure in \S \ref{sec:dist:E1-E2-E3}.

To prove Theorem \ref{t:dist:subgaussRIP_univ_diff_mc} we follow the techniques of \S \ref{sec:prf:dist:subgaussRIP_univ} using the same adjustments as made above.  The parameters in Theorem \ref{t:Krahmer} now satisfy
\eas{
d_F(\cA) &\lesssim \sqrt{\beta}
\\
d_{2 \rightarrow 2}(\cA) &\lesssim \frac{1}{\sqrt{C}} \max_{c=1,\ldots,C} \frac{\| H_c \|_{2 \rightarrow 2}}{\sqrt{m_c}} \leq \frac{\sqrt{\alpha} \cdot \Xi_{\mathrm{distinct}}}{\sqrt{C} \min_{c=1,\ldots,C} \sqrt{m_c}}
\\
\gamma_2(\cA,\| \cdot \|_{2 \rightarrow 2}) &\lesssim \frac{\sqrt{ s \alpha \ln (eN/s) } \cdot \Xi_{\mathrm{distinct}}}{\sqrt{C} \min_{c=1,\ldots,C} \sqrt{m_c}},
}
where $\Xi_{\mathrm{distinct}} =  \alpha^{-1/2}\max_{c=1,\ldots,C} \| H_c \|_{2 \rightarrow 2}$.  
In particular, for the $\gamma_2(\cA,\| \cdot \|_{2 \rightarrow 2})$ estimate, note that $\cN (\cA, \| \cdot \|_{2 \rightarrow 2}, \nu) \leq \cN (B_{s}, ( \sqrt{\alpha/C} \max_{c=1,\ldots,C} \| H_c \|_{2 \rightarrow 2} / \sqrt{m_c \alpha} ) \| \cdot \|_2, \nu)$ for every $\nu > 0$.
To complete the proof, we repeat the procedure of \S \ref{sec:distUniv:E1-E2-E3}.

\section{Proofs II - Identical sampling}

\subsection{Reformulation as a chaos process} \label{sec:prf:idt}
Let $A$ be as in \S \ref{sec:setupIdt}. Similar to \S \ref{sec:Reform_dist}, we first show that the quantity $\sup_{z \in B_s} \left | \| A z \|^2_2 - \bbE \| A z \|^2_2 \right |$ can be viewed as the supremum of a chaos process.  We have
\bes{
\| A z \|^2_2 
= \frac{1}{m} \sum_{c=1}^C \| \tilde{A} H_c U z \|^2_2 
= \Bigg\| \frac{1}{\sqrt{m}} \underbrace{ \left[ \begin{array}{c} z^* U^* H_1^* \\ \vdots \\ z^* U^* H_C^*  \end{array} \right] }_{\mathrm{\hbox{$=: Z \in \bbC^{C \times N}$}}} \tilde{A}^* \Bigg\|_F^2
= \sum_{i=1}^{m/C} \frac{1}{m} \left\| Z \tilde{a}_i \right\|^2_2
= \| A_{z} \xi \|^2_2
}
where $\tilde{a}_i^*$ is the $i\rth$ row of $\tilde{A}$,
\be{
\label{def:Z_idt}
A_{z} := \frac{1}{\sqrt{m}} \left[ \begin{array}{cccc} Z & & \\ & \ddots & \\ & & Z \end{array} \right] \in \bbC^{m \times mN/C} \qquad \mbox{and} \qquad \xi := \left[ \begin{array}{c} \tilde{a}_1 \\ \vdots \\ \tilde{a}_{m/C} \end{array} \right] \in \bbC^{m N / C},
}
where the entries of $\xi$ are i.i.d., zero-mean, unit-variance, and $\phi$-subgaussian random variables.
If we define the set of $m \times m N/C$ matrices
\be{
\cA = \left\{ A_{z} : z \in B_{s} \right\},
}
then it follows that
\bes{
\sup_{x \in B_{s}} \left| \| A x \|^2_2 - \| x \|^2_2 \right| = \sup_{A_z \in \cA} \left| \| A_z \xi \|^2_2 - \bbE \| A_z \xi \|^2_2  \right|.
}

\subsection{Proof of Theorem \ref{t:idt:subgaussRIP}} \label{sec:prf:idt:subgaussRIP}

\subsubsection{Estimation of $d_F(\cA)$} \label{sec:prf:idt:subgaussRIP:univ_dF}
We have
\bes{
\sup_{A_{z} \in \cA} \| A_{z} \|_F = \sup_{z \in B_{s}} \sqrt{ \frac{1}{m} \frac{m}{C} z^* U^* \left( \sum_{c=1}^C H_c^* H_c \right) U z } \leq \sqrt{\beta}
}
by the joint near-isometry condition \R{eq:joint_iso_idt}.  Therefore, $d_F(\cA) \leq \sqrt{\beta}$.

\subsubsection{Estimation of $d_{2 \rightarrow 2}(\cA)$}

We have
\eas{
\sup_{A_{z} \in \cA} \| A_{z} \|_{2 \rightarrow 2} 
&=  \frac{1}{\sqrt{m}} \sup_{z \in B_{s}} \left\| \left[ \begin{array}{ccc} H_1 U z & \cdots & H_c U z  \end{array} \right] \right\|_{2 \rightarrow 2}
\\
&= \frac{1}{\sqrt{m}} \sup_{z \in B_{s}} \left\| \sum_{j=1}^N z_j \left[ \begin{array}{ccc} H_1 U e_j & \cdots & H_c U e_j  \end{array} \right]  \right\|_{2 \rightarrow 2}
\\
&\leq \frac{1}{\sqrt{m}} \sup_{z \in B_{s}} \sum_{j=1}^N | z_j | \left\| \left[ \begin{array}{ccc} H_1 U e_j & \cdots & H_c U e_j  \end{array} \right]  \right\|_{2 \rightarrow 2}
\\
&\leq \sqrt{\frac{\alpha}{m}} \cdot \Gamma_{\mathrm{identical}} \cdot \sup_{z \in B_{s}} \| z \|_1
\\
&= \sqrt{\frac{s \alpha}{m}} \cdot \Gamma_{\mathrm{identical}}
}
where $\Gamma_{\mathrm{identical}} = \alpha^{-1/2}  \max_{j=1,\ldots,N} \left\| \left[ \begin{array}{ccc} H_1 U e_j & \cdots & H_C U e_j \end{array} \right] \right\|_{2 \rightarrow 2}$.

\subsubsection{Estimation of $\gamma_2(\cA, \| \cdot \|_{2 \rightarrow 2})$}

Let $\cF : \bbC^{N} \rightarrow \bbC^{m \times m N/C}$ be the mapping $z \mapsto A_{z}$, where $A_z$ is as in \R{def:Z_idt}.  As shown in the previous section, we have
\bes{
\| \cF(z) \|_{2 \rightarrow 2} \leq \sqrt{\frac{\alpha}{m}} \Gamma_{\mathrm{identical}} \| z \|_1.
}
Following the same argument as in \S \ref{sec:proof:gamma2_dist} (replacing $ \Gamma_{\mathrm{distinct}}$ by $ \Gamma_{\mathrm{identical}}$ wherever necessary), we now deduce the estimate
\bes{
\gamma_2(\cA, \| \cdot \|_{2 \rightarrow 2}) 
\lesssim \sqrt{ \frac{2s\alpha}{m} } \cdot \Gamma_{\mathrm{identical}} \cdot \sqrt{\ln(2m) \ln(2N)} \ln (2s) .
}

\subsubsection{Estimates of $E_1$, $E_2$, and $E_3$}

Given these estimates, we have the following estimates for $E_1$, $E_2$, and $E_3$
\eas{
E_{1} \lesssim & \frac{s \alpha}{m} \Gamma^2_{\mathrm{identical}} L^2 + \sqrt{\frac{s \alpha \beta}{m}} \Gamma_{\mathrm{identical}} L + \sqrt{\frac{s \alpha \beta}{m}} \Gamma_{\mathrm{identical}}
\\
E_2 \lesssim & \frac{s \alpha}{m} \Gamma^2_{\mathrm{identical}} L + \sqrt{\frac{s \alpha \beta}{m}} \Gamma_{\mathrm{identical}}
\\
E_{3} \lesssim & \frac{s \alpha}{m} \Gamma^2_{\mathrm{identical}},
}
where $L = \sqrt{\ln(2m) \ln(2N)} \ln(2s)$.
To complete the proof of Theorem \ref{t:idt:subgaussRIP}, one can repeat the arguments in \S \ref{sec:dist:E1-E2-E3} so as to satisfy \R{E1}, \R{E2}, and \R{E3}.

\subsection{Proof of Theorems \ref{t:idt:subgaussRIP_univ}} \label{sec:prf:idt:subgaussRIP_univ}

We follow the same setup as in \S \ref{sec:prf:idt:subgaussRIP}.  Since $d_F(\cA) \leq \sqrt{\beta}$ (see \S \ref{sec:prf:idt:subgaussRIP:univ_dF}), in this section we need only provide different estimates for the quantities $d_{2 \rightarrow 2}(\cA)$ and $\gamma_2(\cA, \| \cdot \|_{2 \rightarrow 2})$.

\subsubsection{Estimation of $d_{2 \rightarrow 2}(\cA)$}  \label{sec:prf:idt:subgaussRIP:univ_d2}

We estimate $d_{2 \rightarrow 2}(\cA) = \sup_{A_{z} \in \cA} \| A_{z} \|_{2 \rightarrow 2}$ as follows:
\eas{
\sup_{A_{z} \in \cA} \| A_{z} \|_{2 \rightarrow 2} 
&=  \frac{1}{\sqrt{m}} \sup_{z \in B_{s}} \left\| \left[ \begin{array}{ccc} H_1 U z & \cdots & H_c U z  \end{array} \right] \right\|_{2 \rightarrow 2}
\\
&\leq \frac{1}{\sqrt{m}}  \left\| \left[ \begin{array}{ccc} H_1 U & \cdots & H_c U  \end{array} \right] \right\|_{2 \rightarrow 2} \sup_{z \in B_{s}} \left\| \left[ \begin{array}{ccc} z & & \\ & \ddots & \\ & & z  \end{array} \right] \right\|_{2 \rightarrow 2}
\\
&= \sqrt{\frac{\alpha}{m}} \cdot  \Xi_{\mathrm{identical}}
}
where $\Xi_{\mathrm{identical}} =  \alpha^{-1/2}\max_{c=1,\ldots,C} \left\| \left[ \begin{array}{ccc} H_1 & \cdots & H_C \end{array} \right] \right\|_{2 \rightarrow 2}$.

\subsubsection{Estimation of $\gamma_2(\cA, \| \cdot \|_{2 \rightarrow 2})$}

Note that, as shown above
\bes{
\| A_{z} \|_{2 \rightarrow 2} \leq \sqrt{\frac{\alpha}{m}} \cdot \Xi_{\mathrm{identical}} \cdot \| z \|_2,
}
and thus $\| A_z - A_{z'} \|_{2 \rightarrow 2} = \| A_{z-z'} \|_{2 \rightarrow 2} \leq \sqrt{\alpha/m}  \Xi_{\mathrm{identical}} \| z - z' \|_2$.
Therefore, for every $\nu > 0$, $\cN (\cA, \| \cdot \|_{2 \rightarrow 2}, \nu) \leq \cN (B_{s}, \sqrt{\alpha/m}  \Xi_{\mathrm{identical}} \| \cdot \|_2, \nu)$. 
Following the same argument as in \S \ref{sec:prf:dist:subgaussRIP:univ_gamma2} (replacing $\Xi_{\mathrm{distinct}}$ by $\Xi_{\mathrm{identical}}$ wherever necessary), we deduce the estimate
\bes{
\gamma_2(\cA, \| \cdot \|_{2 \rightarrow 2}) 
\lesssim \sqrt{\frac{s \alpha \ln (eN/s)}{m}} \cdot \Xi_{\mathrm{identical}}.
}

\subsubsection{Estimation of $E_1$, $E_2$, and $E_3$} \label{sec:idtUniv:E1-E2-E3}

Given these estimates, we have the following bounds for $E_1$, $E_2$, and $E_3$
\eas{
E_{1} &\lesssim \frac{s \alpha}{m} \ln(\E N/s) \Xi^2_{\mathrm{identical}} + \sqrt{\frac{s \alpha \beta \ln(\E N/s)}{m}} \Xi_{\mathrm{identical}}
\\
E_2 &\lesssim \frac{\sqrt{s \ln(\E N/s)} \alpha}{m} \Xi^2_{\mathrm{identical}}  + \sqrt{\frac{\alpha \beta}{m}} \Xi_{\mathrm{identical}}
\\
E_{3} &\lesssim \frac{\alpha}{m} \Xi^2_{\mathrm{identical}}.
}
To complete the proof of Theorem \ref{t:idt:subgaussRIP_univ}, one can repeat the arguments in \S \ref{sec:dist:E1-E2-E3} so as to satisfy \R{E1}, \R{E2}, and \R{E3}.

\section{Proofs of Propositions \ref{prop:GammaXi:bounds:dist}, \ref{prop:GammaXi:bounds:idt} and \ref{prop:GammaXi:bounds:idt:2}}

\subsection{Proof of Proposition \ref{prop:GammaXi:bounds:dist}} \label{sec:prf:prop:GammaXi:bounds:dist}

Observe that, for any $j$, $\alpha \leq C^{-1} \sum^{C}_{c=1} \| H_c U e_j \|^2_2$ due to \R{eq:joint_iso_dist} and the fact that $U$ is an isometry.  In particular, $\max_{c=1,\ldots,C} \| H_c U e_j \|_2 \geq \sqrt{\alpha}$.  This gives the first inequality.  For the second, we merely notice that $\| H_c U e_j \|_2 \leq \| H_{c} \|_{2\rightarrow2}$, since $U$ is an isometry and $\|e_j \|_2 = 1$.  Finally, for the third we first notice that \R{eq:joint_iso_dist} gives $C^{-1} \sum^{C}_{c=1} \| H_c z \|^2_2 \leq \beta \| z \|^2_2$, $\forall z \in \bbC^N$.  Hence $\| H_{c} z \|_2 \leq \sqrt{C}\sqrt{\beta} \| z \|_2$ and therefore $\| H_{c} \|_{2 \rightarrow 2} \leq \sqrt{C} \sqrt{\beta}$, which gives the result.

We now establish sharpness of the inequalities.  Let $H_{1} = \ldots = H_{C} = U = I$ so that \R{eq:joint_iso_dist} holds with $\alpha = \beta = 1$.  Then 
\bes{
1 = \Gamma_{\mathrm{distinct}} = \Xi_{\mathrm{distinct}},
}
which implies the lower two inequalities are sharp.  For the second two inequalities, let $H_{c} = \sqrt{C} P_{I_c}$, where $I_c$ is the index set
\be{
\label{Ic_def}
I_c = \{ (c-1)n,\ldots,cn \},\quad c=1,\ldots,C-1,\qquad I_{C} = \{ (C-1) n+1,\ldots,N \},
}
and $n = \lfloor N/C \rfloor$.  Then \R{eq:joint_iso_dist} holds with $\alpha = \beta = 1$.  Moreover $\| H_{c} \|_{2 \rightarrow 2} = \sqrt{C}$ and, if $U = I$, then $\max_{j=1,\ldots,N} \| H_c  Ue_j \|_2 = \sqrt{C}$, as required.

\subsection{Proof of Proposition \ref{prop:GammaXi:bounds:idt}}

Consider the first bound in \R{Xi:idt:bound1}.  Observe that
\be{
\label{Hc_block}
\left\| \left[ \begin{array}{ccc} H_1 & \cdots & H_C \end{array} \right] \right\|^2_{2 \rightarrow 2} = \max_{\sum^{C}_{c=1} \| x_c \|^2_2 \neq 0} \frac{\nm{\sum_{c=1}^C H_c x_c }_2^2}{\sum_{c=1}^C \| x_c \|^2_2}.
}
Fix $j$ and let $x_c = z_c U e_j$, where $\| z  \|_2 = 1$.  Then $\sum_{c=1}^C \| x_c \|^2_2 = \| U e_j \|^2_2 \| z \|^2_2 = 1$.  Hence
\bes{
\left\| \left[ \begin{array}{ccc} H_1 & \cdots & H_C \end{array} \right] \right\|^2_{2 \rightarrow 2} \geq \nm{\sum^{C}_{c=1} z_c H_c U e_j }^2_2.
}
Since $z$ was arbitrary we deduce that $\left\| \left[ H_1 \cdots H_C \right] \right\|^2_{2 \rightarrow 2} \geq  \left\| \left[ H_1 U e_j  \cdots H_C U e_j \right] \right\|_{2 \rightarrow 2}^{2}$ which gives the first bound of \R{Xi:idt:bound1}.  For the second bound of \R{Xi:idt:bound1}, we notice that
\bes{
\left\| \left[ \begin{array}{ccc} H_1 & \cdots & H_C \end{array} \right] \right\|^2_{2 \rightarrow 2} \leq \max_{\sum^{C}_{c=1} \| x_c \|^2_2 \neq 0}  \frac{\left (\sum_{c=1}^C \| H_c \|_{2 \rightarrow 2} \| x_c \|_2 \right )^2}{\sum_{c=1}^C \| x_c \|^2_2} \leq \sum_{c=1}^C \| H_c \|^2_{2 \rightarrow 2} \leq C \max_{c=1,\ldots,C} \| H_c \|^2_{2 \rightarrow 2}.
}
Hence $\Xi_{\mathrm{identical}} \leq \sqrt{C}\Xi_{\mathrm{distinct}}$ and the second bound of \R{Xi:idt:bound1} now follows from Proposition \ref{prop:GammaXi:bounds:dist}.  Now consider \R{Gamma:idt:bound}.  The second bound follows immediately from \R{Xi:idt:bound1}.  For the first, we merely observe that 
\bes{
 \left\| \left[ \begin{array}{ccc} H_1 U e_j & \cdots & H_C U e_j \end{array} \right] \right\|_{2 \rightarrow 2} = \max_{\substack{ \| z \|_2 = 1 \\ z \in \bbC^C}} \nm{\sum^{C}_{c=1} z_c H_c U e_j }_2 \geq \| H_c U e_j \|_2,
}
for any $c = 1,\ldots,C$.  Finally, for \R{Xi:idt:bound2} we notice from \R{Hc_block} that
\bes{
\left\| \left[ \begin{array}{ccc} H_1 & \cdots & H_C \end{array} \right] \right\|^2_{2 \rightarrow 2} \geq \max_{\|x\|_2 \neq 0} \frac{\| H_c x \|^2_2}{\| x \|^2_2} = \| H_c \|^2_{2 \rightarrow 2},
}
for any $c = 1,\ldots,C$.  This gives the result.

Now consider sharpness.  Let $U = I$ and $H_c = \sqrt{C} P_{I_c}$, where $I_c$ is as in \R{Ic_def}.  Then \R{eq:joint_iso_idt} holds with $\alpha = \beta = 1$.  Also
\bes{
\Gamma_{\mathrm{distinct}} = \Gamma_{\mathrm{identical}} = \Xi_{\mathrm{distinct}} =  \Xi_{\mathrm{identical}} = \sqrt{C}.
}
Hence it remains only to show the sharpness of the second inequalities in \R{Xi:idt:bound1} and \R{Xi:idt:bound2}.

For simplicity, suppose that $N/C = n \in \bbN$ and write sensor profile matrices in block form as
\bes{
H_{c} = \left \{ [H_{c}]_{a,b} \right \}^{C}_{a,b = 1}, \qquad [H_{c}]_{a,b} \in \bbC^{n \times n},
}
where $[H_{c}]_{a,b}$ denotes $(a,b)^\rth$ sub-matrix of $H_c$.  Let
\bes{
[H_{c}]_{a,b} = \sqrt{C} \delta_{a,1} \delta_{b,c} I_{n},
}
where $I_n$ is the $n \times n$ identity matrix.  In block form, we now have
\bes{
\left [ H^*_{c} H_c \right ]_{a,b} = \sum^{C}_{k=1} \left[ H_{c} \right]_{k,a} \left[ H_c \right]_{k,b} = C \sum^{C}_{k=1} \delta_{k,1} \delta_{a,c} \delta_{b,c} I_{n} = C \delta_{a,b} \delta_{a,c} I_{n}.
}
Thus, $H^*_c H_c$ is the block diagonal matrix equal to $C I_{n}$ in its $c^{\rth}$ diagonal block and zero elsewhere. 
In particular, $C^{-1} \sum_{c=1}^C H^*_c H_c = I$ which implies that these matrices satisfy the joint isometry property \R{eq:joint_iso_idt} with $\alpha=\beta=1$.  Conversely,
\bes{
\left [  H_c H^*_{c} \right ]_{a,b} = \sum^{C}_{k=1} \left[ H_c \right]_{a,k} \left[ H_c \right]_{b,k} = C \sum^{C}_{k=1} \delta_{a,1} \delta_{k,c} \delta_{b,1} \delta_{k,c} I_{n} = C \delta_{a,1} \delta_{b,1} I_n .
}
Hence, $H_c H^*_c$ is the block-diagonal matrix equal to $C I_n$ in its $(1,1)^{\rth}$ block and zero elsewhere.  Thus
\bes{
\nm{\sum_{c=1}^C H_c H^*_c}_{2 \rightarrow 2} = C^2.
}
To deduce the sharpness of the second bound in \R{Xi:idt:bound2}, we now notice that
\bes{
\Xi_{\mathrm{distinct}} = \left\| \left[ \begin{array}{ccc} H_1 & \cdots & H_C \end{array} \right] \right\|_{2 \rightarrow 2} = \sqrt{\nm{\sum_{c=1}^C H_c H^*_c}_{2 \rightarrow 2}}.
}
Hence, $\Xi_{\mathrm{distinct}} = C$.  Similarly, if $j \in I_d$ then
\bes{
[H_c e_j e^*_j H^*_c]_{a,b} = C \delta_{a,1} \delta_{b,1} \delta_{c,d} e_{j'} e^*_{j'},
}
where $j' = j \mod n$.  Hence
\bes{
\nm{\sum^{C}_{c=1} H_c e_j e^*_j H^*_c}_{2\rightarrow 2} = C \| e_{j'} e^*_{j'} \|_{2 \rightarrow 2} = C, 
}
which gives the result.

\subsection{Proof of Proposition \ref{prop:GammaXi:bounds:idt:2}}
In view of Proposition \ref{prop:GammaXi:bounds:idt} we only need to prove that $\Xi_{\mathrm{identical}} \leq \sqrt{\beta/\alpha} \sqrt{C}$.  Since the $H_c $ are normal, we have 
\bes{
\left\| \left[ \begin{array}{ccc} H_1 & \cdots & H_C \end{array} \right] \right\|^2_{2 \rightarrow 2} = \nm{\sum^{C}_{c=1} H_c H^*_c }_{2 \rightarrow 2} = \nm{\sum^{C}_{c=1} H^*_c H_c }_{2 \rightarrow 2} \leq \sqrt{C} \sqrt{\beta},
}
where in the last inequality we use the joint near-isometry property \R{eq:joint_iso_idt}.  This gives the result.  To show that this bound is sharp we may let $H_c = \sqrt{C} P_{I_c}$, where $I_c$ is as in \R{Ic_def}, as before.

\section{Proof of Lemma \ref{l:RozellCovering}}\label{a:RozellCoveringProof}

For small values of $\nu$, we estimate the covering number using a volumetric argument.  
We first introduce the sets $B_{S} = \left\{ z \in \bbC^N, \| z \|_2 \leq 1, \supp(z) \subset S \right\}$, so that
\bes{
B_{s} = \left\{ z \in \bbC^N : \| z \|_0 \leq s,  \| z \|_2 \leq 1 \right\} =  \bigcup_{ \substack{ S \subset \{ 1,\ldots,N \} \\ |S|=s }} B_S.
}
Let $z \in B_{s}$.  Then
\bes{
\| \cF(z) \|_{2 \rightarrow 2} \leq \frac{\theta}{\sqrt{m}} \| z \|_{1} \leq \theta \sqrt{\frac{s}{m}}.
}
Using subadditivity of the covering numbers and treating the $s$-dimensional complex unit ball as the real $2s$-dimensional unit ball, we obtain
\eas{
\cN(\cA, \| \cdot \|_{2 \rightarrow 2}, \nu) 
&= \cN(B_{s}, \| \cF (\cdot ) \|_{2 \rightarrow 2}, \nu) 
\\
&\leq \sum_{\substack{ S \subset \{1,\ldots,N\} \\ |S|=s}} \cN(B_{S}, \theta \sqrt{s/m} \| \cdot \|_2, \nu)
\\
&\leq {N \choose s} \left(1 + 2 \frac{\theta \sqrt{s/m}}{\nu} \right)^{2s} 
\\
&\leq \left( \frac{e N}{s} \right)^s \left(1 + 2 \frac{\theta \sqrt{s/m}}{\nu} \right)^{2s}.
}
Therefore, 
\bes{
\sqrt{ \ln ( \cN(\cA, \| \cdot \|_{2 \rightarrow 2}, \nu) ) } \leq \sqrt{2s} \left( \sqrt{\ln \! \left( \frac{eN}{s} \right)} + \sqrt{\ln \! \left(1 + 2 \sqrt{\frac{s}{m}} \frac{\theta}{\nu} \right)} \right),
}
which gives the bound for small $\nu >0$.

For large $\nu$, we first require the following:
\lem{[Maurey's lemma]
\label{l:maurey}
Let $X$ be a normed vector space and $U \subset X$ be a set of cardinality $N$, and assume that for every $M \geq 1$ and $u_1,\ldots,u_M \in U$ we have 
\bes{
\bbE \nm{\sum^{M}_{i=1} \varepsilon_i u_i }_{X} \leq A \sqrt{M},
}
where $\{\varepsilon_i \}^{M}_{i=1}$ is a Rademacher sequence.  Then for every $\nu > 0$ we have
\bes{
\ln \left ( \cN(\mathrm{conv}(U) , \nm{\cdot}_{X} , \nu) \right ) \lesssim (A/\nu)^2 \ln(N),
}
where $\mathrm{conv}(U)$ denotes the convex hull of $U$.
}
See, for example, \cite[Lem.\ 4.2]{Krahmer:14CPAM}.  We shall also use the following non-commutative Khintchine inequality (see, for example, \cite[Lem.\ 9]{Eftekhari&etal:15ACHA}):
\lem{[Noncommutative Khintchine inequality]
\label{l:khintchine}
Let $A_1,\ldots,A_M$ be a sequence of matrices of the same dimension and rank at most $r$.  Then
\bes{
\bbE \nm{\sum^{M}_{i=1} \varepsilon_i A_i }_{2 \rightarrow 2} \lesssim \sqrt{\max \{ \ln(r) , 1 \}} \sqrt{\sum^{M}_{i=1} \| A_i \|^2_{2\rightarrow 2} },
}
where $\{\varepsilon_i \}^{M}_{i=1}$ is a Rademacher sequence.
}
Now let
\be{
\label{def:l1_star}
\| z \|_1^* = \sum_{j=1}^N ( | \Re z_j | + | \Im z_j | ), \qquad z \in \bbC^N,
} 
which is the usual $\ell_1$-norm after identification of $\bbC^N$ with $\bbR^{2N}$.  By Cauchy-Schwarz inequality, we have the embedding
\bes{
B_{s} \subset \sqrt{2s} B_{\| \cdot \|_1^*} = \left\{ z \in \bbC^N, \| z \|_1^* \leq \sqrt{2s} \right\},
}
where $B_{\| \cdot \|_1^*} = \{ z \in \bbC^N, \| z \|_1^* = 1 \}$.  Therefore
\eas{
 \cN(B_{s}, \nm{\cdot}_{\cF}, \nu) \leq \cN \left( \sqrt{2s} B_{\| \cdot \|_1^*},  \nm{\cdot}_{\cF}, \nu \right )= \cN \left ( B_{\| \cdot \|_1^*},  \nm{\cdot}_{\cF}, \nu/\sqrt{2s} \right ).
}
We shall now use Maurey's lemma (Lemma \ref{l:maurey}).  Let $\nm{\cdot}_{X} = \nm{\cF(\cdot)}_{2 \rightarrow 2}$ and consider the set
\bes{
U = \{ \pm e_j , \pm \I e_j : j =1,\ldots,N \},
}
so that $\mathrm{conv}(U) = B_{\nm{\cdot}^*_1}$.  Now let $u_1,\ldots,u_M \in U$.  Then
\bes{
\bbE \nm{\sum^{M}_{i=1} \varepsilon_i u_i }_{X} = \bbE \nm{\sum^{M}_{i=1} \varepsilon_i \cF(u_i) }_{2 \rightarrow 2}.
}
To estimate this term we use a non-commutative Khintchine inequality (Lemma \ref{l:khintchine}).  This gives
\bes{
\bbE \nm{\sum^{M}_{i=1} \varepsilon_i \cF(u_i) }_{2 \rightarrow 2} \lesssim \sqrt{\max \{ \ln(m) , 1 \} } \sqrt{\sum^{M}_{i=1} \| \cF(u_i) \|^2_{2\rightarrow 2}  } \leq \frac{\theta\sqrt{\max \{ \ln(m) , 1 \} }  }{\sqrt{m}}\sqrt{M} = A \sqrt{M},
}
where in the penultimate step we use \R{Fnormineq} and the fact that $\|u_i \|_{1} = 1$ since $u_i$ is a canonical vector.  Applying Maurey's lemma with this value of $A$ now gives the bound for large $\nu$.

\section*{Acknowledgements}
The work of IYC at the University of Michigan was supported in part by a W. M. Keck Foundation grant.
BA wishes to acknowledge the support of Alfred P. Sloan Research Foundation and the Natural Sciences and Engineering Research Council of Canada through grant 611675.  BA and IYC both acknowledge the support of the National Science Foundation through DMS grant 1318894.  The authors would like to thank Jeffrey A. Fessler, Felix Krahmer, Richard Kueng, Hassan Mansour, Rayan Saab, and Mike Wakin for useful comments and suggestions.

\bibliographystyle{IEEEtran}
\bibliography{referenceBibs_Bobby}

\end{document}